\documentclass[preprint,notoc,hyper]{JHEP3}
\usepackage{graphicx,etoolbox,indentfirst}
\newcommand{\Ord}{\mathcal{O}}
\def\Ordals#1{\ifstrequal{#1}{1}{\mathcal{O}\left(\alpha_s\right)}{\mathcal{O}\left(\alpha_s^#1\right)}}

\newcommand{\el}{\nonumber \\ &&}

\newcommand{\elale}{\nonumber \\ &=&}

\newcommand{\eqref}[1]{(\ref{#1})}
\newcommand{\text}[1]{{\rm #1}}
\newenvironment{changemargin}[2]{%
  \begin{list}{}{%
    \setlength{\topsep}{0pt}%
    \setlength{\leftmargin}{#1}%
    \setlength{\rightmargin}{#2}%
    \setlength{\listparindent}{\parindent}%
    \setlength{\itemindent}{\parindent}%
    \setlength{\parsep}{\parskip}%
  }%
  \item[]}{\end{list}}
\def\roman#1{\sbox0{#1}\dimen0=\dimexpr\wd0+1pt\relax
  \makebox[\dimen0]{\rlap{\vrule width\dimen0 height 0.06ex depth 0.06ex}%
    \rlap{\vrule width\dimen0 height\dimexpr\ht0+0.03ex\relax 
            depth\dimexpr-\ht0+0.09ex\relax}%
    \kern.5pt#1\kern.5pt}}

\newcommand{\be}{\begin{equation}} 
\newcommand{\ee}{\end{equation}} 
\newcommand{\cmb}{\begin{changemargin}}
\newcommand{\cme}{\end{changemargin}}
\newcommand{\bea}{\begin{eqnarray}} 
\newcommand{\eea}{\end{eqnarray}} 
\newcommand{\D}{\delta}

\newcommand{\G}{\Gamma}

\newcommand{\tauo}{\tau_\omega}

\newcommand{\dd}{\mathrm{d}}
\newcommand{\als}{\alpha_s}
\newcommand{\ep}{\epsilon}

\def\spa#1.#2{\langle#1\,#2\rangle}
\def\spb#1.#2{[#1\,#2]}
\def\sandmm#1.#2.#3{%
\left\langle\smash{#1}{\rphantom1}\right|{#2}%
\left|\smash{#3}{\rphantom1}\right]}
\def\spab#1.#2.#3{\sandmm#1.#2.#3}
\def\spba#1.#2.#3{\sandpp#1.#2.#3}
\def\spaa#1.#2.#3.#4{\sandmp#1.{#2#3}.#4}
\def\spbb#1.#2.#3.#4{\sandpm#1.{#2#3}.#4}
\def\spash#1.#2{\spa{\smash{#1}}.{\smash{#2}}}
\def\spbsh#1.#2{\spb{\smash{#1}}.{\smash{#2}}}

\def\ksl{\not{\hbox{\kern-2.3pt $k$}}}

\def\e{\epsilon}

\def\Ord{{\cal O}}

\def\li#1{{\mathop{\rm Li}\nolimits}_#1}

\newcommand{\sym}{\mathcal{S}}
\newcommand{\cp}{\Delta}

\preprint{MITP/13-048, SLAC-PUB-15689}
\title{The Complete Two-Loop Integrated Jet Thrust Distribution In Soft-Collinear Effective Theory}

\author{Andreas von Manteuffel and Robert M. Schabinger\\
The PRISMA Cluster of Excellence and\\
Mainz Institute of Theoretical Physics\\
Johannes Gutenberg Universit\"{a}t,\\
55099 Mainz, Deutschland}
\author{Hua Xing Zhu\\
SLAC National Accelerator Laboratory,\\
Stanford University,\\
Stanford, CA 94309, USA}
        
\abstract{
In this work, we complete the calculation of the soft part of the two-loop integrated jet thrust distribution in $e^+ e^-$ annihilation.
This jet mass observable is based on the thrust cone jet algorithm, which involves
a veto scale for out--of--jet radiation.
The previously uncomputed part of our result depends in a complicated way on the jet cone size, $r$, and at intermediate stages of
the calculation we actually encounter a new class of multiple polylogarithms.
We employ an extension of the coproduct calculus to systematically exploit functional relations
and represent our results concisely.
In contrast to the individual contributions, the sum of all global terms can be
expressed in terms of classical polylogarithms.
Our explicit two-loop calculation enables us to clarify the small $r$ picture
discussed in earlier work.
In particular, we show that the resummation of the logarithms of $r$ that appear
in the previously uncomputed part of the two-loop integrated jet thrust distribution is inextricably linked to the resummation of the non-global logarithms. 
Furthermore, we find that the logarithms of $r$ which cannot be absorbed into the non-global logarithms in the way advocated in earlier work have coefficients fixed by the two-loop cusp anomalous dimension.
We also show that in many cases one can straightforwardly predict potentially large logarithmic contributions to the integrated jet thrust distribution at $L$ loops
by making use of analogous contributions to the simpler integrated hemisphere soft function.
}

\begin{document}
\tableofcontents
\newpage

\section{Introduction} 

\label{sec:intro}
\begin{figure}[t]
\begin{center}
  \includegraphics[width=0.7\textwidth]{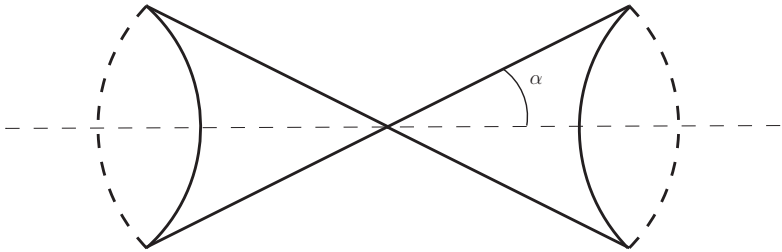}
  \caption{The thrust cone jet algorithm used to define the $\tauo$ observable enforces dijet kinematics by imposing a veto on the out-of-jet radiation with
  total energy greater than $\omega$.
  The cone size, $r$, is defined to be $\tan^2\left(\alpha\over 2\right)$.}
\end{center}
\end{figure}

Jet physics has been of considerable importance since the early days of Quantum Chromodynamics (QCD) and the advent of the Large Hadron Collider has led to a renewed interest
in the subject. The study of jet mass distributions has recently
been a particularly hot topic, as evidenced by the sheer number of noteworthy publications over the last few years, in both
QCD and soft-collinear effective theory~(SCET) (see~{\it e.g.}~\cite{Stewart:2010tn,Kelley:2010qs,Banfi:2010xy,
Monni:2011gb,Li:2012bw,Dasgupta:2012hg,Chien:2012ur,Kelley:2012kj,Gallicchio:2012ez,Ferroglia:2012uy,Becher:2012za,Jouttenus:2013hs}).
Another contribution of particular interest to us is the recent study~\cite{Kelley:2011aa} of the two-loop jet mass
distribution by some of us and others.
In that work, a soft gluon resummation study of a hadron collider-applicable jet algorithm is performed,
without the additional complications of initial state radiation and non-trivial color correlations. This was accomplished by using an inclusive jet algorithm with
a veto on the out-of-jet radiation to study the resummation properties of an appropriate observable at, however, an $e^+ e^-$ collider. 

The jet algorithm employed in \cite{Kelley:2011aa} is similar to the Sterman-Weinberg approach~\cite{Sterman:1977wj}.
As depicted in Figure 1, jets are defined by two cones of size $r$ with a common vertex at the collision point and a common axis determined by the thrust axis in the event.
The radiation in the out-of-jet region is controlled by vetoing on events which have total out-of-jet energy greater than $\omega$, the veto scale. Using this thrust cone (TC) jet algorithm~\cite{Almeida:1982em}, it is
possible to define a more exclusive analog of the thrust event shape called the jet thrust~\cite{Kelley:2011tj}.
This observable is a dijet $e^+ e^-$ event shape by virtue 
of the fact that, in any given event, the thrust cone jet algorithm ensures that exactly two jets are defined. 
The jet thrust
\begin{equation}
\label{eq:jetomega}
\tau_\omega = \frac{M^2_1+ M^2_2}{Q^2}\,\Theta(\omega-\lambda)
\end{equation}
is zero for events which fail the jet veto and
the normalized sum of the squared jet masses, $M_1^2$ and $M_2^2$, otherwise.
Here, $\lambda$ is the total energy of the out-of-jet radiation in the event and $Q$ is the center of mass energy.
Without the $\Theta$ function this definition of jet thrust reproduces the
formula for hemisphere thrust, which in turn coincides with standard thrust
in the endpoint region.
Many collider observables are defined in terms of particular limits of jet mass and our expectation is that a detailed study of the resummation properties of the $\tau_\omega$
observable will shed light on an entire class of collider observables, of which jet thrust is just one example. 

By construction, the jet thrust distribution $\dd \sigma / \dd \tau_\omega$ has nice factorization properties in the endpoint region and one can write down a factorization formula 
for it~\cite{Kelley:2011tj}, completely analogous to the factorization formula for thrust~\cite{Catani:1992ua,Schwartz:2007ib,Fleming:2007qr,Becher:2008cf},
\begin{eqnarray}
  \label{eq:factauo1}
  \frac{1}{\sigma_0}\frac{\dd \sigma}{\dd \tau_\omega}
 &=& H\left(Q^2,\mu\right) \int \! \dd k_L\, 
\dd k_R\, \dd M_L^2\, \dd M_R^2\,
J(M^2_L - Q k_L,\mu) J(M^2_R-Q k_R,\mu) 
\el
\times
S_{TC}(k_L,k_R,\omega,r,\mu)\delta\left( \tau_\omega - \frac{M^2_L + M^2_R}{Q^2} \right)\,,
\end{eqnarray}
up to power corrections of the general form $\tau_\omega^i \ln^j(r)$ and $(\omega/Q)^i \ln^j(r)$, where $i \geq 1$ and $j \geq 0$.\footnote{In the language of SCET, both $\tau_\omega$ and $\omega/Q$ are soft:
$\tau_\omega \sim \omega/Q \sim \lambda^2$. Given the form of the power corrections to the factorization theorem, we see that $r$ should not be exponentially small relative to $\tau_\omega$ and $\omega/Q$.
This condition will always be satisfied in practice and, as was noted in reference [13], one can consistently interpolate between narrow and fat (even hemisphere) jets by varying $r$.}
The integration is over the soft momenta, $k_L$ and $k_R$, and the squared jet masses, $M_L^2$ and $M_R^2$,
for the left and right jets. The normalization is with respect to the total Born-level cross section $\sigma_0$ and the factorization scale is given by $\mu$.
Physically the factorization formula indicates that the singular jet thrust distribution 
is a convolution of collinear radiation, described by the jet function\footnote{We use the fully-inclusive jet function (see Appendix \ref{app:refeqs} for further details) because this is justified in 
the endpoint region for measured jets~\cite{Ellis:2010rwa}. Roughly speaking, the $r$ dependence of the measured jet function is expected to enter through finite-angle radiation which is power-suppressed in the collinear approximation.
However, it is worth pointing out that algorithm-dependent jet functions have $r$ dependence (see {\it e.g.}~\cite{Ellis:2010rwa,Jouttenus:2009ns,Ellis:2009su}) which shows up in terms
non-singular in the event shape.}, $J(q^2, \mu)$, and soft radiation, described by the soft function, $S_{TC}(k_L,k_R,\omega,r,\mu)$.
The virtual corrections, encoded in the hard function, $H(Q^2, \mu)$, normalize the distribution. 

The main conceptual difference between the hemisphere soft function,  $S_{hemi}(k_L,k_R,\mu)$, appearing in the factorization formula
for hemisphere thrust\footnote{In references~\cite{Kelley:2011ng}
and~\cite{Hornig:2011iu} the computation of the $\Ordals 2$ hemisphere soft function is discussed in detail.} and 
the thrust cone soft function, $S_{TC}(k_L,k_R,\omega,r,\mu)$, is that there is no out-of-jet region to complicate the calculation in the hemisphere case. A natural way to impose the
jet veto is to define an auxiliary soft function valid for fixed out-of-jet energy. One can then calculate
$S_{TC}(k_L,k_R,\omega,r,\mu)$ by first computing the auxiliary soft
function and then integrating $\lambda$ from zero to $\omega$ at the end. In fact, the authors of reference \cite{Kelley:2011aa} employed the modified factorization formula
\begin{eqnarray}
  \label{eq:factauo2}
  \frac{1}{\sigma_0}\frac{\dd \sigma}{\dd \tau_\omega}
 &=& H\left(Q^2,\mu\right) \int \! \dd k_L\, 
\dd k_R\, \dd M_L^2\, \dd M_R^2\,
J(M^2_L - Q k_L,\mu) J(M^2_R-Q k_R,\mu)
\el
\times
\int^\omega_0 \! \dd \lambda \,
S(k_L,k_R,\lambda,r,\mu)
\delta\left( \tau_\omega - \frac{M^2_L + M^2_R}{Q^2} \right)
\end{eqnarray}
to set up their partial calculation of the soft contributions to the integrated jet thrust distribution.
In the limit $r \to 1$ the thrust cone soft function approaches the hemisphere
soft function.
Together with the known results for the hemisphere case, this provides
us with a useful cross check of our results.

Although the operator definition of $S(k_L,k_R,\lambda,r,\mu)$ (Eq.\ (2.5) of reference~\cite{Kelley:2011aa})
looks very similar to the operator definition of $S_{hemi}(k_L,k_R,\mu)$ used in reference~\cite{Kelley:2011ng} (Eq.\ (9) of that work), $S(k_L,k_R,\lambda,r,\mu)$ 
is considerably more complicated due to the fact that it depends on the additional scale $\lambda$ and the jet size $r$. 
Actually, there is another technical
problem that makes the resummation properties of $S(k_L,k_R,\lambda,r,\mu)$ difficult to study. The issue is that $S(k_L,k_R,\lambda,r,\mu)$ 
depends in a complicated way on singular distributions. It turns out to be much easier in practice
to calculate the $\Ord(\als^2)$ soft contributions to the \emph{integrated} jet thrust distribution
and only then attempt to study the result. Following~\cite{Kelley:2011aa}, we focus in this work on the calculation of the $\Ordals 2$ term in the perturbative expansion of
the integrated jet thrust soft function for the color structures which first appear at this order in perturbation theory,  $C_F n_f T_F$ and $C_A C_F$. 
This is sensible because we expect the potentially large logarithms in the $C_F^2$ color structure to be controlled by the non-Abelian exponentiation theorem~\cite{Gatheral:1983cz,Frenkel:1984pz}
and the two-loop hard and integrated jet functions have already been calculated~\cite{Gonsalves:1983nq,Kramer:1986sg,Matsuura:1987wt,Becher:2006qw} (see Appendix \ref{app:refeqs} for explicit formulae). Concretely,
we use the relations
\be
\label{eq:treehardandjet}
H\left(Q^2,\mu\right) = 1 + \Ord(\als) \qquad {\rm and} \qquad J\left(q^2,\mu\right) = \D\left(q^2\right) + \Ord(\als)
\ee
and obtain
\begin{eqnarray}
  \label{eq:intdist1}
&&K_{TC}^{(2)}(\tauo,\omega,r,\mu) \equiv
  \int_0^{\tauo} d\tauo^\prime \frac{1}{\sigma_0}\frac{\dd \sigma^{(2)}}{\dd \tau_\omega}\left(\tauo^\prime\right)\bigg|_{\rm soft}\nonumber
\\
&&\qquad=\left({\als\over 4\pi}\right)^2 \int_0^{\tauo} \dd\tauo^\prime \int \! \dd k_L\, 
\dd k_R
\int^\omega_0 \! \dd \lambda \,
S^{(2)}(k_L,k_R,\lambda,r,\mu)
\delta\left( \tau_\omega^\prime - \frac{k_L + k_R}{Q} \right).
\end{eqnarray}
for the soft part of the two-loop integrated jet thrust distribution. In the first line of Eq.\ (\ref{eq:intdist1}), $\frac{1}{\sigma_0}\frac{\dd \sigma^{(2)}}{\dd \tau_\omega}(\tau_\omega)$ is the two-loop part of the perturbative jet thrust
distribution. In what follows, we will often speak about the analogous $L$-loop construction. In such cases, $K_{TC}^{(L)}(\tauo,\omega,r,\mu)$ is understood
to be the soft part of the $L$-loop integrated jet thrust distribution.

Proceeding in this fashion makes it clear why the resummation of the jet thrust observable is significantly more complicated than the resummation of 
ordinary thrust. In the soft part of the integrated thrust distribution, one finds potentially large logarithms
of the form $\ln\left({\mu\over \tau Q}\right)$, all of which can be resummed in the framework of SCET.
On the other hand, in the soft part of the integrated jet thrust distribution, one finds potentially large logarithms of the
form $\ln\left({\mu\over \tauo Q}\right)$, $\ln\left({\tauo Q \over 2 \omega}\right)$, and $\ln(r)$.\footnote{It is worth pointing out that, actually, there is another issue which we do not address. 
As is well-known, QCD has an intrinsic, dynamically-generated scale, $\Lambda_{\rm QCD}$, which gives rise to effects which cannot be studied in QCD perturbation theory. In this work, we assume that
the non-perturbative effects are power suppressed under refactorization of the soft function. Further analysis of this assumption is beyond the scope of the present paper but should be carried out in future work.}
The $\mu$-dependent logarithms are controlled by the SCET factorization theorem for jet thrust, Eq.\ (\ref{eq:factauo2}) above. 
However, it is not clear that the logarithms of the ratio ${\tauo Q \over 2 \omega}$ can be resummed analytically. In fact, it has been known for more than a decade that such so-called non-global
logarithms are present in a large number of collider observables~\cite{Dasgupta:2001sh,Dasgupta:2002bw,Banfi:2002hw,Appleby:2002ke}. Non-global logarithms typically arise from
the presence of sharp boundaries between regions of phase space with different characteristic energy scales and their study
has been a topic of recent interest~\cite{Kelley:2011tj,Kelley:2011ng,
Hornig:2011iu,KhelifaKerfa:2011zu,Hornig:2011tg,Banfi:2010pa}. So far, using numerical methods, only the resummation of the leading non-global logarithms has been performed~\cite{Dasgupta:2001sh,Hatta:2013iba}.

The complete $\mathcal{O}(\alpha^2_s)$ soft non-global structure\footnote{The non-global structure of the integrated $\tauo$ distribution is the complete set of terms
which depend on ${\tauo Q \over 2 \omega}$ and are not controlled by the SCET factorization theorem, Eq.\ (\ref{eq:factauo1}). There are a large number of such terms in the integrated $\tauo$ distribution
which vanish in the limits $\tauo Q \gg 2 \omega$ and $\tauo Q  \ll 2 \omega$.}
of the integrated $\tauo$ distribution was calculated in \cite{Kelley:2011aa}
and the non-global logarithms were identified analytically for generic $r$.
The paper presents the first complete analysis of non-global logarithms in the integrated distribution of a jet mass observable
defined using a realistic inclusive jet algorithm with a jet veto on the out-of-jet radiation.
Although this analysis is interesting in its own right,
the authors' decision to neglect terms in the integrated distribution which depend only on $r$ or on no parameters at all in some cases inhibited their ability to 
arrive at definitive conclusions.
In this paper we pick up where \cite{Kelley:2011aa} left off and, with the aim of clarifying the structure of the potentially large logarithms of $r$, complete
their calculation of the soft contributions to the integrated $\tauo$ distribution. 

Unfortunately, while completing the calculation of \cite{Kelley:2011aa} is in principle straightforward given a thorough understanding of the original work,
it requires modern techniques to write the final result 
in a compact way in terms of known, linearly independent, transcendental functions.
We perform our analysis in terms of multiple polylogarithms~\cite{Goncharov1,Gehrmann:2000zt}.
These functions are iterated integrals of the type
$G(w_1,\dots,w_n;x) = \int_0^x {dt \over t - w_1} G(w_2,\dots,w_n;t)$
and may depend on multiple variables.
The dependency on external variables enters both via the integration boundary
and via the integrating factors themselves. It is mostly for this reason that multiple polylogarithms typically satisfy a large number of functional relations. 
In fact, in all but the simplest cases, a systematic, computer algebra-based, treatment of these
functional identities is indispensable. The symbol calculus allows for algorithms \cite{Duhr:2011zq},
which become particularly powerful in their coproduct-based extension \cite{Goncharov2,Brown:2011ik,Duhr:2012fh}.
The power of the coproduct formalism has already been proven in many applications, in both QCD and $\mathcal{N} = 4$
super Yang-Mills
theory~\cite{Goncharov:2010jf,Dixon:2011pw,Dixon:2011nj,Dixon:2012yy,Drummond:2012bg,Chavez:2012kn,vonManteuffel:2012je,Gehrmann:2013vga,Anastasiou:2013srw,Golden:2013xva,vonManteuffel:2013uoa,Gehrmann:2013cxs,Henn:2013woa,Dixon:2013eka}.
We employ a variant of the coproduct formalism to remove redundancies in analytical expressions.

At intermediate stages of the calculation, new functions appear which are best described by a slight generalization of the standard multiple polylogarithms usually discussed in the literature.
While, in the standard case, the integration measure is $\mathrm{d}t/(t-w)$ for some complex parameter $w$, we also consider denominators of higher degree in the variable of integration.
In this work, we restrict ourselves to the univariate case, where the integration measures are independent of further variables.
A subset of such polylogarithms are the cyclotomic polylogarithms~\cite{Ablinger:2011te,vonManteuffel:2013uoa}, where the denominators are cyclotomic polynomials.
In our analysis, we also need other types of polynomials.
However, we find that it is possible to work entirely in terms of functions which have rational symbols.
The generalized multiple polylogarithms we consider have integration measures of the form $\mathrm{d}t f'(t)/f(t) = \mathrm{d}\ln \left(f(t)\right)$, where $f(t)$ is an irreducible rational polynomial.

The modern techniques discussed above allow one to reveal structure in the results
by systematically eliminating redundancies.
In particular, we observe that a number of branch cut singularities cancel in a non-trivial way at various stages of the calculation.
For the global contributions to the integrated $\tauo$ distribution, from each soft phase space region, separately, we find that all generalized multiple polylogarithms drop
out and that each individual result can be expressed in terms of harmonic polylogarithms~\cite{Remiddi:1999ew} alone.
The sum of all contributions is structurally even simpler and can be expressed in a compact way in terms of classical polylogarithms.

An interesting universality property in the small $r$ limit was observed in \cite{Kelley:2011aa} from a partial
calculation of the integrated $\tauo$ distribution.
In a precise way, it was shown that the non-global structure present in the integrated $\tauo$
distribution can be very accurately modeled using the integrated hemisphere soft function for small but realistic values of the jet cone size. Although the authors of that work noted that it seems
natural to modify the definition of the non-global logarithms in the small $r$ limit and consider potentially large logarithms of the form
$\ln\left({\tauo Q\over 2 r \omega}\right)$, they were unable to provide more than circumstantial evidence for this picture. The limiting factor, of course, was
their systematic neglect of terms independent of all dimensionful scales. It is not obvious, for
example, that there are not ``dangling'' logarithms of $r$ that cannot be naturally absorbed
into the non-global logarithms, thus potentially spoiling the simple picture advocated in that work. It is interesting to clarify the structure of the $\ln(r)$ terms
that dominate in this limit and our explicit two-loop calculation makes it straightforward to do so. In fact, we will show that, although dangling $\ln(r)$ terms do appear
in the full $\Ord\left(\als^2\right)$ result, their presence might not be an issue due to the fact that their coefficients are
simply $- \Gamma_1$, where, as usual $\Gamma_{L-1}$ denotes the $\Ordals L$ term in the perturbative expansion of the cusp anomalous dimension.

Actually, we can learn a great deal
about the $\ln(r)$ terms without doing any calculations beyond those necessary to understand the integrated thrust distribution defined using the hemisphere jet algorithm. The 
most striking new example of this phenomenon concerns the contributions to the integrated $\tauo$ distribution from 
the region of phase space where a maximal number of soft partons (equal to the loop order, $L$) are outside of all jets (what we will call the all-out contributions).
We will show that, to all orders in perturbation theory,
one can derive an exact relationship between the $\ln(r)$ terms in the all-out contributions and the contributions to the integrated hemisphere soft function
where a maximal number of soft partons go into the same hemisphere (what we will call the pure
same-hemisphere contributions). Furthermore, as a by-product of our analysis, 
we prove two relationships conjectured in \cite{Kelley:2011aa} to all orders in perturbation theory: one between the $\Ordals L$ contributions to the integrated $\tauo$ distribution where $L$
soft partons get clustered into the same jet (what we will call the all-in contributions) and the pure same-hemisphere contributions, and another between the in-out contributions to the integrated $\tauo$
distribution and contributions to the integrated hemisphere soft function where $n_{in}$ soft partons go into one hemisphere and $n_{out}$ soft partons go into the other.

This article is organized as follows. Section \ref{sec:secdec} revisits the calculation of the integrated $\tauo$ distribution performed in Section 3 and Appendix A of \cite{Kelley:2011aa}.
In several cases, we are able to derive the desired results with no non-trivial calculation whatsoever by building on some of the ideas
discussed in \cite{Kelley:2011aa}. Whenever the opportunity presents itself, we generalize the discussion to $L$ loops. For the contributions that require non-trivial analysis, we describe the sector decomposition~\cite{Binoth:2000ps}
of the integrals over the event shape and out-of-jet energy in some detail.
In Section \ref{sec:coprod} we explain how the contributions obtained from the sector decompositions are integrated in terms of multiple polylogarithms,
and we define a new class of multiple polylogarithms needed for the analysis.
We explain in detail a variant of the coproduct calculus
which we use to systematically exploit functional relations between multiple polylogarithms and project onto specific basis functions.
Without significant loss of continuity, readers less interested in the technical details of our calculation may safely skip Sections \ref{sec:oppinin} through \ref{sec:gencoprodcalc}.
In Section \ref{sec:niceres}, we present our final, simplified, result for the soft part of the $\Ord\left(\als^2\right)$ integrated $\tauo$ distribution.

In Section \ref{sec:smallR}, we study the small $r$ limit of the soft part of the $\Ord\left(\als^2\right)$ integrated $\tauo$
distribution from several points of view.
We begin by straightforwardly taking the small $r$ limit for the individual contributions.
This analysis makes it readily apparent that, at $\Ordals2$, the $\ln(r)$ terms in the all-out contributions
can be derived from the all-in contributions by simply making the replacements $r \rightarrow 1/r$ and $\tauo Q \rightarrow 2 \omega$. 
We then show that this observed correspondence can be easily explained and generalized to all orders in perturbation theory. Furthermore, we note that 
the observed correspondence extends naturally
to one between the $\ln(r)$ terms in the all-out contributions to the integrated $\tauo$ distribution 
and the pure same-hemisphere contributions to the integrated hemisphere soft function. 
Next, we study the small $r$ picture advocated in \cite{Kelley:2011aa}. In previous work it was suggested that there is a preferred way to take the small $r$ limit of the integrated distribution.
We show that, in fact, working in the way suggested in reference \cite{Kelley:2011aa} makes it easy to see the structure of the potentially large logarithms of $r$. In the small $r$ limit,
we find that the $\ln(r)$ terms can either be naturally absorbed into the non-global logarithms or are $\ln^2(r)$ times minus the $\Ordals2$ term in the perturbative expansion of the cusp anomalous dimension.
In Section \ref{sec:conclusions}, we present our conclusions and outlook.

In Appendix \ref{app:chi}, we collect two lengthy contributions to the non-global part of the integrated distribution
which would have been inconvenient to define in the text where they first appear.
Appendix \ref{app:indcontrib} contains our results for the contributions to the integrated distribution
from the individual soft phase space regions.
Finally, in Appendix \ref{app:refeqs}, we collect certain useful $\Ordals 1$ and $\Ordals2$ formulae
for the convenience of the reader.

\section{Derivation of the Soft Part of the $\Ord\left(\als^2\right)$ Integrated $\tauo$ Distribution}
\label{sec:secdec}

\subsection{The General Structure of the Calculation}
\begin{figure}[!htp]
\begin{center}
  \includegraphics[width=\textwidth]{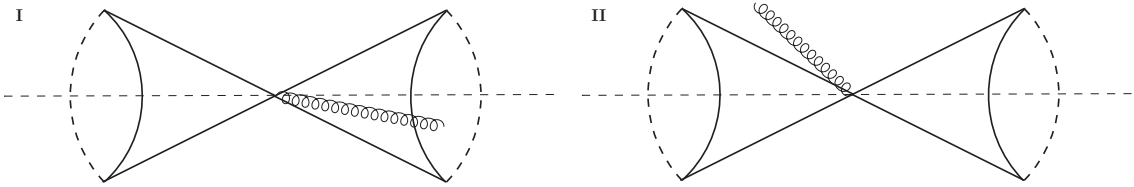}
  \caption{Final state phase space configurations with a single soft parton. Panel {I} shows a single soft gluon going into the right jet and panel {II} shows a single soft gluon going out of all jets.}
\end{center}
\end{figure}
\begin{figure}[!htp]
\begin{center}
  \includegraphics[width=\textwidth,height=\textwidth,keepaspectratio]{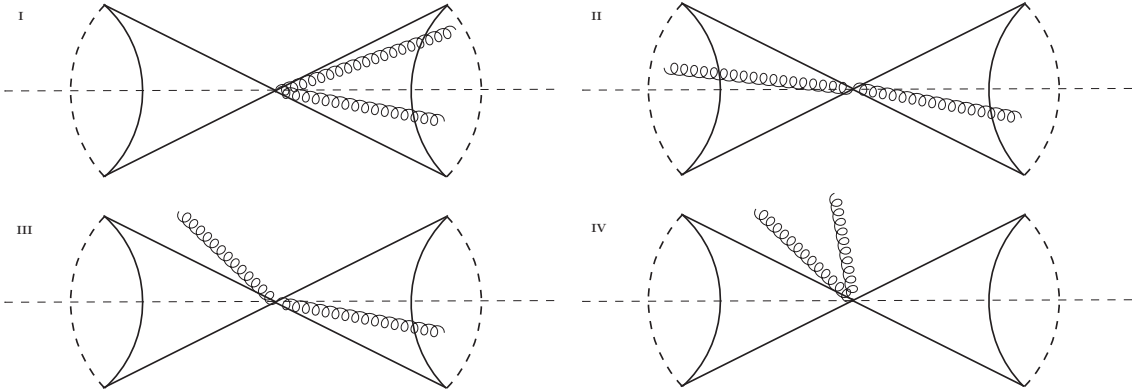}
  \caption{Final state phase space configurations containing two soft partons. Panel {I} shows two soft gluons going into the right jet, panel {II} shows one soft gluon going into the right jet and another going into the left jet,
  panel {III} shows one soft gluon going into the right jet and one soft gluon going out of all jets, and, finally, panel {IV} shows two soft gluons going out of all jets. In the above, it is in all cases possible
  to replace the two soft gluons with a soft quark-antiquark pair.}
\end{center}
\end{figure}
In this section, we briefly outline our calculation.  For further details, we refer the interested 
reader to reference~\cite{Kelley:2011aa}. In all, there are eight distinct contributions to the finite part of the $\Ordals 2$ integrated distribution
which must be considered. Let us begin with the contributions where a single soft gluon crosses
the final state cut. There are four such single-parton contributions (see Figure 2):
\begin{enumerate}
 \item {\rm The in-jet charge renormalization contributions, where a
 single soft gluon goes into a jet and the charge renormalization constant is expanded to $\Ordals 1$.}
 \item {\rm The in-jet real-virtual contributions, where a
 single soft gluon radiated off the gluon line in the one-loop vertex correction goes into a jet.}
 \item {\rm The out-of-jet charge renormalization contributions, where a
 single soft gluon goes out of all jets and the charge renormalization constant is expanded to $\Ordals 1$.}
 \item {\rm The out-of-jet real-virtual contributions, where a
 single soft gluon radiated off the gluon line in the one-loop vertex correction goes out of all jets.}
\end{enumerate}
These contributions are all essentially trivial since, as explained in Section \ref{sec:trivial}, they can easily be derived from available all-orders-in-$\ep$ one-loop results. 

The contributions with two soft partons in the final state are, for the most part, significantly more complicated. Once again, there are four independent contributions to deal with (see Figure 3):
\begin{enumerate}
 \item {\rm The same-side in-in contributions, also referred to in this work as the $\Ordals 2$ all-in contributions, where both soft partons go into the same jet.}
 \item {\rm The opposite-side in-in contributions, where one soft parton goes into the ${\bf n}$ jet and the other soft parton goes into the ${\bf \bar{n}}$ jet.}
 \item {\rm The in-out contributions, where one soft parton goes into a jet and the other soft parton goes out of all jets.}
 \item {\rm The out-out contributions, also referred to in this work as the $\Ordals 2$ all-out contributions, where both soft partons go out of all jets.}
\end{enumerate}
The $\Ordals 2$ all-in contributions, are special in that, as explained in Section \ref{sec:allinpredict},
they are precisely related to the pure same-hemisphere contributions to the integrated thrust distribution. The opposite-side in-in, in-out, and $\Ordals 2$ all-out contributions require non-trivial
analysis and will be discussed in Sections \ref{sec:oppinin}, \ref{sec:inout}, and \ref{sec:outout} respectively.

The goal of Section~\ref{sec:secdec} is to set up and prepare explicit expressions for all required contributions.
The subsequent integrations in terms of multiple polylogarithms will be discussed in Section~\ref{sec:coprod}.

\subsection{Trivial Contributions}
\label{sec:trivial}
In this section, we discuss those contributions to the soft part of the $\Ord\left(\als^2\right)$ integrated $\tauo$ distribution which can be derived from known results with no
non-trivial calculation. These contributions can either be exactly related to pieces of the integrated hemisphere soft function or are easily obtained from known all-orders-in-$\ep$
one-loop results. In the first, all-in, category, we have the $\Ordals2$
real-real contributions where two soft partons get clustered into the same jet, in-jet 
real-virtual interference contributions where a single soft gluon gets clustered into a jet, and the in-jet charge renormalization contributions, which are nothing but the
$\Ordals1$ in-jet contributions multiplied by the $\Ordals1$ term in the expansion of the charge renormalization constant. In the second, out-of-jet, category, we have the out-of-jet
real-virtual interference contributions where a single soft gluon goes into the out-of-jet region and the out-of-jet charge renormalization contributions, which are simply the 
$\Ordals1$ out-of-jet contributions multiplied by the $\Ordals1$ term in the expansion of the charge renormalization constant. 

\subsubsection{All-In Contributions}
\label{sec:allinpredict}
In this section, we will show how to treat all of the trivial contributions in the all-in category
by first identifying and then 
exploiting an exact relation between the all-in contributions at $\Ord\left(\als^L\right)$ and the pure same-hemisphere contributions to the integrated thrust distribution at
$\Ord\left(\als^L\right)$. Let us begin by determining the exact $k_L$ and $k_R$ dependence of the pure right-hemisphere real-real part of the $L$-loop hemisphere soft function,
\begin{eqnarray}
\label{eq:hemisoft1}
{\mathop{S}^{\Rightarrow}}^{\,(L)}_{hemi}(k_L,k_R,\mu)
&=&
\left( \frac{\mu^2 e^{\gamma_E}}{4\pi}\right)^{\!{4 - d \over 2} L}\!(-2\pi i)^L
\prod_{i = 1}^L \left(\int\frac{\dd^d q_i}{(2\pi)^d} \delta\left(q_i^2\right)
\Theta\left(q^+_i\right)\Theta\left(q^-_i\right) \Theta(q_i^- - q_i^+)\right)
\el
\times I^{(L)}\left(q_i^+,q_i^-,{\bf q}_T^{(i)}\right) \delta\left(k_L\right)\delta\left(k_R - \sum_{i = 1}^L q_i^+\right) 
\elale
\left( \frac{\mu^2 e^{\gamma_E}}{4\pi}\right)^{\ep L}(-\pi i)^L \prod_{i = 1}^L 
\left(\int \dd q_i^- \dd q_i^+ \Theta\left(q^+_i\right)\Theta\left(q^-_i\right) \Theta(q_i^- - q_i^+)\right) 
\el
\times
\prod_{i = 1}^L \left({1\over 2}\int\frac{\dd |{\bf q}_T^{(i)}|^2}{(2\pi)^{4 - 2 \ep}} \left(|{\bf q}_T^{(i)}|^2\right)^{-\ep} \delta\left(q_i^- q_i^+ - |{\bf q}_T^{(i)}|^2\right)
\right) \int \dd {\bf \Omega}_\ep
\el
\times I^{(L)}\left(q_i^+,q_i^-,|{\bf q}_T^{(i)}|^2,{\bf \Omega}\right) \delta\left(k_L\right)\delta\left(k_R - \sum_{i = 1}^L q_i^+\right) \,,
\end{eqnarray}
where $\dd {\bf \Omega}_\ep$ represents the angular measure for the $L(L-1)/2$ integrations over the angles specifying the relative orientations of the ${\bf q}_T^{(i)}$
and, as usual, $S^{(L)}_{hemi}(k_L,k_R,\mu)$ is understood to be the coefficient of $\left(\als\over 4\pi\right)^{L}$ in the perturbative expansion of $S_{hemi}(k_L,k_R,\mu)$.
The second line of Eq.\ (\ref{eq:hemisoft1}) makes it clear that we can use the $\delta\left(q_i^2\right)$
to do the integrals over the $|{\bf q}_T^{(i)}|^2$ and make the functional dependence on the light cone components of the $q_i$ manifest. We have
\begin{eqnarray}
\label{eq:hemisoft2}
{\mathop{S}^{\Rightarrow}}^{\,(L)}_{hemi}(k_L,k_R,\mu)
&=&
\left( \frac{\mu^2 e^{\gamma_E}}{4\pi}\right)^{\ep L}(-\pi i)^L \prod_{i = 1}^L 
\left(\int \dd q_i^- \dd q_i^+ \Theta\left(q^+_i\right)\Theta\left(q^-_i\right) \Theta(q_i^- - q_i^+)\right) 
\el
\hspace{-5ex}
\times\prod_{i = 1}^L \left(\frac{ \left(q_i^- q_i^+\right)^{-\ep}}{2 (2\pi)^{4 - 2 \ep}}\right) \int \dd {\bf \Omega}_\ep
I^{(L)}\left(q_i^+,q_i^-,{\bf \Omega}\right) \delta\left(k_L\right)\delta\left(k_R - \sum_{i = 1}^L q_i^+\right) .
\end{eqnarray}
At this stage, we can simply scale $k_R$ out of the remaining integrals by making the change of variables
\be
q_i^+ = p_i^+ k_R \qquad {\rm and} \qquad q_i^- = p_i^- k_R\,.
\ee
This will work because the exact $q_i^+$ and $q_i^-$ dependence of $I^{(L)}\left(q_i^+,q_i^-,{\bf \Omega}\right)$ is unimportant; we are rescaling
all of the light cone coordinates so all that matters is that 
$I^{(L)}\left(q_i^+,q_i^-,{\bf \Omega}\right)$ has a fixed mass dimension. In fact, it is easy to see that
\be
I^{(L)}\left(q_i^+,q_i^-,{\bf \Omega}\right) = k_R^{-2 L} I^{(L)}\left(p_i^+,p_i^-,{\bf \Omega}\right)
\ee
by analyzing the SCET Feynman rules. Altogether, we find that
\begin{eqnarray}
\label{eq:hemisoft3}
{\mathop{S}^{\Rightarrow}}^{\,(L)}_{hemi}(k_L,k_R,\mu)
&=&
\left( \frac{\mu^2 e^{\gamma_E}}{4\pi}\right)^{\ep L}\frac{(-\pi i)^L}{k_R^{1+2 \ep L}}
\prod_{i = 1}^L 
\left(\int \dd p_i^- \dd p_i^+ \Theta\left(p^+_i\right)\Theta\left(p^-_i\right) \Theta(p_i^- - p_i^+)\right) 
\el
\times\prod_{i = 1}^L \left(\frac{\left(p_i^- p_i^+\right)^{-\ep}}{2 (2\pi)^{4 - 2 \ep}} \right) \int \dd {\bf \Omega}_\ep
I^{(L)}\left(p_i^+,p_i^-,{\bf \Omega}\right) \delta\left(k_L\right)\delta\left(1 - \sum_{i = 1}^L p_i^+\right) 
\elale
k_R^{-1-2 \ep L}\D\left(k_L\right)\mu^{2 \ep L}g_{hemi}^{(L)}(\ep)\,,
\end{eqnarray}
where the function $g_{hemi}^{(L)}(\ep)$ is defined by the last equation.
Of course, the $n$-$\bar{n}$ symmetry immediately implies that the pure left-hemisphere real-real part of the $L$-loop hemisphere soft function can be obtained from Eq.\ (\ref{eq:hemisoft3})
by exchanging $k_L$ and $k_R$. 
Finally, for later reference, we write down the pure $\Ord\left(\als^L\right)$ same-hemisphere contributions
to the integrated thrust distribution. By symmetry, we have
\begin{eqnarray}
\label{eq:inthemi1}
{\mathop{K}^{\Rightarrow}}^{\,(L)}_{hemi}(\tau,\mu) + {\mathop{K}^{\Leftarrow}}^{\,(L)}_{hemi}(\tau,\mu)
&=& 2 \int_0^\tau \dd \tau^\prime \int \dd k_L \dd k_R \left(k_R^{-1-2 \ep L}\D\left(k_L\right)\mu^{2 \ep L}g_{hemi}^{(L)}(\ep)\right)
\el
\times\D\left(\tau^\prime - {k_L + k_R\over Q}\right)\,.
\end{eqnarray}

Now, it is a simple consequence of the geometry depicted in Figure 1 that the plane separating the two hemispheres used to define the hemisphere soft function can
be mapped by a Lorentz boost along the thrust axis onto one of the cones used to define the thrust cone soft function (see Figure 4). To see this, 
consider the action of such a boost on spacetime points which lie on the plane 
perpendicular to the thrust axis at the collision point\footnote{This is exactly the coordinate system which is implicitly used to
define both the hemisphere and thrust cone jet algorithms and their associated soft functions.}. In other words,
\bea
\label{eq:boost}
\left(\begin{array}{c} x_c^0 \\ x_c^{thr} \end{array}\right) &=& \left(\begin{array}{cc} \cosh y  & -\sinh y \\ -\sinh y & \cosh y \end{array}\right)
\left(\begin{array}{c} x_h^0 \\ 0 \end{array}\right)
=\left(\begin{array}{c} \cosh y \\ -\sinh y \end{array}\right)x_h^0\,,
\eea
for some $y$. The fact that this is possible is not particularly surprising given our setup and, in fact,
similar reasoning has been successfully applied in several other collider physics studies (see {\it e.g.}~\cite{Marchesini:2003nh,
Stewart:2009yx}).
Fortunately, it turns out that the rapidity is a simple function of the 
jet radius. From Figure 1 and the definition of rapidity, we
see that
\be
\label{eq:rap}
y = -\ln\left(\tan\left({\alpha \over 2}\right)\right) = \ln\left({1\over \sqrt{r}}\right)\,.
\ee
Together, Eqs.\ (\ref{eq:boost}) and (\ref{eq:rap}) determine a transformation of the light cone coordinates such that
\be
\label{eq:trans1}
{\mathop{S}^{\Rightarrow}}^{\,(L)}_{TC}(k_L^\prime,k_R^\prime,\omega,r,\mu) = {\mathop{S}^{\Rightarrow}}^{\,(L)}_{hemi}\left(k_L(k_L^\prime,r),k_R(k_R^\prime,r),\mu\right)\,,
\ee
where $\displaystyle {\mathop{S}^{\Rightarrow}}^{\,(L)}_{TC}(k_L^\prime,k_R^\prime,\omega,r,\mu)$ is the pure right-jet part of the $L$-loop thrust cone soft function. 
\begin{figure}[t]
\begin{center}
  \includegraphics[width=0.8\textwidth]{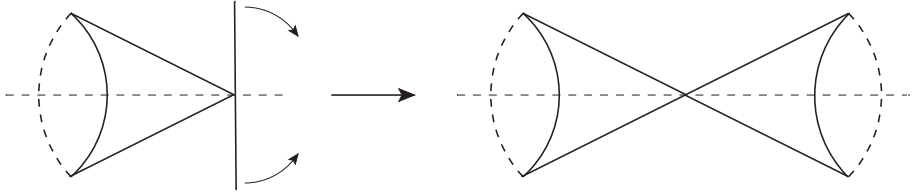}
  \caption{A hemisphere boundary can be mapped onto a cone boundary by choosing an appropriate Lorentz boost.}
\end{center}
\end{figure}
Explicitly, we have
\be
k_R^\prime = \left(\cosh y - \sinh y\right) k_R = e^{-y} k_R \qquad {\rm and} \qquad k_L^\prime = \left(\cosh y + \sinh y\right) k_L = e^y k_L\,,
\ee
from which it follows that
\be
\label{eqs:krkltrans}
k_R(k_R^\prime,r) = {k_R^\prime\over \sqrt{r}} \qquad {\rm and} \qquad k_L(k_L^\prime,r) = \sqrt{r} k_L^\prime\,.
\ee
Plugging Eqs.\ (\ref{eqs:krkltrans}) into Eq.\ (\ref{eq:trans1}) gives
\be
\label{eq:trans2}
{\mathop{S}^{\Rightarrow}}^{\,(L)}_{TC}(k_L^\prime,k_R^\prime,\omega,r,\mu) = {\mathop{S}^{\Rightarrow}}^{\,(L)}_{hemi}\left(\sqrt{r} k_L^\prime,{k_R^\prime\over \sqrt{r}},\mu\right)
\,.
\ee
Finally, we can derive an expression for the $\Ord\left(\als^L\right)$ all-in contributions to the integrated $\tauo$ distribution in terms of the $\Ord\left(\als^L\right)$ 
pure same-hemisphere contributions to the integrated thrust distribution. 
From Eqs.\ (\ref{eq:hemisoft3}), (\ref{eq:inthemi1}), and (\ref{eq:trans2}) we see that
\begin{eqnarray}
\label{eq:intcone}
K^{{\rm all-in}\,(L)}_{TC}(\tauo,\omega,r,\mu)
&=& 2 \int_0^{\tauo} \dd \tauo^\prime \int \dd k_L \dd k_R {\mathop{S}^{\Rightarrow}}^{\,(L)}_{TC}(k_L,k_R,\omega,r,\mu)\D\left(\tauo^\prime - {k_L + k_R\over Q}\right)
\el\hspace{-15ex}=
2 \int_0^{\tauo} \dd \tauo^\prime \int \dd k_L \dd k_R \left(\left({k_R \over \sqrt{r}}\right)^{-1-2 \ep L}\D(\sqrt{r} k_L)\mu^{2 \ep L} g_{hemi}^{(L)}(\ep)\right)
\D\left(\tauo^\prime - {k_L + k_R\over Q}\right)
\el\hspace{-15ex}=
2 r^{\ep L} \int_0^{\tauo} \dd \tauo^\prime \int \dd k_L^\prime \dd k_R \left(k_R^{-1-2 \ep L}\D(k_L^\prime)\mu^{2 \ep L}g_{hemi}^{(L)}(\ep)\right)
\D\left(\tauo^\prime - {k_L^\prime + k_R\over Q}\right)
\el\hspace{-15ex}=
r^{\ep L} \left({\mathop{K}^{\Rightarrow}}^{\,(L)}_{hemi}(\tauo,\mu) + {\mathop{K}^{\Leftarrow}}^{\,(L)}_{hemi}(\tauo,\mu)\right)
\,.
\end{eqnarray}
This analysis provides a simple explanation for the known rescaling properties\footnote{For $L = 1~{\rm or}~2$,
it was observed in reference~\cite{Kelley:2011aa} that $I^{(L)}\left(q_i^+,q_i^-,{\bf \Omega}\right)$ maps to $r^L I^{(L)}\left(q_i^+,q_i^-,{\bf \Omega}\right)$ under the rescaling
$q_i^- \rightarrow q_i^-/r$.} of \\$I^{(1)}\left(q_i^+,q_i^-,{\bf \Omega}\right)$ and
$I^{(2)}\left(q_i^+,q_i^-,{\bf \Omega}\right)$. Even better, it is now clear that the observed trend will persist to all orders in perturbation theory.

For our present purposes, we will only need to make use of Eq.\ (\ref{eq:intcone}) for $L = 1$ and $L = 2$. In fact, 
for $L = 1$, Eq.\ (\ref{eq:intcone}) immediately implies that the in-in
charge renormalization contributions are given by
\begin{eqnarray}
\label{eq:intconerenin1}
K^{{\rm in\,Ren}}_{TC}(\tauo,\omega,r,\mu)
&=& -{\beta_0 \over \ep} \left({\mathop{K}^{\Rightarrow}}^{\,(1)}_{hemi}(\tauo,\mu) + {\mathop{K}^{\Leftarrow}}^{\,(1)}_{hemi}(\tauo,\mu)\right) r^\ep
\el\hspace{-19.5ex}=
-{2\left(\frac{11}{3}C_A - \frac{4}{3} n_f T_F\right)r^\ep \over \ep}
\int_0^{\tauo} \dd \tauo^\prime \int \dd k_L \dd k_R \,k_R^{-1-2 \ep}\D\left(k_L\right)\mu^{2 \ep}g_{hemi}^{(1)}(\ep)
\D\left(\tauo^\prime - {k_L + k_R\over Q}\right)
\el\hspace{-19.5ex}=
-{2\left(\frac{11}{3}C_A - \frac{4}{3} n_f T_F\right)r^\ep \over \ep} \left(-{1\over 2\ep}\left({\mu \over \tauo Q}\right)^{2\ep}\right)
\left(\frac{4 C_F e^{\gamma_E \e}}{\ep \,\G(1-\e)}\right)\,,
\end{eqnarray}
where $\beta_0=\frac{11}{3}C_A - \frac{4}{3} n_f T_F$ is the leading-order $\beta$-function and we have used Eq.\ (3.8) of reference~\cite{Kelley:2011aa} 
(see also Eq.\ (A13) of reference~\cite{Fleming:2007xt}) in deriving
the third line of Eq.\ (\ref{eq:intconerenin1}).

In the same vein, for $L = 2$, 
Eq.\ (\ref{eq:intcone}) immediately implies that the real-real same-side in-in contributions are given by
\bea
\label{eq:intconeallinals2}
K^{{\rm all-in}\,(2)}_{TC}(\tauo,\omega,r,\mu)
&=&  \left({\mathop{K}^{\Rightarrow}}^{\,(2)}_{hemi}(\tauo,\mu) + {\mathop{K}^{\Leftarrow}}^{\,(2)}_{hemi}(\tauo,\mu)\right) r^{2 \ep}
\el\hspace{-21ex} =
{2 r^{2\ep}}
\int_0^{\tauo} \dd \tauo^\prime \int \dd k_L \dd k_R \,k_R^{-1-4 \ep}\D\left(k_L\right)\mu^{4 \ep}
g_{hemi}^{(2)}(\ep)\D\left(\tauo^\prime - {k_L + k_R\over Q}\right)
\el\hspace{-21ex} =
2 r^{2\ep} {(-1)\over 4\ep}\left({\mu \over \tauo Q}\right)^{4\ep}
\Bigg[C_A C_F\Bigg(
\frac{4}{\e^3}+\frac{22}{3\e^2} +\left(\frac{134}{9}-\frac{4\pi^2}{3}
\right)\frac{1}{\e}
+ \frac{772}{27}+\frac{11\pi^2}{9}-{116 \zeta_3\over 3}
\el\hspace{-21ex}
+ \ep\left( \frac{4784}{81} + \frac{67\pi^2}{27} - \frac{137\pi^4}{90}
+ \frac{484\zeta_3}{9} \right) + \Ord\left(\e^2\right)\!\Bigg)
\el\hspace{-21ex}
+ C_F n_f T_F
\Bigg(\! -\frac{8}{3\e^2} - \frac{40}{9\e} - \frac{152}{27} - \frac{4\pi^2}{9} - \e \left(\frac{952}{81} + \frac{20\pi^2}{27} + \frac{176\zeta_3}{9}\right) + \Ord\left(\e^2\right)\!\Bigg)\Bigg],
\eea
where we have used Eq.\ (3.16) of reference~\cite{Kelley:2011aa}  in deriving the third line of Eq.\ (\ref{eq:intconeallinals2}).

At first sight, it is not obvious that Eq.\ (\ref{eq:intcone}) 
has any relevance to the in-jet real-virtual contributions because it was obviously derived with the all-in $\Ordals L$ real-real contributions in mind. 
However, after examining the line of reasoning that led to Eq.\ (\ref{eq:intcone}), it becomes clear that the
argument goes through with minor modifications for the all-in $\Ordals 2$ real-virtual contributions, where a gluon is radiated off the gluon line in the one-loop virtual correction. This time, we can write
\begin{eqnarray}
\label{eq:intconervin1}
K^{{\rm in\,R-V}}_{TC}(\tauo,\omega,r,\mu)
&=& -2 \cos(\pi \ep) \G^2(\e)\G^2(1-\ep) C_A 
\left({\mathop{K}^{\Rightarrow}}^{\,(1)}_{hemi}(\tauo,\mu) + {\mathop{K}^{\Leftarrow}}^{\,(1)}_{hemi}(\tauo,\mu)\right)_{\ep \rightarrow 2\ep} r^{2 \ep}
\el\hspace{-12ex}=
-4 r^{2 \ep}\cos(\pi \ep) \G^2(\e)\G^2(1-\ep) C_A \int_0^{\tauo} \dd \tauo^\prime \int \dd k_L \dd k_R
 k_R^{-1-4 \ep}\D\left(k_L\right)\mu^{4 \ep}g_{hemi}^{(1)}(2\ep)
\el\hspace{-12ex}\quad\times
\D\left(\tauo^\prime - {k_L + k_R\over Q}\right)
\el\hspace{-12ex}=
-4 r^{2 \ep}\cos(\pi \ep) \G^2(\e)\G^2(1-\ep) C_A
\left(-{1\over 4\ep}\left({\mu \over \tauo Q}\right)^{4\ep}\right) \left(\frac{4 C_F e^{2\gamma_E \e}}{2\ep \,\G(1-2\e)}\right)\,,
\end{eqnarray}
where we have used a simple all-orders-in-$\ep$ relationship to derive the result (compare the epsilon expansion of Eq.\ (20) in reference~\cite{Kelley:2011ng} to Eq.\ (\ref{eq:intconervin2}) below). 

\subsubsection{Out-Of-Jet Contributions}
In this section, we will treat all of the trivial contributions in the out-of-jet category. For both the out-of-jet charge renormalization and out-of-jet
real-virtual interference contributions, we proceed by recognizing that the first lines of Eqs.\ (\ref{eq:intconerenin1}) and (\ref{eq:intconervin1}) have a very distinctive structure. Both the in-jet charge renormalization
and in-jet real-virtual interference contributions are equal to the in-jet $\Ordals 1$ integrated $\tauo$ distribution times 
prefactors {\it independent} of the final state phase space cuts. In fact, the out-of-jet charge renormalization and out-of-jet
real-virtual interference contributions have essentially the same structure; all one needs to do is replace the in-jet $\Ordals 1$ integrated $\tauo$ distribution with the out-of-jet $\Ordals 1$ integrated $\tauo$ distribution.

Of course, it is obvious that the out-of-jet charge renormalization contributions can be treated in this way. We can immediately write
\begin{eqnarray}
\label{eq:intconerenout1}
\hspace{-4ex}K^{{\rm out\,Ren}}_{TC}(\tauo,\omega,r,\mu)
&=& -{\beta_0 \over \ep} K^{{\rm all-out}\,(1)}_{TC}(\tauo,\omega,r,\mu)
\el\hspace{-21ex}=
-{\left(\frac{11}{3}C_A - \frac{4}{3} n_f T_F\right) \over \ep}
\int_0^{\tauo}\! \dd \tauo^\prime \int\! \dd k_L \dd k_R \int_0^\omega\!\! \dd \lambda
\left(\frac{8 C_F e^{\gamma_E \e}\mu^{2\ep}}{(1-\ep)\ep\,\G(1-\e) (1+r)}\right)
\left(\frac{1+r}{r}\right)^{\ep}
\el\hspace{-21ex}
\Bigg[r^\ep
\bigg( \ep\, _{2}F_1\left(1-\ep,1+\ep,2-\ep;\frac{1}{1+r}\right)
+ (\ep-1)\left(1+r\right)\, _2F_1\left( -\ep,\ep,1-\ep;\frac{1}{1+r}\right)\!\bigg)
\el\hspace{-21ex}
-\bigg( \ep\, r\, _2F_1\left(1-\ep,1+\ep,2-\ep;\frac{r}{1+r}\right)
+ (\ep-1)(1+r)\, _2F_1\left(-\ep,\ep,1-\ep;\frac{r}{1+r}\right)\!\bigg)\Bigg]
\el\hspace{-21ex}
 (2\lambda)^{-1-2\ep} \D\left(k_L\right)\D(k_R)\D\left(\tauo^\prime - {k_L + k_R\over Q}\right)
\el\hspace{-21ex}=
-{\left(\frac{11}{3}C_A - \frac{4}{3} n_f T_F\right) \over \ep}
{(-1)\over 4 \ep}\left({\mu \over 2 \omega}\right)^{2\ep}
 \left( 
\frac{8 C_F e^{\gamma_E \e}}{(1-\ep)\ep\,\G(1-\e) (1+r)}\right)
\left( \frac{1+r}{r}\right)^{\ep}
\el\hspace{-21ex}
\Bigg[r^\ep
\bigg( \ep\, _{2}F_1\left(1-\ep,1+\ep,2-\ep;\frac{1}{1+r}\right)
+ (\ep-1)\left(1+r\right)\, _2F_1\left( -\ep,\ep,1-\ep;\frac{1}{1+r}\right)\!\bigg)
\el\hspace{-21ex}
  -
\bigg( \ep\, r\, _2F_1\left(1-\ep,1+\ep,2-\ep;\frac{r}{1+r}\right)
+ (\ep-1)(1+r)\, _2F_1\left(-\ep,\ep,1-\ep;\frac{r}{1+r}\right)\!\bigg)\Bigg],
\end{eqnarray}
where we have used Eq.\ (3.9) of reference~\cite{Kelley:2011aa} in deriving
the second line of Eq.\ (\ref{eq:intconerenout1}).

In a similar fashion, we can derive an expression for the $\Ordals 2$ out-of-jet real-virtual interference. We find
\begin{eqnarray}
\label{eq:intconervout1}
K^{{\rm out\,R-V}}_{TC}(\tauo,\omega,r,\mu)
&=& -2 \cos(\pi \ep) \G^2(\ep)\G^2(1-\ep) C_A K^{{\rm all-out}\,(1)}_{TC}(\tauo,\omega,r,\mu)_{\ep \rightarrow 2\ep}
\el\hspace{-23ex}
=-2 \cos(\pi \ep) \G^2(\ep)\G^2(1-\ep) C_A \left(-{1\over 8 \ep}\left({\mu \over 2 \omega}\right)^{4\ep}\right)
\left( 
\frac{8 C_F e^{2\gamma_E \ep}\mu^{4\ep}}{(1-2\ep)2\ep\,\G(1-2\ep) (1+r)}\right)
\left( \frac{1+r}{r}\right)^{2\ep}
\el\hspace{-23ex}
\Bigg[r^{2\ep}
\bigg( 2\ep\, _{2}F_1\!\left(\!1-2\ep,1+2\ep,2-2\ep;\frac{1}{1+r}\right)
+ (2\ep-1)\left(1+r\right)\, _2F_1\!\left(\! -2\ep,2\ep,1-2\ep;\frac{1}{1+r}\right)\!\bigg)
\el\hspace{-23ex}
- \bigg( 2\ep\, r\, _2F_1\!\left(\!1-2\ep,1+2\ep,2-2\ep;\frac{r}{1+r}\right)
+ (2\ep-1)(1+r)\, _2F_1\!\left(\!-2\ep,2\ep,1-2\ep;\frac{r}{1+r}\right)\!\bigg)\Bigg].
\nonumber\\[-0.2ex] &&
\end{eqnarray}

\subsection{Opposite-Side In-In Contributions}
\label{sec:oppinin}
In this section we discuss the calculation of the contributions to the integrated jet thrust distribution where
one soft parton gets clustered into the left jet and one soft parton gets clustered into the right jet. The relevant part of the auxiliary soft function, $S^{{\rm opp\,in-in}}(k_L,k_R,\lambda,r,\mu)$,
can be expressed as
\bea
S^{{\rm opp\,in-in}}(k_L,k_R,\lambda,r,\mu) &=&
\frac{\mu^{4\e}}{(k_L k_R)^{1+2\ep}}\,\delta(\lambda)
\left[
C_A C_F\,\mbox{\large $f$}_{C_A}\!\left(\!\frac{k_L}{k_R},r\!\right)
+ C_F n_f T_F\,\mbox{\large $f$}_{n_f}\!\left(\!\frac{k_L}{k_R},r\!\right)
\right]\!,
\nonumber\\[-0.2ex]&&
\eea
where $\mbox{\large $f$}_{C_A}\!\left(z,r\right)$ and $\mbox{\large $f$}_{n_f}\!\left(z,r\right)$ are functions with $\ep$ expansions that begin at $\Ord\left(\e^0\right)$. In other words,
\bea
\mbox{\large $f$}_{C_A}\!\left(z,r\right) &=& \mbox{\large $f$}_{C_A}^{(0)}\left(z,r\right) + \mbox{\large $f$}_{C_A}^{(1)}\left(z,r\right) \e + \mbox{\large $f$}_{C_A}^{(2)}\left(z,r\right) \e^2 + \Ord\left(\e^3\right)
\nonumber\\
\mbox{\large $f$}_{n_f}\!\left(z,r\right) &=& \mbox{\large $f$}_{n_f}^{(0)}\left(z,r\right) + \mbox{\large $f$}_{n_f}^{(1)}\left(z,r\right) \e + \Ord\left(\e^2\right)\,.
\label{eq:fmoms}
\eea

Let us now analyze the opposite-side in-in contributions to the integrated distribution,
\bea
K_{TC}^{{\rm opp\,in-in}}(\tauo,\omega,r,\mu) &=& \int_0^{\tauo} \dd \tauo^\prime \int \dd k_L \dd k_R \int_0^\omega \dd \lambda \frac{\mu^{4\e}}{(k_L k_R)^{1+2\ep}}\,\delta(\lambda)
\el\hspace{-5ex}\quad\times
\Bigg[C_A C_F\,\mbox{\large $f$}_{C_A}\!\left(\frac{k_L}{k_R},r\right)+ C_F n_f T_F\,\mbox{\large $f$}_{n_f}\!\left(\frac{k_L}{k_R},r\right)\Bigg]\D\left(\tauo^\prime - {k_L + k_R\over Q}\right)
\el\hspace{-5ex}
=\int_0^{\tauo Q} \dd x \int_0^x \dd k_L \frac{\mu^{4\e}}{(k_L (x - k_L))^{1+2\ep}}
\el\hspace{-5ex}\quad\times
\Bigg[C_A C_F\,\mbox{\large $f$}_{C_A}\!\left(\frac{k_L}{x - k_L},r\right)+ C_F n_f T_F\,\mbox{\large $f$}_{n_f}\!\left(\frac{k_L}{x - k_L},r\right)\Bigg]\,.
\eea
To start, note that we can factorize the iterated integral by making the change of variables $k_L = x y$:
\bea
\hspace{-4ex}
K^{{\rm opp\,in-in}}(\tauo,\omega,r,\mu) &=& \int_0^{\tauo Q} \dd x \frac{\mu^{4\e}}{x^{1+4\ep}}\int_0^1 \dd y \frac{1}{(y (1 - y))^{1+2\ep}}
\el
\times\Bigg[C_A C_F\,\mbox{\large $f$}_{C_A}\!\left(\frac{y}{1 - y},r\right)+ C_F n_f T_F\,\mbox{\large $f$}_{n_f}\!\left(\frac{y}{1 - y},r\right)\Bigg]
\,.
\eea
The $x$ integral is trivial and the integral over $y$ can be easily sector decomposed. In this case, it makes sense to split the $y$ integral at $y = 1/2$ because,
due to the fact that $\mbox{\large $f$}(z,r) = \mbox{\large $f$}(1/z,r)$, the two integrals that result will actually be equal:
\bea
\hspace{-3ex}K_{TC}^{{\rm opp\,in-in}}(\tauo,\omega,r,\mu) &=&
\left(-{1\over 4 \ep}\left({\mu \over \tauo Q}\right)^{4\ep}\right) 2 \int_0^{1/2} \dd y \frac{1}{(y (1 - y))^{1+2\ep}}
\el
\times
\Bigg[C_A C_F\,\mbox{\large $f$}_{C_A}\!\left(\frac{y}{1 - y},r\right)+ C_F n_f T_F\,\mbox{\large $f$}_{n_f}\!\left(\frac{y}{1 - y},r\right)\Bigg]
\,.
\eea

At this stage, all of the phase space singularities have been mapped to zero and we can simply shift the integration to the
unit interval ($y = 1/2 z$) and then expand the factor $z^{-1-2\ep}$ under the integral sign using the relation
\bea
\label{eq:secdecrel}
\int_0^1 \dd z {p(z)\over z^{1 + n \ep}} &=& -{p(0)\over n \ep} + \int_0^1 \dd z {p(z)-p(0)\over z} - n \ep \int_0^1 \dd z {p(z)-p(0)\over z}\ln(z) 
\el
+  {n^2\ep^2\over 2!}\int_0^1 \dd z {p(z)-p(0)\over z}\ln^2(z) + \Ord\left(\e^3\right)
\eea
for $n = 2$. Going through these steps and expanding in $\ep$, we find a contribution of 
\bea
\bar{K}_{TC}^{{\rm opp\,in-in}}(\tauo,\omega,r,\mu) &=& C_A C_F
\left[\vphantom{\Bigg[}\right.{\mbox{\large $f$}_{C_A}^{(2)}\big(0,r\big)\over 4}
-{1\over 2} \int_0^1  {\dd z\over z}\Bigg({2\mbox{\large $f$}_{C_A}^{(1)}\big({z \over 2-z},r\big)\over 2-z} - \mbox{\large $f$}_{C_A}^{(1)}(0,r)\Bigg) 
\el\hspace{-23ex}
+ {\mbox{\large $f$}_{C_A}^{(1)}\big(0,r\big)\ln(2)\over 2}
- \int_0^1  {\dd z\over z}\Bigg({2\mbox{\large $f$}_{C_A}^{(0)}\big({z \over 2-z},r\big) \ln\left({4\over 2-z}\right)\over 2-z}
- \mbox{\large $f$}_{C_A}^{(0)}\big(0,r\big)\ln(2)\Bigg)
+ {\mbox{\large $f$}_{C_A}^{(0)}\big(0,r\big)\ln^2(2)\over 2}
\el\hspace{-23ex}
+ \int_0^1  \dd z {\ln(z) \over z}\Bigg({2\mbox{\large $f$}_{C_A}^{(0)}\big({z \over 2-z},r\big)\over 2-z} - \mbox{\large $f$}_{C_A}^{(0)}\big(0,r\big)\Bigg) 
+ \mbox{\large $f$}_{C_A}^{(1)}\big(0,r\big)\ln \left(\!\frac{\mu }{\tau_\omega Q}\!\right)
+2 \mbox{\large $f$}_{C_A}^{(0)}\big(0,r\big)\ln^2 \left(\!\frac{\mu }{\tau_\omega Q}\!\right)
\el\hspace{-23ex}
-2\ln \left(\!\frac{\mu }{\tau_\omega Q}\!\right)\int_0^1  {\dd z\over z}\left({2\mbox{\large $f$}_{C_A}^{(0)}\big({z \over 2-z},r\big)\over 2-z} - \mbox{\large $f$}_{C_A}^{(0)}\big(0,r\big)\right) 
+ 2 \mbox{\large $f$}_{C_A}^{(0)}\big(0,r\big)\ln (2)\ln \left(\!\frac{\mu }{\tau_\omega Q}\!\right)\Bigg]
\el\hspace{-23ex}
+C_F n_f T_F\Bigg[-{1\over 2} \int_0^1  {\dd z\over z}{2\mbox{\large $f$}_{n_f}^{(1)}\big({z \over 2-z},r\big)\over 2-z}
- \int_0^1  {\dd z\over z}{2\mbox{\large $f$}_{n_f}^{(0)}\big({z \over 2-z},r\big) \ln\left({4\over 2-z}\right)\over 2-z}
\el\hspace{-23ex}
+\int_0^1  \dd z {\ln(z) \over z}{2\mbox{\large $f$}_{n_f}^{(0)}\big({z \over 2-z},r\big)\over 2-z}
-2\ln \left(\!\frac{\mu }{\tau_\omega Q}\!\right)\int_0^1  {\dd z\over z}{2\mbox{\large $f$}_{n_f}^{(0)}\big({z \over 2-z},r\big)\over 2-z}
\left.\vphantom{\Bigg[}\right]
\label{eq:secdecompinin}
\eea
to the finite part of the integrated $\tauo$ distribution.
We have also made use of the fact that $\mbox{\large $f$}_{n_f}(0,r) = 0$.
Here and in the following we use the symbol $\bar{K}_{TC}$ to denote the finite part of $K_{TC}$.

\subsection{In-Out Contributions}
\label{sec:inout}
In this section we discuss the calculation of the contributions to the integrated jet thrust distribution where
one soft parton gets clustered into either the left jet or the right jet and the other soft parton is out of all jets. The relevant part of the auxiliary soft function, $S^{{\rm in-out}\,(2)}(k_L,k_R,\lambda,r,\mu)$,
can be expressed as
\bea
S^{{\rm in-out}\,(2)}(k_L,k_R,\lambda,r,\mu) &=&
\frac{\mu^{4\ep}}{(2\lambda)^{1+2\ep}}\Bigg\{\frac{\delta\left(k_L\right)}{k_R^{1+2\ep}}
\left[ C_A C_F\, \mbox{\large $g$}_{C_A}\!\left(\frac{k_R}{2\lambda},r\right)
+C_F n_f T_F\, \mbox{\large $g$}_{n_f}\!\left(\frac{k_R}{2\lambda},r\right)\right]
\el\quad
+ (k_L \leftrightarrow k_R)\Big\},
\eea
where $\mbox{\large $g$}_{C_A}\!\left(z,r\right)$ and $\mbox{\large $g$}_{n_f}\!\left(z,r\right)$ are functions with $\ep$ expansions that begin at $\Ord\left(\e^0\right)$. In other words,
\bea
\mbox{\large $g$}_{C_A}\!\left(z,r\right) &=& \mbox{\large $g$}_{C_A}^{(0)}\left(z,r\right) + \mbox{\large $g$}_{C_A}^{(1)}\left(z,r\right) \e + \mbox{\large $g$}_{C_A}^{(2)}\left(z,r\right) \e^2 + \Ord\left(\e^3\right)
\\ \nonumber
\mbox{\large $g$}_{n_f}\!\left(z,r\right) &=& \mbox{\large $g$}_{n_f}^{(0)}\left(z,r\right) + \mbox{\large $g$}_{n_f}^{(1)}\left(z,r\right) \e + \Ord\left(\e^2\right)\,.
\label{eq:gmoms}
\eea

Let us now analyze the in-out contributions to the integrated distribution,
\bea
K_{TC}^{{\rm in-out}\,(2)}(\tauo,\omega,r,\mu) &=& 2 \int_0^{\tauo} \dd \tauo^\prime \int \dd k_L \dd k_R \int_0^\omega \dd \lambda \frac{\mu^{4\ep}}{(2\lambda)^{1+2\ep}}\frac{\delta\left(k_L\right)}{k_R^{1+2\ep}}
\el\hspace{-20ex}
\quad\times
\left[ C_A C_F\, \mbox{\large $g$}_{C_A}\!\left(\frac{k_R}{2\lambda},r\right)+C_F n_f T_F\, \mbox{\large $g$}_{n_f}\!\left(\frac{k_R}{2\lambda},r\right)\right]\D\left(\tauo^\prime - {k_L + k_R\over Q}\right)
\el\hspace{-20ex}
= \int_0^{2\omega} \dd y \int_0^{\tauo Q} \dd x \frac{\mu^{4\e}}{(x y)^{1+2\ep}}
\left[ C_A C_F\, \mbox{\large $g$}_{C_A}\!\left(\frac{x}{y},r\right)+C_F n_f T_F\, \mbox{\large $g$}_{n_f}\!\left(\frac{x}{y},r\right)\right].
\eea
As a first step, we split the $x$ integration at $x = y$:
\bea
\label{eq:inoutsetup1}
K_{TC}^{{\rm in-out}\,(2)}(\tauo,\omega,r,\mu) &=&  \int_0^{2\omega}\! \dd y \int_0^{y}\! \dd x \frac{\mu^{4\e}}{(x y)^{1+2\ep}}
\left[ C_A C_F\, \mbox{\large $g$}_{C_A}\!\left(\frac{x}{y},r\right)+C_F n_f T_F\, \mbox{\large $g$}_{n_f}\!\left(\frac{x}{y},r\right)\right]
\el\hspace{-18ex}
+ \int_0^{2\omega} \dd y \int_y^{\tauo Q} \dd x \frac{\mu^{4\e}}{(x y)^{1+2\ep}}
\left[ C_A C_F\, \mbox{\large $g$}_{C_A}\!\left(\frac{x}{y},r\right)+C_F n_f T_F\, \mbox{\large $g$}_{n_f}\!\left(\frac{x}{y},r\right)\right]\,.
\eea
Now, reverse the order of integration in the second integral on the right-hand side of Eq.\ (\ref{eq:inoutsetup1}). This gives
\bea
\label{eq:inoutsetup2}
K_{TC}^{{\rm in-out}\,(2)}(\tauo,\omega,r,\mu) &=&  \int_0^{2\omega}\! \dd y \int_0^{y}\! \dd x \frac{\mu^{4\e}}{(x y)^{1+2\ep}}
\left[ C_A C_F\, \mbox{\large $g$}_{C_A}\!\left(\frac{x}{y},r\right)+C_F n_f T_F\, \mbox{\large $g$}_{n_f}\!\left(\frac{x}{y},r\right)\right]
\el\hspace{-16ex}
+ \int_0^{2\omega}\! \dd x \int_0^x\! \dd y \frac{\mu^{4\e}}{(x y)^{1+2\ep}}
\left[ C_A C_F\, \mbox{\large $g$}_{C_A}\!\left(\frac{x}{y},r\right)+C_F n_f T_F\, \mbox{\large $g$}_{n_f}\!\left(\frac{x}{y},r\right)\right]
\el\hspace{-16ex}
+ \int_{2\omega}^{\tauo Q}\! \dd x \int_0^{2\omega}\! \dd y \frac{\mu^{4\e}}{(x y)^{1+2\ep}}
\left[ C_A C_F\, \mbox{\large $g$}_{C_A}\!\left(\frac{x}{y},r\right)+C_F n_f T_F\, \mbox{\large $g$}_{n_f}\!\left(\frac{x}{y},r\right)\right]\,.
\eea
At this point, we found it useful to make a number of variable changes in the integrals on the right-hand side of Eq.\ (\ref{eq:inoutsetup2}). Let $x = y z_1$ in the first integral, $y = x z_1$ in the second,
and $y = 2 \omega z_1$ and $x = 2 \omega z_2$ in the third.
Carrying out these steps, we find that 
\bea
\label{eq:inoutsetup3}
K_{TC}^{{\rm in-out}\,(2)}(\tauo,\omega,r,\mu) &=&
-{1\over 4\ep} \left(\frac{\mu}{2\omega}\right)^{4\e}\int_0^1 \frac{\dd z_1}{z_1^{1+2\ep}}
\Bigg[C_A C_F\,\mbox{\large $g$}_{C_A}\!\left(z_1,r\right) + C_F n_f T_F\,\mbox{\large $g$}_{n_f}\!\left(z_1,r\right)\Bigg]
\el\hspace{-23ex}\quad
-{1\over 4\ep} \left(\frac{\mu}{2\omega}\right)^{4\e} \int_0^1 \frac{\dd z_1}{z_1^{1+2\ep}}
\Bigg[C_A C_F\,\mbox{\large $g$}_{C_A}\!\left(\frac{1}{z_1},r\right) + C_F n_f T_F\,\mbox{\large $g$}_{n_f}\!\left(\frac{1}{z_1},r\right)\Bigg]
\el\hspace{-23ex}\quad
+ \left({\mu \over 2\omega}\right)^{4\ep}\int_{1}^{{\tauo Q\over 2\omega}} {\dd z_2\over z_2^{1+2\ep}} \int_0^1 {\dd z_1\over z_1^{1+2\ep}}
\Bigg[C_A C_F\,\mbox{\large $g$}_{C_A}\!\left(\frac{z_2}{z_1},r\right) + C_F n_f T_F\,\mbox{\large $g$}_{n_f}\!\left(\frac{z_2}{z_1},r\right)\Bigg].
\eea

Now that all of the remaining phase space singularities in the above have been mapped to $z_1 = 0$, we can straightforwardly expand the factor $z_1^{-1-2\ep}$ in each term 
on the right-hand side of Eq.\ (\ref{eq:inoutsetup3}) under the integral sign using Eq.\ (\ref{eq:secdecrel}). Carrying out the $\ep$ expansion, we find a contribution of 
\bea
\bar{K}_{TC}^{{\rm in-out}\,(2)}(\tauo,\omega,r,\mu) &=& 
C_A C_F
\Bigg[ \mbox{\large $g$}_{C_A}^{(1)}\big(0,r\big)\ln \left(\frac{\mu }{2\omega}\right) 
-\ln\left(\frac{\mu }{2\omega}\right)\int_0^1{\dd z_1\over z_1}\left(\mbox{\large $g$}_{C_A}^{(0)}\left(z_1,r\right) \right.
\el\hspace{-23ex}
\left.+ \mbox{\large $g$}_{C_A}^{(0)}\left(1/z_1,r\right) 
- 2 \mbox{\large $g$}_{C_A}^{(0)}\big(0,r\big)\right)+ \mbox{\large $g$}_{C_A}^{(0)}\big(0,r\big)\ln^2 \left(\frac{\mu}{2\omega}\right)
+\mbox{\large $g$}_{C_A}^{(0)}\big(0,r\big)\ln^2\left(\!\frac{\mu }{\tau_\omega Q}\!\right) 
\el\hspace{-23ex}
 - \frac{1}{2}\mbox{\large $g$}_{C_A}^{(1)}\big(0,r\big)\ln \left(\frac{\tau_\omega Q}{2\omega}\right) - \frac{1}{2}\mbox{\large $g$}_{C_A}^{(0)}\big(0,r\big)\ln^2 \left(\frac{\tau_\omega Q}{2\omega}\right)
 +\int_1^{\tau_\omega Q \over 2 \omega} {\dd z_2 \over z_2}\int_0^1  {\dd z_1 \over z_1}\left(\mbox{\large $g$}_{C_A}^{(0)}\left({z_2 \over z_1},r\right)
\right.
\el\hspace{-23ex}
\left.
  - \mbox{\large $g$}_{C_A}^{(0)}\big(0,r\big)\right) + {\mbox{\large $g$}_{C_A}^{(2)}\big(0,r\big)\over 4} - {1\over 4}\int_0^1 {\dd z_1 \over z_1} \left(\mbox{\large $g$}_{C_A}^{(1)}\left(z_1,r\right) 
  + \mbox{\large $g$}_{C_A}^{(1)}\left(1/z_1,r\right) - 2 \mbox{\large $g$}_{C_A}^{(1)}\big(0,r\big)\right)
\el\hspace{-23ex}
+ {1\over 2}\int_0^1 {\dd z_1 \over z_1} \left(\mbox{\large $g$}_{C_A}^{(0)}\left(z_1,r\right) 
  + \mbox{\large $g$}_{C_A}^{(0)}\left(1/z_1,r\right) - 2 \mbox{\large $g$}_{C_A}^{(0)}\big(0,r\big)\right) \ln\left(z_1\right)\Bigg]
+
C_F n_f T_F
\Bigg[
-\ln\left(\frac{\mu }{2\omega}\right)
\el\hspace{-23ex}
\times\int_0^1{\dd z_1\over z_1}\left(\mbox{\large $g$}_{n_f}^{(0)}\left(z_1,r\right) + \mbox{\large $g$}_{n_f}^{(0)}\left(1/z_1,r\right) \right)
  + {1\over 2}\int_0^1 {\dd z_1 \over z_1} \left(\mbox{\large $g$}_{n_f}^{(0)}\left(z_1,r\right)+ \mbox{\large $g$}_{n_f}^{(0)}\left(1/z_1,r\right)\right)\ln\left(z_1\right)
\el\hspace{-23ex}
  +\int_1^{\tau_\omega Q \over 2 \omega} {\dd z_2 \over z_2}\int_0^1  {\dd z_1 \over z_1}\mbox{\large $g$}_{n_f}^{(0)}\left({z_2 \over z_1},r\right)
 - {1\over 4}\int_0^1 {\dd z_1 \over z_1} \left(\mbox{\large $g$}_{n_f}^{(1)}\left(z_1,r\right)  + \mbox{\large $g$}_{n_f}^{(1)}\left(1/z_1,r\right) \right)\Bigg]
\label{eq:secdecompinout}
\eea
to the finite part of the integrated $\tauo$ distribution.
Here, we made use of the relations $\mbox{\large $g$}_{C_A}(0,r) = \mbox{\large $g$}_{C_A}(\infty,r)$, $\mbox{\large $g$}_{n_f}(0,r) = \mbox{\large $g$}_{n_f}(\infty,r)$, and $\mbox{\large $g$}_{n_f}(0,r) = 0$.

\subsection{All-Out Contributions}
\label{sec:outout}
In this section we discuss the calculation of the contributions to the integrated jet thrust distribution where
both soft partons go into the out-of-jet region. In this case, the integrals over $k_L$, $k_R$, $\tauo^\prime$, and $\lambda$ are trivial.
However, the integrals defining $S^{{\rm all-out}\,(2)}(k_L,k_R,\lambda,r,\mu)$ have phase space singularities themselves in this case and one must perform
a non-trivial sector decomposition to extract them. This sector decomposition is very similar to the one used in reference~\cite{Kelley:2011ng} to calculate the same-side in-in contributions
to the hemisphere soft function. $S^{{\rm all-out}\,(2)}(k_L,k_R,\lambda,r,\mu)$ can be expressed as
\bea
S^{{\rm all-out}\,(2)}(k_L,k_R,\lambda,r,\mu) &=&
\frac{\mu^{4\ep}}{(2\lambda)^{1+4\ep}}\D\left(k_L\right)\D(k_R)
\el \hspace*{-2 cm}
\times\left[ C_A C_F\left( {\mbox{\large $h$}_{C_A}^{(-2)}\left(r\right)\over \ep^2}
 + {\mbox{\large $h$}_{C_A}^{(-1)}\left(r\right)\over \ep} + \mbox{\large $h$}_{C_A}^{(0)}\left(r\right) + \mbox{\large $h$}_{C_A}^{(1)}\left(r\right) \ep + \Ord\left(\ep^2\right)\vphantom{{h_{C_A}^{(-2)}\left(r\right)\over \ep^2}}\right)
\right.
\el \hspace*{-2 cm}
\left. 
+ C_F n_f T_F \left({\mbox{\large $h$}_{n_f}^{(-1)}\left(r\right)\over \ep} + \mbox{\large $h$}_{n_f}^{(0)}\left(r\right)
+ \mbox{\large $h$}_{n_f}^{(1)}\left(r\right) \ep + \Ord\left(\ep^2\right)\right)\right],
\eea
which implies that
\bea
K_{TC}^{{\rm all-out}\,(2)}(\tauo,\omega,r,\mu) &=&
\left(-{1\over 8\ep} \left(\frac{\mu}{2\omega}\right)^{4\e}\right)
\el \hspace*{-2 cm}
\times\left[ C_A C_F\left( {\mbox{\large $h$}_{C_A}^{(-2)}\left(r\right)\over \ep^2}
 + {\mbox{\large $h$}_{C_A}^{(-1)}\left(r\right)\over \ep} + \mbox{\large $h$}_{C_A}^{(0)}\left(r\right) + \mbox{\large $h$}_{C_A}^{(1)}\left(r\right) \ep + \Ord\left(\ep^2\right)\vphantom{{h_{C_A}^{(-2)}\left(r\right)\over \ep^2}}\right)
\right.
\el \hspace*{-2 cm}
\left. 
+ C_F n_f T_F \left({\mbox{\large $h$}_{n_f}^{(-1)}\left(r\right)\over \ep} + \mbox{\large $h$}_{n_f}^{(0)}\left(r\right)
+ \mbox{\large $h$}_{n_f}^{(1)}\left(r\right) \ep + \Ord\left(\ep^2\right)\right)\right].
\eea

Carrying out the $\ep$ expansion, we find a contribution of 
\bea
\bar{K}_{TC}^{{\rm all-out}\,(2)}(\tauo,\omega,r,\mu) &=&
\el\hspace{-21ex}
C_A C_F
\Bigg[ -{4\over 3} \mbox{\large $h$}_{C_A}^{(-2)}\left(r\right)\ln^3 \left(\frac{\mu }{2\omega}\right) - \mbox{\large $h$}_{C_A}^{(-1)}\left(r\right)\ln^2 \left(\frac{\mu }{2\omega}\right) 
- {1\over 2}\mbox{\large $h$}_{C_A}^{(0)}\left(r\right)\ln \left(\frac{\mu }{2\omega}\right) - {1\over 8}\mbox{\large $h$}_{C_A}^{(1)}\left(r\right)\Bigg]
\el\hspace{-21ex}
+C_F n_f T_F
\Bigg[- \mbox{\large $h$}_{n_f}^{(-1)}\left(r\right)\ln^2 \left(\frac{\mu }{2\omega}\right)
- {1\over 2}\mbox{\large $h$}_{n_f}^{(0)}\left(r\right)\ln \left(\frac{\mu }{2\omega}\right)
  - {1\over 8}\mbox{\large $h$}_{n_f}^{(1)}\left(r\right)
  \Bigg]
\label{eq:secdecompoutout}
\eea
to the finite part of the integrated $\tauo$ distribution.

\section{Computation in Terms of Multiple Polylogarithms}
\label{sec:coprod}
\subsection{Integration of the Terms Produced by the Sector Decomposition}
In this section, we describe how we evaluate the large number of integrals produced by the sector decompositions~\cite{Binoth:2000ps} discussed in Section \ref{sec:secdec}.
We employ two independent methods for this purpose, which are described in the following. The first method uses {\tt Mathematica}'s {\tt Integrate} function, extends its domain of applicability
in a way appropriate for the problem at hand, and, finally, makes use of the coproduct formalism~\cite{Duhr:2011zq,Duhr:2012fh} to simplify the result obtained. 
The second method employs dedicated integration routines without reference to {\tt Mathematica}'s {\tt Integrate}
function or the coproduct.

Our first method relies on the fact that, at the stage of the calculation where only one non-trivial integration remained, each integral
in the problem can be expressed as a single-parameter integral over classical polylogarithms of 
transcendentality weight less than or equal to three ({\it e.g.} structures like $\ln(x)$, $\ln^2(x)$, $\pi^2 \ln(x)$, $\ln(x) \li2(y)$, and $\li3(x)$).
By judiciously combining integrands with similar analytical structures, we find that we can arrive at a sum of single-parameter integrals with integrands written solely in terms of classical polylogarithms.
This is not always straightforward to do, due to numerous issues with the function {\tt Integrate}. 
At this stage, we find it convenient to cast the integrals into a form in which we can differentiate their integrands with respect to $r$.
Taking a  derivative reduces all integrands to a form which we have solved already, namely
$$\int{\ln(a + b x)\ln(c + d x)\over e + f x}\dd x$$
or forms that differ only by a series of integrations by parts. 
It is worth pointing out that, at times, before attempting to integrate by parts, we make use of the properties
of logarithms and the small number of functional relations satisfied by the dilogarithm~\cite{Kirillov:1994en}.
After performing the final definite integral, we are left with explicit expressions for $r$ derivatives of the quantities of interest.
At this stage we rewrite our results in terms of multiple polylogarithms.
Once we identified all generalized weights (defined in Section \ref{sec:genweights} below) it is trivial to integrate with respect to $r$ and 
obtain the expressions of interest up to an overall constant built out of zeta values. This constant is then determined using the {\tt Mathematica} implementation of the LLL algorithm~\cite{LLL}.
The $C_F n_f T_F$ color structure is simpler than the $C_A C_F$ color structure because it can be expressed in terms of classical polylogarithms of at most transcendentality weight three.
In particular, no numerical methods are used to fix the $C_F n_f T_F$ constants.
However, for both non-trivial color structures, the resulting expressions suffer from ambiguities due to functional relations which are not entirely straightforward to resolve. 
Our solution was to develop an extension of the coproduct calculus applicable to all of the functions encountered at this stage of the calculation (see Section \ref{sec:gencoprodcalc}
below for a detailed discussion).

Our second and preferred method employs a dedicated integration algorithm for multiple polylogarithms independent of the coproduct calculus.
In this approach, a number of integration constants expressed in terms of $G$ functions need to be rewritten in terms of known fundamental constants.
To accomplish this, we use fits against high-precision numerical evaluations
of multiple polylogarithms obtained with the implementation \cite{Vollinga:2004sn} in the {\tt GiNaC} CAS \cite{Bauer:2000cp}.
In this way, our second method allows for a fully automated evaluation of the integrals produced by the sector decomposition.
These integration routines as well as the extended coproduct calculus are implemented in an in-house {\tt Mathematica} package~\cite{AndreasGPL}.
While the first method described above was our default, we used the second method to verify our results for all difficult integrals
and found full agreement.

\subsection{Multiple Polylogarithms With Generalized Weights}
\label{sec:genweights}
Recall that, as was mentioned in the introduction, multiple
polylogarithms~\cite{Goncharov1} are a recursively defined class of iterated integrals. Take $G(;x) = 1$ for all $x$. Then multiple
polylogarithms with $n$ weights are defined as follows. If we have a set of $n$ complex numbers called weights, $\{w_1,\ldots,w_n\}$, at least one of which is different from zero, then 
\be
G(w_1,\dots,w_n;x) = \int_0^x {dt \over t - w_1} G(w_2,\dots,w_n;t).
\ee
If, on the other hand $w_i = 0$ for all $i$, then
\be
G(0,\dots,0;x) = \frac{1}{n!}\ln^n(x)\,.
\ee
The weights, $w_i$, as well as the argument $x$, are considered as rational functions of the indeterminates. 
For the univariate case, we generalize the integration measure
\begin{equation}
 \frac{\mathrm{d}t}{t-w} \rightarrow \frac{f'(t)}{f(t)} \mathrm{d}t = \mathrm{d}\ln(f(t))
\end{equation}
where $f(x)$ is an irreducible rational polynomial.
We use square brackets and a dummy variable $o$ to denote generalized weights:
\begin{equation}
G([f(o)], w_2, \ldots, w_n; x) = \int_0^x \mathrm{d}t \frac{ f'(t)}{f(t)} G(w_2,\ldots, w_n; t)
\end{equation}
The linear case, $[o-w]$, reduces to the standard weight, $w$.

In our calculation, we encounter multiple polylogarithms with both standard and non-standard weights.
Specifically, multiple polylogarithms of argument $r$ appear with combinations of both standard weights,
$$\{-2, -1, -1/2, 0, 1/2, 1, 2 \},$$
and generalized weights,
$$\{ \big[o^2+1\big], \big[o^2+o+1\big], \big[o^2+o-1\big], \big[o^2-o+1\big], \big[o^2-o-1\big], \big[o^3+o^2-1\big] \}.$$
The first three generalized weights are based on cyclotomic polynomials and generate cyclotomic
polylogarithms~\cite{Ablinger:2011te,vonManteuffel:2013uoa}.
The remaining generalized weights generate multiple polylogarithms beyond that class of functions,
and, to the best of our knowledge, they have not been studied by other authors.

Let us discuss the relation of our generalized weights to those used in Goncharov's formalism.
As an explicit example we consider
\begin{equation}
\label{genlogexample}
G([o^2+o+1]; x) = \int_0^x \mathrm{d}t \frac{2 t + 1}{t^2 + t + 1} = \ln(x^2+x+1)
\end{equation}
Traditionally, quadratic denominators have been factorized in the complex domain:
\begin{eqnarray}
\int_0^x \mathrm{d}t \frac{2 t + 1}{t^2 + t + 1}
  &=& \int_0^x \mathrm{d}t \frac{1}{t - (-1+\sqrt{3} i)/2} + \int_0^x \mathrm{d}t  \frac{1}{t - (-1-\sqrt{3} i)/2}\nonumber\\
  &=& G((-1+\sqrt{3} i)/2; x) + G((-1-\sqrt{3} i)/2; x)
\end{eqnarray}
As a result, such quadratic denominators give rise to pairs of complex conjugated weights and corresponding pairs of multiple polylogarithms.
Polynomials in the denominator of general degree $m\ge2$ can be treated as well in a completely analogous fashion.
From
\begin{equation}
\frac{f'(t)}{f(t)} = \frac{1}{t-w_1} + \cdots + \frac{1}{t-w_n}\qquad\text{for}\qquad f(t) = (t - w_1)\cdots (t-w_n)\,,
\end{equation}
we see that this leads to
\begin{equation}
\label{genfactorized}
G(\ldots,[f(o)],\ldots; x) = G(\ldots,w_1,\ldots; x) + \ldots + G(\ldots,w_n,\ldots; x)\,.
\end{equation}

Such a treatment has the disadvantage of introducing root objects and may give rise to spurious imaginary parts.
In contrast, our notation avoids splitting the real-valued function \eqref{genlogexample} defined on the unit interval into complex-valued components. Furthermore, our approach has the clear 
advantage that the intermediate expression swell which would occur
if one proceeded by splitting real-valued functions is avoided.
As an alternative, in~\cite{Ablinger:2011te}, integration measures $n(t)/f(t)\mathrm{d}t$ are defined
where in general $n(t) \neq f'(t)$.
In their notation, our integral \eqref{genlogexample} splits into two cyclotomic polylogarithms,
$G([o^2+o+1]; x) = 2 C^1_3(x) + C^0_3(x)$.
In different physics applications we studied, we observed that integrals indeed appear with integration measures of the form $f'(t)/f(t)\mathrm{d}t $
and our notation directly reflects this. An important feature of this structure is the possibility to define a root-free symbol and coproduct in a straightforward way.
This will be discussed in the following.

\subsection{Coproduct Calculus for Generalized Weights}
\label{sec:gencoprodcalc}
In order to identify functional relations and project onto specific basis functions, we employ a variant of
the coproduct calculus formulated in~\cite{Duhr:2011zq,Duhr:2012fh}. It is convenient to consider multiple polylogarithms using the functions
\begin{equation}
I(a_0;a_1,\ldots,a_n;a_{n+1}) = \int_{a_0}^{a_{n+1}} \mathrm{d}t \frac{1}{t-a_n} I(a_0;a_1,\ldots,a_{n-1};t)
\end{equation}
which can easily be related to the $G$ functions. The coproduct for standard multiple polylogarithms is defined via the polygon construction of reference~\cite{Goncharov2},
\begin{eqnarray}
\label{coproduct}
&&\cp\big(I(a_0;a_1,\ldots,a_n;a_{n+1})\big)
= \\
&&\sum_{0=i_0 < \ldots < i_{k+1} = n + 1}
  I(a_0; a_{i_1},\ldots,a_{i_k};a_{n+1}) \otimes
  \prod_{p=0}^k
  I(a_{i_p};a_{i_p+1},\ldots,a_{i_{p+1}-1}; a_{i_{p+1}}).\nonumber
\end{eqnarray}
This prescription is supplemented by an appropriate regularization procedure to handle endpoint singularities
in the emerging $I$ functions. The symbol follows from the maximal iteration of the coproduct modulo $\pi$.

We wish to define the coproduct for univariate polylogarithms $I(0;a_1,\ldots,a_n;z)$
with generalized or standard weights $a_1,\ldots,a_n$
according to a similar prescription.
The weights are considered to be independent of further parameters.
Our goal is to avoid any reference to roots of irreducible polynomials
or to general algebraic numbers, which lack a unique factorization.
We achieve this by systematically dropping certain contributions
which can not be directly generalized.
When applying the coproduct to the left-hand side of \eqref{genfactorized} we require
consistency with what is produced by the standard rules for the right-hand
side of \eqref{genfactorized} up to constants in the second slot.
However, in contrast to the case of standard (rational) weights, it is no longer straightforward
to keep constants like $\ln (2)$ everywhere in the coproduct and symbol.

Here, our solution is to treat such constants in a way analogous to the way in which reference \cite{Duhr:2012fh} handles the constant $\pi$.
In practice, this means that we exclude all constants from the symbol algebra part of our simplification routine and postpone their treatment to the numerical fitting step whenever
generalized weights are involved.
Needless to say, a corresponding prescription must be used also for
the other functions involved in the considered identities, such as multivariate
polylogarithms with standard weights.
From these considerations we obtain the following rules for the coproduct and symbol
for univariate polylogarithms with generalized or standard weights.
We apply \eqref{coproduct} taken as a formal expression and replace generalized
weights in the boundaries of $I$ functions by 0.
We drop terms with constants of weight $>1$ in the second slot of the coproduct.
For iterations of the coproduct this means we keep constants only in the first slot,
which is sufficient for our purposes.
Finally, the symbol $\sym$ of a multiple polylogarithm with generalized weight is defined,
such that it is compatible with the maximal iteration of the coproduct and
\begin{equation}
\label{eq:laststep}
\sym\left(I(0; [f(o)]; x)\right) = f(x)\,.
\end{equation}
Our normalization sets symbols to zero whenever one of their slots contains a pure number.
In terms of multiple polylogarithms, Eq.\ (\ref{eq:laststep}) is nothing but the statement that
\begin{equation}
\sym\left(G([f(o)]; x)\right) = \sym\left(\ln\left( f(x) \right) \right) = f(x)\,.
\end{equation}
These prescriptions allowed us to successfully use the coproduct calculus in different
contexts involving multiple polylogarithms with generalized weights.

Finally, we describe our treatment of products when integrating the symbol.
Reference \cite{Duhr:2011zq} explains how one can effectively isolate contributions to the symbol which originate from products of multiple polylogarithms.
The idea is to construct projection operators which annihilate certain classes of multiple
polylogarithms which can be rewritten as a shuffle of lower weight multiple polylogarithms. In this fashion, one can construct a filtration which allows one to consider {\it e.g.} terms of the form $\li3(x)\ln(y)$ independently
of those of the form $\li2(x) \li2(y)$. We observe that, actually, it is possible to exploit this filtration at the level of the symbol to obtain a unique form for product terms upon integration.
In this way, we reduce the integration of the symbol to the integration of product-free terms.
In other words, in our approach, products need not be specified explicitly when constructing a complete set of basis functions;
they can be constructed in a systematic way on the fly from the product-free basis functions.
In this way, inefficiencies due to the rapid combinatorial growth of the number of possible product terms are avoided.

We apply these techniques
to our results and obtain
concise analytical expressions in terms of $G$ functions.
The individual contributions to the integrated $\tauo$ distribution are given
in Appendix~\ref{app:indcontrib}.
When performing the integrations we observe interesting structural cancellations.
At intermediate stages of the calculation, the solutions of certain integrals indeed require a relatively large set
of weights, both standard and generalized.
These terms are required due to the way our phase space integrals are split up by the sector decomposition procedure.
It may be possible to perform the sector decomposition in a way which prevents this enlargement of the function space but, to the best of our knowledge, no such strategy is known.
Once the sum of all contributions for a specific phase space region is taken, we find that all contributions involving generalized weights cancel.
Indeed, only multiple polylogarithms with weights drawn from the set $\{-1,0,1\}$ appear in the global part of the integrated $\tauo$ distribution.

\section{Exact Result}
\label{sec:niceres}

We combine the individual soft contributions to the finite part of the integrated $\tauo$ distribution according to
\bea
\bar{K}_{TC}^{(2)}(\tauo,\omega,r,\mu) &=& \bar{K}^{{\rm in\,Ren}}_{TC}(\tauo,\omega,r,\mu)+\bar{K}^{{\rm all-in}\,(2)}_{TC}(\tauo,\omega,r,\mu)+\bar{K}^{{\rm in\,R-V}}_{TC}(\tauo,\omega,r,\mu)
\el
+\bar{K}^{{\rm out\,Ren}}_{TC}(\tauo,\omega,r,\mu)+\bar{K}^{{\rm out\,R-V}}_{TC}(\tauo,\omega,r,\mu)+\bar{K}_{TC}^{{\rm opp\,in-in}}(\tauo,\omega,r,\mu)
\el
+\bar{K}_{TC}^{{\rm in-out}\,(2)}(\tauo,\omega,r,\mu)+\bar{K}_{TC}^{{\rm all-out}\,(2)}(\tauo,\omega,r,\mu)
\eea
and obtain the summed result in terms of $G$ functions.
In contrast to the individual contributions, the sum of all purely $r$ dependent
terms can be written in terms of classical polylogarithms.
Employing the coproduct calculus, we find
\bea
\label{eq:finalpoly}
\bar{K}_{TC}^{(2)}(\tauo,\omega,r,\mu) &=& C_A C_F \left[
-\frac{176}{9} \ln ^3\left(\frac{\mu }{Q \tau_{\omega }}\right)
+\left(\frac{8 \pi^2}{3}-\frac{536}{9}\right) \ln ^2\left(\frac{\mu }{Q \tau_{\omega }}\right)
+\left(-\frac{88}{3} \ln^2(r)
\right.\right.\el\hspace{-19.5ex}\left.\left.
-\frac{176 \li2(-r)}{3}
+56 \zeta_3-\frac{1616}{27}\right) \ln \left(\frac{\mu }{Q \tau_{\omega }}\right)
+\frac{176}{3} \ln (r) \ln \left(\frac{\mu }{Q \tau_{\omega }}\right) \ln \left(\frac{Q \tau _{\omega }}{2 \omega }\right)
+\left(32 \li2(r)
\right.\right.\el\hspace{-19.5ex}\left.\left.
+32 \li2(-r)+\frac{88 \ln (r)}{3}-\frac{8 \pi ^2}{3}\right) \ln ^2\left(\frac{Q \tau _{\omega }}{2 \omega }\right)
+\left(-176 \li2(-r)-\frac{352 \li2(r)}{3}-64 \li3(1-r)
\right.\right.\el\hspace{-19.5ex}\left.\left.
+224 \li3(-r)+160 \li3(r)+64
   \li3\left(\frac{1}{r+1}\right)+32 \li3(1-r^2)+128 \li2(-r) \ln (r+1)
\right.\right.\el\hspace{-19.5ex}\left.\left.
+64 \li2(r) \ln (r+1)+64 \li2(-r) \ln (1-r)-128 \li2(-r) \ln (r)-96 \li2(r) \ln
   (r)+\frac{8}{3 (r-1)}
\right.\right.\el\hspace{-19.5ex}\left.\left.
-\frac{8}{3 (r+1)}-\frac{32}{3} \ln ^3(r+1)+64 \ln (r) \ln ^2(r+1)-\frac{44 \ln ^2(r)}{3}+64 \ln (1-r) \ln (r) \ln (r+1)
\right.\right.\el\hspace{-19.5ex}\left.\left.
-\frac{176}{3} \ln (r) \ln (r+1)+\frac{16}{3} \pi ^2 \ln
   (r+1)+\frac{16}{3} \pi ^2 \ln (1-r)-\frac{176}{3} \ln (1-r) \ln (r)-\frac{8 \ln (r)}{3 (r-1)}
\right.\right.\el\hspace{-19.5ex}\left.\left.
+\frac{8 \ln (r)}{3 (r+1)}-\frac{8 \ln (r)}{3 (r-1)^2}-\frac{8 \ln (r)}{3 (r+1)^2}+\frac{8}{3} \pi ^2 \ln
   (r)+\frac{536 \ln (r)}{9}-48 \zeta_3+\frac{44 \pi ^2}{9}+\frac{8}{3}\right)
\right.\el\hspace{-19.5ex}\left.
\times \ln \left(\frac{Q \tau _{\omega }}{2 \omega }\right)
  -4 \ln^4(1-r)
  +\frac{16}{3} \ln (r) \ln^3(1-r)+\frac{32}{3} \ln (2) \ln^3(1-r)
 +40 \ln^2(r) \ln^2(1-r)
\right.\el\hspace{-19.5ex}\left.
-\frac{176}{3} \ln (r) \ln^2(1-r)-16 \ln^2(2) \ln^2(1-r)-\frac{32}{3} \ln^3(r+1) \ln(1-r)
-88 \ln^2(r) \ln (1-r)
\right.\el\hspace{-19.5ex}\left.
+64 \ln (r) \ln^2(r+1) \ln 
(1-r)-\frac{16 \ln (r) \ln (1-r)}{3 (1-r)}
-\frac{16 \ln (r) \ln 
(1-r)}{3 (r+1)}
+\frac{16 \ln (r) \ln (1-r)}{3 (1-r)^2}
\right.\el\hspace{-19.5ex}\left.
+\frac{16 
\ln (r) \ln (1-r)}{3 (r+1)^2}-\frac{40}{3} \pi^2 \ln (r) \ln(1-r)
+\frac{536}{9} \ln (r) \ln (1-r)
+\frac{176}{3} \ln (r) \ln(r+1)
\right.\el\hspace{-19.5ex}\left.
\times \ln (1-r)
+\frac{32}{3} \pi^2 \ln (r+1) \ln (1-r)
+64 \ln(r+1) \li2(-r) \ln (1-r)
+\frac{176}{3} \li2(-r)
\right.\el\hspace{-19.5ex}\left.
 \times\ln (1-r)+80 \ln(r) \li2(r) \ln (1-r)
-\frac{176}{3} \li2(r) \ln (1-r)
+32 \li3(-r) \ln (1-r)
\right.\el\hspace{-19.5ex}\left.
+64 \li3\left(\frac{1}{r+1}\right) \ln (1-r)+\frac{16 \ln (1-r)}{3 (1-r)}
-32 \zeta_3 \ln (1-r)
+\frac{32}{3} \ln^3(2) 
\ln (1-r)
\right.\el\hspace{-19.5ex}\left.
-\frac{16}{3} \pi^2 \ln (2) \ln (1-r)+\frac{44}{3} \pi^2 \ln (1-r)
-\frac{8}{3} \ln (1-r)
-\frac{88}{3} \ln^4(r+1)
-\frac{88}{3} \ln^3(r+1)
\right.\el\hspace{-19.5ex}\left.
+\frac{352}{3} \ln (r) \ln^3(r+1)
+\frac{32}{3} \ln (2)\ln^3(r+1)
+\frac{4 \ln^2(r)}{1-r}
+\frac{4 \ln^2(r)}{r+1}
-\frac{4 \ln^2(r)}{(1-r)^2}
-\frac{4 \ln^2(r)}{(r+1)^2}
\right.\el\hspace{-19.5ex}\left.
-\frac{4}{3} \pi^2 \ln^2(r)-\frac{268 \ln^2(r)}{9}
-80 \ln^2(r) \ln^2(r+1)+88 \ln (r) 
\ln^2(r+1)
-16 \ln^2(2) \ln^2(r+1)
\right.\el\hspace{-19.5ex}\left.
+\frac{64}{3} \pi^2 \ln 
^2(r+1)
-32 \li2^2(-r)+40 \li2^2(r)-\frac{44 \ln (r)}{9 
(1-r)}-\frac{308 \ln (r)}{9 (r+1)}
+\frac{20 \ln (r)}{9 
(1-r)^2}+\frac{212 \ln (r)}{9 (r+1)^2}
\right.\el\hspace{-19.5ex}\left.
+\frac{88}{9} \pi^2 \ln (r)
+\frac{32}{3} \ln (r)-88 \ln^2(r) \ln (r+1)+\frac{8 \ln (r) \ln 
(r+1)}{3 (1-r)}
-\frac{8 \ln (r) \ln (r+1)}{3 (r+1)}
\right.\el\hspace{-19.5ex}\left.
-\frac{8 \ln (r) \ln (r+1)}{3 (1-r)^2}
+\frac{8 \ln (r) \ln (r+1)}{3 
(r+1)^2}-\frac{80}{3} \pi^2 \ln (r) \ln (r+1)
+\frac{536}{9} \ln (r) \ln (r+1)
\right.\el\hspace{-19.5ex}\left.
+\frac{8 \ln (r+1)}{1-r}
+\frac{8 \ln (r+1)}{3 (r+1)}+
\frac{32}{3} \ln^3(2) \ln (r+1)
-\frac{16}{3} \pi^2 \ln (2) \ln (r+1)
+\frac{44}{3} \pi^2 \ln (r+1)
\right.\el\hspace{-19.5ex}\left.
-\frac{16}{3} \ln (r+1)+16 \ln 
^2(r) \li2(-r)
+96 \ln^2(r+1) \li2(-r)
-\frac{616}{3} \ln (r) 
\li2(-r)
-160 \ln (r)
\right.\el\hspace{-19.5ex}\left.
\times \ln (r+1) \li2(-r)+\frac{352}{3} \ln (r+1) 
\li2(-r)
+\frac{32 \li2(-r)}{3 (1-r)}-\frac{32 \li2(-r)}{3 
(1-r)^2}
-\frac{64}{3} \pi^2 \li2(-r)
\right.\el\hspace{-19.5ex}\left.
+\frac{536 \li2(-r)}{9}-16 \ln 
^2(r) \li2(r)+32 \ln^2(r+1) \li2(r)
-176 \ln (r) 
\li2(r)+\frac{176}{3} \ln (r+1)
\right.\el\hspace{-19.5ex}\left.
\times \li2(r)+32 \li2(-r) \li2(r)-\frac{16 
\li2(r)}{3 (r+1)}+\frac{16 \li2(r)}{3 (r+1)^2}
-\frac{64 \pi^2 
\li2(r)}{3}+\frac{1072 \li2(r)}{9}
\right.\el\hspace{-19.5ex}\left.
+80 \ln (r) \li3(1-r)-64 \ln 
(r+1) \li3(1-r)
-176 \li3(1-r)+128 \ln (r+1) \li3(-r)
\right.\el\hspace{-19.5ex}\left.
+\frac{616 \li3(-r)}{3}
+80 \ln (r) \li3(r)
+32 \ln (r+1) \li3(r)+\frac{352 
\li3(r)}{3}-160 \ln (r) \li3\left(\frac{1}{r+1}\right)
\right.\el\hspace{-19.5ex}\left.
+128 \ln (r+1) \li3\left(\frac{1}{r+1}\right)
+176 \li3\left(\frac{1}{r+1}\right)
+32 \ln (r+1) \li3(1-r^2)-\frac{88}{3} \li3(1-r^2)
\right.\el\hspace{-19.5ex}\left.
-64 \li4\left(\frac{1-r}{2}\right)
+96 \li4(1-r)-256 
\li4(-r)-176 \li4(r)-8 \li4\left(-\frac{4 r}{(1-r)^2}\right)
\right.\el\hspace{-19.5ex}\left.
+64 \li4\left(-\frac{2 r}{1-r}\right)
-32 \li4\left(-\frac{r}{1-r}\right)-8 
\li4\left(\frac{4 r}{(r+1)^2}\right)+32 \li4\left(\frac{1}{r+1}\right)
\right.\el\hspace{-19.5ex}\left.
+32 \li4\left(-\frac{1-r}{r+1}\right)
-32 
\li4\left(\frac{1-r}{r+1}\right)-160 \li4\left(\frac{r}{r+1}\right)
+64 \li4\left(\frac{2 r}{r+1}\right)-64 \li4\left(\frac{r+1}{2}
\right)
\right.\el\hspace{-19.5ex}\left.
-16 \li4(1-r^2)+\frac{8 \pi^2}{9 (1-r)}+\frac{20}{9 
(1-r)}+\frac{8 \pi^2}{9 
(r+1)}+\frac{116}{9 (r+1)}
-\frac{8 \pi^2}{9 (1-r)^2}-\frac{8 \pi^2}{9 (r+1)^2}
\right.\el\hspace{-19.5ex}\left.
+64 \ln (r) \zeta_3
-48 \ln (r+1) \zeta_3
-\frac{418 \zeta_3}{9}+128 \li4\left(\frac{1}{2}\right)
+\frac{83 \pi ^4}{45}-\frac{1139 \pi^2}{54}-\frac{2860}{81}
\right.\el\hspace{-19.5ex}\left.
+\mbox{\large $\Lambda$}_{C_A}\left({\tauo Q \over 2\omega},r\right)\right]
\el\hspace{-19.5ex}
+C_F n_f T_F \left[
\frac{64}{9} \ln ^3\left(\!\frac{\mu }{Q \tau_{\omega }}\!\right)
+\frac{160}{9} \ln ^2\left(\!\frac{\mu }{Q \tau_{\omega } }\!\right)
+\left(\frac{64 \li2(-r)}{3}+\frac{32 \ln ^2(r)}{3}+\frac{448}{27}\right) \ln\left(\!\frac{\mu }{Q \tau_{\omega }}\!\right)
\right.\el\hspace{-19.5ex}\left.
-\frac{64}{3} \ln (r) \ln \left(\!\frac{\mu }{Q \tau_{\omega }}\!\right) \ln \left(\frac{Q \tau _{\omega }}{2 \omega }\right)
-\frac{32}{3} \ln (r) \ln ^2\left(\frac{Q \tau _{\omega }}{2 \omega }\right)
+\left(64 \li2(-r)+\frac{128 \li2(r)}{3}
\right.\right.\el\hspace{-19.5ex}\left.\left.
-\frac{16}{3 (r-1)}+\frac{16}{3
   (r+1)}+\frac{16 \ln ^2(r)}{3}+\frac{64}{3} \ln (1-r) \ln (r)+\frac{64}{3} \ln (r+1) \ln (r)+\frac{16 \ln (r)}{3 (r-1)}
\right.\right.\el\hspace{-19.5ex}\left.\left.
-\frac{16 \ln (r)}{3 (r+1)}+\frac{16 \ln (r)}{3 (r-1)^2}+\frac{16 \ln (r)}{3
   (r+1)^2}-\frac{160 \ln (r)}{9}-\frac{16 \pi ^2}{9}-\frac{16}{3}\right) \ln \left(\frac{Q \tau _{\omega }}{2 \omega }\right)
  +\frac{32}{3} \ln^3(r+1)
\right.\el\hspace{-19.5ex}\left.
 -32 \ln (r) \ln^2(r+1)+32 \ln^2(r) \ln (r+1)
-\frac{16 \ln (r) \ln (r+1)}{3 (1-r)}-\frac{64}{3} \ln (1-r) \ln (r) 
\ln (r+1)
\right.\el\hspace{-19.5ex}\left.
 +\frac{16 \ln (r) \ln (r+1)}{3 (r+1)}+\frac{16 \ln (r) \ln (r+1)}{3 (1-r)^2}
 -\frac{16 \ln (r) \ln (r+1)}{3 (r+1)^2}
-\frac{128}{3} \li2(-r) \ln (r+1)
\right.\el\hspace{-19.5ex}\left.
 -\frac{208}{9} \ln (r) \ln (r+1)
-\frac{64}{3} \li2(r) \ln(r+1)
-\frac{16 \ln (r+1)}{1-r}-\frac{16 \ln (r+1)}{3 (r+1)}
-\frac{16}{3} \pi^2 \ln (r+1)
\right.\el\hspace{-19.5ex}\left.
+\frac{32}{3} \ln (r+1)+32 \ln (1-r) \ln^2(r)
+\frac{64}{3} \ln^2(1-r) \ln (r)
-\frac{8 \ln^2(r)}{1-r}-\frac{8 \ln ^2(r)}{r+1}
+\frac{8 \ln^2(r)}{(1-r)^2}
\right.\el\hspace{-19.5ex}\left.
+\frac{8 \ln^2(r)}{(r+1)^2}
+\frac{80 \ln^2(r)}{9}
-\frac{32 \ln (1-r)}{3 (1-r)}-\frac{16}{3} \pi^2 \ln (1-r)
+\frac{16}{3} \ln (1-r)
+\frac{32 \ln (1-r) \ln(r)}{3 (1-r)}
\right.\el\hspace{-19.5ex}\left.
+\frac{32 \ln (1-r) \ln (r)}{3 (r+1)}
-\frac{32 \ln (1-r) \ln (r)}{3 (1-r)^2}
-\frac{32 \ln (1-r) \ln (r)}{3 (r+1)^2}
-\frac{208}{9} \ln (1-r) \ln (r)
\right.\el\hspace{-19.5ex}\left.
+\frac{88 \ln (r)}{9 (1-r)}
+\frac{616 \ln (r)}{9 (r+1)}
-\frac{40 \ln (r)}{9 (1-r)^2}
-\frac{424 \ln (r)}{9 (r+1)^2}
-\frac{32}{9} \pi^2 \ln (r)
-\frac{64 \ln (r)}{3}
-\frac{64}{3} \ln (1-r)
\right.\el\hspace{-19.5ex}\left.
\times \li2(-r)
+\frac{224}{3} \ln (r) \li2(-r)
-\frac{64 \li2(-r)}{3 (1-r)}
+\frac{64 \li2(-r)}{3 (1-r)^2}
-\frac{256 \li2(-r)}{9}
+\frac{64}{3} \ln (1-r)
\right.\el\hspace{-19.5ex}\left.
\times \li2(r)
+64 \ln (r) \li2(r)
+\frac{32 \li2(r)}{3 (r+1)}
-\frac{32 \li2(r)}{3 (r+1)^2}
-\frac{416 \li2(r)}{9}
+64 \li3(1-r)
-\frac{128 \li3(r)}{3}
\right.\el\hspace{-19.5ex}\left.
-\frac{224 \li3(-r)}{3}
-64 \li3\left(\frac{1}{r+1}\right)
+\frac{32}{3} \li3(1-r^2)-\frac{16 \pi^2}{9 
(1-r)}-\frac{40}{9 (1-r)}
+\frac{1520}{81}
\right.\el\hspace{-19.5ex}\left.
-\frac{16 \pi^2}{9 (r+1)}-\frac{232}{9 (r+1)}
+\frac{16 \pi^2}{9 (1-r)^2}+\frac{16 \pi^2}{9 (r+1)^2}+\frac{152 \zeta_3}{9}+\frac{218 \pi^2}{27}
+\mbox{\large $\Lambda$}_{n_f}\left({\tauo Q \over 2\omega},r\right)\right]
\eea
for the finite part of the two-loop integrated jet thrust distribution. 
The functions $\mbox{\large $\Lambda$}_{C_A}\left(x,r\right)$ and $\mbox{\large $\Lambda$}_{n_f}\left(x,r\right)$ are defined in Appendix \ref{app:chi}. It is straightforward to check explicitly using
Eq.\ (\ref{eq:finalpoly}) above and the expressions for the hard and integrated jet functions collected in Appendix \ref{app:refeqs} that our $\Ordals 2$ result
is consistent with the prediction of the SCET-based resummation of the jet thrust distribution discussed in reference~\cite{Kelley:2011tj}. 
As another non-trivial cross-check, we confirmed that our result is equal to the soft part of the $\Ordals 2$ integrated thrust distribution in the hemisphere limit ($r \rightarrow 1$). In the next section, we analyze the 
small $r$ limit in detail and show that it has a non-trivial structure.

\section{The Small $r$ Limit}
\label{sec:smallR}
In this section, we study the small $r$ limit of the two-loop integrated jet thrust distribution from several points of view. 
Although a compact, closed-form, expression valid for arbitrary $r$ is certainly interesting in its own right, there is physics in the result which is obscured by Eq.\ (\ref{eq:finalpoly}) as it stands;
it is clear that $\bar{K}_{TC}^{(2)}(\tauo,\omega,r,\mu)$ will simplify considerably in the small $r$ limit. Although we could start by taking the small $r$ limit of Eq.\ (\ref{eq:finalpoly}) directly,
we prefer instead to consider the small $r$ limit of each individual contribution to $\bar{K}_{TC}^{(2)}(\tauo,\omega,r,\mu)$ in turn.
Proceeding in this fashion makes it clear that there is a connection between the $\ln(r)$ terms in the all-out contributions and the $\ln(r)$ terms in the all-in contributions which demands an explanation.
We provide such an explanation at the end of Section \ref{sec:smallRres} and, furthermore, make it clear that an analogous result holds at $\Ordals L$. In Section \ref{sec:lnrstruct}, we show that, in fact,
the small $r$ limit of Eq.\ (\ref{eq:finalpoly})
has non-trivial structure as well and that this only becomes apparent if one takes the small $r$ limit in the way advocated by the authors of reference~\cite{Kelley:2011aa}.

\subsection{Contributions in the Small $r$ Limit}
\label{sec:smallRres}
We begin by taking the small $r$ limit of each contribution to the soft part of the $\Ordals 2$ integrated jet thrust distribution,
$\bar{K}^{{\rm in\,Ren}}_{TC}$, $\bar{K}^{{\rm all-in}\,(2)}_{TC}$,
$\bar{K}^{{\rm in\,R-V}}_{TC}$, $\bar{K}^{{\rm out\,Ren}}_{TC}$,
$\bar{K}^{{\rm out\,R-V}}_{TC}$, $\bar{K}_{TC}^{{\rm opp\,in-in}}$,
$\bar{K}_{TC}^{{\rm in-out}\,(2)}$,
and $\bar{K}_{TC}^{{\rm all-out}\,(2)}$, in turn.
The limit $r \to 0$ is performed for fixed $\tauo$, $\omega$, $\mu$ and $Q$.
We start from the expressions given in Appendix~\ref{app:indcontrib}
and keep logarithmic terms proportional to $\ln^k(r)$, $k=0,\ldots,4$, neglecting
all power-suppressed contributions.
For the univariate contributions involving $r$ alone we gave results in terms
of $G$ functions with argument $r$, such that the limit $r \to 0$ can be
immediately obtained.
On the other hand, the small $r$ asymptotics of
$\mbox{\large $\Lambda$}_{C_A}\left(x,r\right)$ and
$\mbox{\large $\Lambda$}_{n_f}\left(x,r\right)$ are non-trivial to extract.
In the small $r$ limit, these functions become independent of $x$ and are in fact proportional to the coefficients of the sub-leading non-global logarithms (see~\cite{Kelley:2011aa}):
\bea
\mbox{\large $\Lambda$}_{C_A}\left(x,r\rightarrow 0\right) &=& -2\left(-{8\over 3} + {88\pi^2\over 9} - 16 \zeta_3\right)\ln(r)
\label{eq:smallRchica}
\\
\mbox{\large $\Lambda$}_{n_f}\left(x,r\rightarrow 0\right) &=& -2\left({16 \over 3} - {32\pi^2 \over 9}\right)\ln(r)\,.
\label{eq:smallRchinf}
\eea

Consequently, we find the following:
\bea
\label{eq:smallrinren}
\hspace{18ex}&&\hspace{-20ex}
\bar{K}^{{\rm in\,Ren}}_{TC}(\tauo,\omega,r\rightarrow 0,\mu) = C_A C_F \left[\frac{176}{9} \ln ^3\left(\frac{\mu }{Q \tau _{\omega }}\right)
+\frac{88}{3} \ln (r) \ln ^2\left(\frac{\mu }{Q \tau _{\omega }}\right)+\left(\frac{44 \ln ^2(r)}{3}
\right.\right.
\el\hspace{-19ex}
\left.\left.
-\frac{22 \pi^2}{9}\right) \ln \left(\frac{\mu }{Q \tau _{\omega }}\right)+\frac{22 \ln ^3(r)}{9}-\frac{11}{9} \pi ^2 \ln (r)-\frac{44 \zeta _3}{9}\right]
   +C_F n_f T_F \left[-\frac{64}{9} \ln ^3\left(\frac{\mu }{Q \tau _{\omega}}\right)
\right.
\el\hspace{-19ex}
\left.
   -\frac{32}{3} \ln (r) \ln ^2\left(\frac{\mu }{Q \tau _{\omega }}\right)+\left(\frac{8 \pi ^2}{9}-\frac{16 \ln ^2(r)}{3}\right) \ln \left(\frac{\mu }{Q \tau _{\omega }}\right)
   -\frac{8 \ln^3(r)}{9}+\frac{4}{9} \pi ^2 \ln (r)+\frac{16 \zeta _3}{9}\right]\,,
\el
\\
\label{eq:smallrallin}
\hspace{18ex}&&\hspace{-20ex}
\bar{K}^{{\rm all-in}\,(2)}_{TC}(\tauo,\omega,r\rightarrow 0,\mu) = C_A C_F \left[-\frac{64}{3} \ln ^4\left(\frac{\mu }{Q \tau _{\omega }}\right)
+\left(-\frac{128 \ln (r)}{3}-\frac{352}{9}\right) \ln ^3\left(\frac{\mu }{Q \tau _{\omega }}\right)
\right.
\el\hspace{-19ex}
\left.
+\left(-32 \ln ^2(r)-\frac{176 \ln (r)}{3}+\frac{16 \pi ^2}{3}-\frac{536}{9}\right) \ln ^2\left(\frac{\mu }{Q \tau _{\omega }}\right)
+\left(\frac{232 \zeta _3}{3}-\frac{32 \ln ^3(r)}{3}-\frac{88 \ln ^2(r)}{3}
\right.\right.
\el\hspace{-19ex}
\left.\left.
+\frac{16}{3} \pi ^2 \ln (r)-\frac{536 \ln (r)}{9}-\frac{22 \pi ^2}{9}-\frac{1544}{27}\right) \ln \left(\frac{\mu }{Q \tau _{\omega }}\right)
+\frac{116}{3} \zeta _3 \ln (r)
-\frac{4 \ln ^4(r)}{3}
\right.
\el\hspace{-19ex}
\left.
-\frac{44 \ln ^3(r)}{9}+\frac{4}{3}\pi ^2 \ln ^2(r)-\frac{134 \ln ^2(r)}{9}-\frac{11}{9} \pi ^2 \ln (r)-\frac{772 \ln (r)}{27}+\frac{137 \pi ^4}{180}-\frac{67 \pi ^2}{54}-\frac{2392}{81}
  \right.
  \el\hspace{-19ex}
  \left.
   -\frac{242 \zeta _3}{9}\right]+C_F n_f T_F \left[\frac{128}{9}\ln ^3\left(\frac{\mu }{Q \tau _{\omega }}\right)+\left(\frac{64 \ln (r)}{3}+\frac{160}{9}\right) \ln ^2\left(\frac{\mu }{Q \tau _{\omega }}\right)
   \right.
\el\hspace{-19ex}
  \left.
   +\left(\frac{32 \ln ^2(r)}{3}+\frac{160 \ln (r)}{9}+\frac{8 \pi ^2}{9}+\frac{304}{27}\right) \ln \left(\frac{\mu }{Q \tau _{\omega }}\right)+\frac{16 \ln ^3(r)}{9}+\frac{40 \ln ^2(r)}{9}+\frac{4}{9} \pi ^2 \ln (r)
   \right.
\el\hspace{-19ex}
   \left.
   +\frac{152 \ln (r)}{27}+\frac{10 \pi^2}{27}+\frac{476}{81}+\frac{88 \zeta _3}{9}\right]\,,
\\[3ex]
\label{eq:smallrinrv}
\hspace{18ex}&&\hspace{-20ex}
\bar{K}^{{\rm in\,R-V}}_{TC}(\tauo,\omega,r\rightarrow 0,\mu) = C_A C_F \left[\frac{64}{3} \ln ^4\left(\frac{\mu }{Q \tau _{\omega }}\right)+\left(-\frac{64 \zeta _3}{3}
+\frac{32 \ln ^3(r)}{3}
-8 \pi ^2 \ln (r)\vphantom{-\frac{64 \zeta _3}{3}}\right)
\right.
\el\hspace{-19ex}
\left.
\times \ln \left(\frac{\mu }{Q \tau _{\omega }}\right)
+\frac{128}{3} \ln (r) \ln^3\left(\frac{\mu }{Q \tau _{\omega }}\right)+\left(32 \ln ^2(r)-8 \pi ^2\right) \ln ^2\left(\frac{\mu }{Q \tau _{\omega }}\right)
-\frac{32}{3} \zeta _3 \ln (r)
\right.
\el\hspace{-19ex}
\left.
+\frac{4 \ln ^4(r)}{3}-2 \pi ^2 \ln ^2(r)+\frac{\pi^4}{60}\right]\,,
\\[3ex]
\label{eq:smallroutren}
\hspace{18ex}&&\hspace{-20ex}
\bar{K}^{{\rm out\,Ren}}_{TC}(\tauo,\omega,r\rightarrow 0,\mu) = C_A C_F \left[\frac{88 \zeta _3}{3}-\frac{88}{3} \ln (r) \ln ^2\left(\frac{\mu }{2 \omega }\right)+\left(\frac{44 \ln ^2(r)}{3}
+\frac{44 \pi ^2}{9}\right)\right.
\el\hspace{-19ex}
\left.
\times \ln \left(\frac{\mu }{2 \omega }\right)-\frac{22 \ln^3(r)}{9}+\frac{11}{9} \pi ^2 \ln (r)\right]
+C_F n_f T_F \left[-\frac{32 \zeta _3}{3}+\frac{32}{3} \ln (r) \ln ^2\left(\frac{\mu }{2 \omega }\right)
\right.
\el\hspace{-19ex}
\left.
+\left(-\frac{16}{3} \ln ^2(r)-\frac{16 \pi ^2}{9}\right) \ln\left(\frac{\mu }{2 \omega }\right)+\frac{8 \ln ^3(r)}{9}-\frac{4}{9} \pi ^2 \ln (r)\right]\,,
\\[3ex]
\label{eq:smallroutrv}
\hspace{18ex}&&\hspace{-20ex}
\bar{K}^{{\rm out\,R-V}}_{TC}(\tauo,\omega,r\rightarrow 0,\mu) = C_A C_F \left[-\frac{128}{3} \ln (r) \ln ^3\left(\frac{\mu }{2 \omega}\right)
+\left(128 \zeta _3-\frac{32 \ln ^3(r)}{3}
\right.\right.
\el\hspace{-19ex}
\left.\left.
+8 \pi ^2 \ln (r)\vphantom{-\frac{32 \ln ^3(r)}{3}}\right)\ln \left(\frac{\mu }{2 \omega }\right)
+\frac{32}{3} \zeta _3 \ln (r)+\left(32 \ln ^2(r)+\frac{32 \pi ^2}{3}\right) \ln ^2\left(\frac{\mu }{2 \omega }\right)
+\frac{4 \ln ^4(r)}{3}
\right.
\el\hspace{-19ex}
\left.
-2 \pi ^2 \ln ^2(r)+\frac{2 \pi ^4}{15}\right]\,,
\\[3ex]
\label{eq:smallroppinin}
\hspace{18ex}&&\hspace{-20ex}
\bar{K}_{TC}^{{\rm opp\,in-in}}(\tauo,\omega,r\rightarrow 0,\mu) = 0\,,
\\[3ex]
\label{eq:smallrinout}
\hspace{18ex}&&\hspace{-20ex}
\bar{K}_{TC}^{{\rm in-out}\,(2)}(\tauo,\omega,r\rightarrow 0,\mu) = C_A C_F \left[-\frac{8}{3} \pi ^2 \ln ^2\left(\frac{Q \tau _{\omega }}{2 \omega }\right)
-16 \zeta _3 \ln \left(\frac{Q \tau _{\omega }}{2 \omega }\right)
\right.
\el\hspace{-19ex}
\left.
+\frac{16}{3} \pi ^2 \ln ^2\left(\frac{\mu }{Q \tau _{\omega }}\right)
+\frac{16}{3} \pi ^2 \ln ^2\left(\frac{\mu }{2 \omega }\right)+\left(\frac{176 \pi ^2}{9}
-\frac{16}{3}\right) \ln \left(\frac{\mu }{2 \omega }\right)
+\frac{88 \zeta _3}{3}+16 \zeta _3 \ln (r)
\right.
\el\hspace{-19ex}
\left.
-\frac{88}{9} \pi ^2 \ln (r)
+\frac{8 \ln (r)}{3}
+\frac{34 \pi ^4}{45}-\frac{268 \pi ^2}{27}+\frac{80}{9}\right]
+C_F n_f T_F \left[\left(\frac{32}{3}-\frac{64 \pi ^2}{9}\right) \ln \left(\frac{\mu }{2 \omega }\right)
\right.
\el\hspace{-19ex}
\left.
+\frac{32}{9} \pi ^2 \ln (r)-\frac{16 \ln (r)}{3}
+\frac{128 \pi ^2}{27}-\frac{136}{9}-\frac{32 \zeta_3}{3}\right]\,,
\\[3ex]
\label{eq:smallrallout}
\hspace{18ex}&&\hspace{-20ex}
\bar{K}_{TC}^{{\rm all-out}\,(2)}(\tauo,\omega,r\rightarrow 0,\mu) = C_A C_F \left[\frac{128}{3} \ln (r) \ln ^3\left(\frac{\mu }{2 \omega }\right)
+\left(-32 \ln ^2(r)+\frac{176 \ln (r)}{3}
\right.\right.
\el\hspace{-19ex}
\left.\left.
-16 \pi ^2\vphantom{\frac{128}{3}}\right) \ln ^2\left(\frac{\mu }{2 \omega}\right)
+\left(-128 \zeta _3+\frac{32 \ln ^3(r)}{3}-\frac{88 \ln ^2(r)}{3}
-\frac{16}{3} \pi ^2 \ln (r)+\frac{536 \ln (r)}{9}-\frac{176 \pi^2}{9}
\right.\right.
\el\hspace{-19ex}
\left.\left.
+\frac{8}{3}\right)\ln \left(\frac{\mu }{2 \omega }\right) -\frac{116}{3} \zeta _3 \ln (r)
-\frac{4 \ln ^4(r)}{3}+\frac{44 \ln ^3(r)}{9}+\frac{4}{3} \pi ^2 \ln ^2(r)
-\frac{134 \ln ^2(r)}{9}
\right.
\el\hspace{-19ex}
\left.
+\frac{11}{9} \pi ^2 \ln (r)+\frac{772 \ln (r)}{27}+\frac{34 \pi ^4}{45}-\frac{268 \pi^2}{27}+\frac{4}{9}-\frac{308 \zeta _3}{3}\right]
+C_F n_f T_F \left[-\frac{64}{3} \ln (r) \ln ^2\left(\frac{\mu }{2 \omega }\right)
\right.
\el\hspace{-19ex}
\left.
+\left(\frac{32 \ln ^2(r)}{3}-\frac{160 \ln (r)}{9}+\frac{64 \pi^2}{9}
-\frac{16}{3}\right) \ln \left(\frac{\mu }{2 \omega }\right)-\frac{16 \ln ^3(r)}{9}+\frac{40 \ln ^2(r)}{9}-\frac{4}{9} \pi ^2 \ln (r)
\right.
\el\hspace{-19ex}
\left.
-\frac{152 \ln (r)}{27}+\frac{80 \pi ^2}{27}-\frac{20}{9}+\frac{112 \zeta _3}{3}\right]\,.
\eea

Here, we have chosen to present the small $r$ limit of each contribution individually because it highlights a remarkable feature of the results which would otherwise have remained hidden. 
Comparing Eqs.\ (\ref{eq:smallrinren}) and (\ref{eq:smallroutren}), (\ref{eq:smallrinrv}) and (\ref{eq:smallroutrv}), and, finally, (\ref{eq:smallrallin}) and (\ref{eq:smallrallout}), 
we see that the $\ln(r)$ terms in the out-of-jet contributions
can be obtained from the in-jet contributions by making the replacements $r\rightarrow 1/r$ and $\tauo Q \rightarrow 2\omega$. 
This is interesting because the in-jet contributions are almost trivial to calculate
once one has the ordinary thrust distribution under control (see Section \ref{sec:allinpredict}) and, by comparison, the calculation of the out-of-jet contributions is much more difficult. 
The correspondence between $\bar{K}^{{\rm in\,Ren}}_{TC}(\tauo,\omega,r\rightarrow 0,\mu)$ and $\bar{K}^{{\rm out\,Ren}}_{TC}(\tauo,\omega,r\rightarrow 0,\mu)$ and
between $\bar{K}^{{\rm in\,R-V}}_{TC}(\tauo,\omega,r\rightarrow 0,\mu)$ and $\bar{K}^{{\rm out\,R-V}}_{TC}(\tauo,\omega,r\rightarrow 0,\mu)$ could be written off as a one-loop accident but this is 
clearly not the case since the $\ln(r)$ terms in $\bar{K}_{TC}^{{\rm all-out}\,(2)}(\tauo,\omega,r\rightarrow 0,\mu)$ can be predicted as well. In fact, we can provide a explanation for this phenomenon 
by identifying the configurations of the soft parton momenta which give a dominate contribution to $K_{TC}^{{\rm all-out}\,(2)}(\tauo,\omega,r,\mu)$ in the small $r$ limit. 
Actually, our argument is independent of the loop order and
we can trivially generalize the discussion from the $\Ordals 2$ all-out contributions, $K_{TC}^{{\rm all-out}\,(2)}(\tauo,\omega,r,\mu)$, to the $\Ordals L$ all-out contributions,
\begin{eqnarray}
\label{eq:alloutL}
K_{TC}^{{\rm all-out}\,(L)}(\tauo,\omega,r,\mu)
&=& \left( \frac{\mu^2 e^{\gamma_E}}{4\pi}\right)^{{4 - d\over 2}L}\int_0^{\tau_\omega}\dd \tau_\omega^\prime \int_0^\omega
\dd \lambda \int \dd k_L \dd k_R 
\el\hspace{-20ex}\quad
\times (-2\pi i)^L\prod_{i = 1}^L \left(\int\frac{\dd^d q_i}{(2\pi)^d} \delta\left(q_i^2\right) \Theta\left(q^+_i\right)\Theta\left(q^-_i\right)
\Theta\left(q_i^- - r q_i^+\right) \Theta\left(q_i^+ - r q_i^-\right)\right)
\el\hspace{-20ex}\quad
\times \,I^{(L)}\left(q_i^+,q_i^-,{\bf q}_T^{(i)}\right)\delta\left(k_L\right) \delta\left(k_R\right) \delta\left(\lambda - \sum_{i = 1}^L {q_i^- + q_i^+\over 2}\right) 
\delta\left(\tau_\omega^\prime - {k_L + k_R\over Q}\right)
\el\hspace{-20ex}= 
\left( \frac{\mu^2 e^{\gamma_E}}{4\pi}\right)^{{4 - d\over 2}L} (-2\pi i)^L
\el\hspace{-20ex}\quad
\times\int_0^{2 \omega}\dd y\prod_{i = 1}^L \left(\int\frac{\dd^d q_i}{(2\pi)^d} \delta\left(q_i^2\right) \Theta\left(q^+_i\right)\Theta\left(q^-_i\right)
\Theta\left(q_i^- - r q_i^+\right) \Theta\left(q_i^+ - r q_i^-\right)\right) 
\el\hspace{-20ex}\quad
\times I^{(L)}\left(q_i^+,q_i^-,{\bf q}_T^{(i)}\right)\delta\left(y - \sum_{i = 1}^L \Big(q_i^- + q_i^+\Big)\right) \,.
\end{eqnarray}

By examining the structure of the all-out integrand in Eq.\ (\ref{eq:alloutL}), we see that, when $r$ is very small, the integrals over the $q_i$ are dominated by configurations where either
 $q_i^+ \gg q_i^-$ or $q_i^- \gg q_i^+$. In other words, the soft parton momenta actually exhibit
collinear scaling in the extreme small $r$ limit.
Starting from the last line of Eq.\ (\ref{eq:alloutL}) above, we see that
\bea
\label{eq:smallalloutL1}
K^{{\rm all-out}\,(L)}_{TC}(\tau_\omega,\omega,r\Rrightarrow 0,\mu)
&=& 2 K^{{\rm all-out}\,(L)}_{TC}(\tau_\omega,\omega,r,\mu)\Big|_{q_i^+ \gg q_i^-}
\el\hspace{-25ex}
= 2\left( \frac{\mu^2 e^{\gamma_E}}{4\pi}\right)^{{4 - d\over 2}L} (-2\pi i)^L
\int_0^{2 \omega}\!\dd y \prod_{i = 1}^L \left(\int\frac{\dd^d q_i}{(2\pi)^d} \delta\left(q_i^2\right) \Theta\left(q^+_i\right)\Theta\left(q^-_i\right)\Theta\left(q_i^- - r q_i^+\right)\right)\,
\el\hspace{-25ex}\quad
\times I^{(L)}\left(q_i^+,q_i^-,{\bf q}_T^{(i)}\right) \delta\left(y - \sum_{i = 1}^L q_i^+\right)
\el\hspace{-25ex}=
2 \left( \frac{\mu^2 e^{\gamma_E}}{4\pi}\right)^{{4 - d\over 2}L} (-2\pi i)^L\int_0^{2 \omega}
\!\dd y 
\prod_{i = 1}^L \left(\int\frac{\dd^d q_i}{(2\pi)^d} \delta\left(q_i^2\right) \Theta\left(q^+_i\right)\Theta\left(q^-_i\right)
\Theta\left(q_i^-/r - q_i^+\right)\right)\,
\el\hspace{-25ex}\quad
\times I^{(L)}\left(q_i^+,q_i^-,{\bf q}_T^{(i)}\right) \delta\left(y - \sum_{i = 1}^L q_i^+\right)
\el\hspace{-25ex}
= K^{{\rm all-in}\,(L)}_{TC}(\tau_\omega,\omega,1/r,\mu)\Big|_{\tauo Q \rightarrow 2 \omega}\,,
\eea
where we have used the fact that, by symmetry, we can simply take the $q_i^+ \gg q_i^-$ limit of $2 K_{TC}^{{\rm all-out}\,(L)}(\tauo,\omega,r,\mu)$ to account for the configurations where $q_i^- \gg q_i^+$. Thus,
the observed one- and two-loop correspondences between $\ln(r)$ terms have a simple explanation which generalizes to all loop orders.
Note that, due to the way in which we take the small $r$ limit, we cannot hope to correctly
predict the constant terms in $K_{TC}^{{\rm all-out}\,(L)}(\tauo,\omega,r,\mu)$. 
To emphasize this point, we use a $\Rrightarrow$ to denote the extreme small $r$ limit of $K^{{\rm all-out}\,(L)}_{TC}(\tau_\omega,\omega,r,\mu)$, where constants as well as power-suppressed terms are neglected.

Furthermore, we note that Eq.\ (\ref{eq:intcone}) now immediately implies that the 
$\ln(r)$ terms in the all-out contributions to the integrated jet thrust distribution are fixed by the pure same-hemisphere contributions to the integrated thrust distribution:
\bea
K^{{\rm all-out}\,(L)}_{TC}(\tau_\omega,\omega,r\Rrightarrow 0,\mu) = r^{-\ep L} \left({\mathop{K}^{\Rightarrow}}^{\,(L)}_{hemi}(\tauo,\mu) + {\mathop{K}^{\Leftarrow}}^{\,(L)}_{hemi}(\tauo,\mu)\right)_{\tauo Q \rightarrow 2 \omega}\,.
\eea

\subsection{The Structure of the $\ln(r)$ Terms}
\label{sec:lnrstruct}
We begin by presenting the small $r$ limit of Eq.\ (\ref{eq:finalpoly}) in a convenient form:
\bea
\label{eq:smallRtot}
\bar{K}^{(2)}_{TC}(\tau_\omega, \omega, r\rightarrow 0, \mu) &=& C_A C_F \left[-\frac{176}{9} \ln^3\!\left(\frac{\mu }{Q \tau_{\omega}}\right)
+\left(-\frac{88 \ln (r)}{3}+\frac{8 \pi^2}{3}-\frac{536}{9}\right)
\right.\el\hspace{-20ex}\left.
\times\ln^2\!\left(\frac{\mu }{Q \tau_{\omega}}\right)
+\left(-\frac{44}{3} \ln^2(r) +\frac{8}{3} \pi^2 \ln (r)-\frac{536 \ln (r)}{9}+56 \zeta_3+\frac{44 \pi^2}{9}
-\frac{1616}{27}\right)
\right.\el\hspace{-20ex}\left.
\times\ln \left(\frac{\mu }{Q \tau_{\omega}}\right)
+\left(-\frac{44}{3} \ln^2(r)-\frac{8}{3} \pi^2 \ln(r)+\frac{536 \ln (r)}{9}-\frac{44 \pi ^2}{9}\right) \ln\left(\frac{\mu }{2 \omega }\right)
+\frac{88}{3} \ln (r)
\right.\el\hspace{-20ex}\left.
\times \ln^2\left(\frac{\mu }{2 \omega }\right)
-\frac{8}{3} \pi ^2 \ln^2\left(\frac{Q \tau_{\omega }}{2 r \omega }\right)+\left(-16 \zeta_3-\frac{8}{3}+\frac{88 \pi ^2}{9}\right)
\ln \left(\frac{Q \tau_{\omega }}{2 r \omega }\right)
+\frac{4}{3} \pi^2 \ln^2(r)
\right.\el\hspace{-20ex}\left.
-\frac{268 \ln^2(r)}{9} -\frac{682 \zeta_3}{9}+\frac{109 \pi^4}{45}-\frac{1139 \pi ^2}{54}-\frac{1636}{81}\right]
+ C_F n_f T_F\left[\frac{64}{9} \ln^3\!\left(\frac{\mu }{Q \tau_{\omega }}\right)
\right.\el\hspace{-20ex}\left.
+\left(\frac{32 \ln (r)}{3}
+\frac{160}{9}\right)\ln^2\!\left(\frac{\mu }{Q \tau_{\omega }}\right)
+\left(\frac{16 \ln^2(r)}{3}+\frac{160 \ln (r)}{9}-\frac{16 \pi ^2}{9}+\frac{448}{27}\right) \ln \left(\frac{\mu }{Q \tau_{\omega }}\right)
\right.\el\hspace{-20ex}\left.
-\frac{32}{3} \ln(r) \ln^2\left(\frac{\mu }{2 \omega }\right)
+\left(\frac{16 \ln^2(r)}{3}-\frac{160 \ln (r)}{9}+\frac{16 \pi ^2}{9}\right) \ln\left(\frac{\mu }{2 \omega }\right)+\left(\frac{16}{3}
-\frac{32 \pi ^2}{9}\right)
\right.\el\hspace{-20ex}\left.
\times \ln\left(\frac{Q \tau_{\omega }}{2 r \omega }\right)
+\frac{80 \ln^2(r)}{9}+\frac{248 \zeta_3}{9}+\frac{218 \pi ^2}{27}-\frac{928}{81}\right]\,.
\eea
The terms preceding the $-{8\pi^2\over 3} \ln^2\left(Q \tauo\over 2 r \omega\right)$ term in the $C_A C_F$ color structure and the\\ $\left({16\over 3} - {32 \pi^2\over 9}\right) \ln\left(Q \tauo\over 2 r \omega\right)$ term in the
$C_F n_f T_F$ color structure correspond precisely to the terms predicted by the factorization theorem of reference \cite{Kelley:2011tj} in the small $r$ limit, provided that one makes the choices
\begin{equation}
c^{(1)}_{in} = C_F \left[-\pi^2 - 2\ln^2(r)\right]
\end{equation}
and
\begin{equation}
 c^{(1)}_{out} = C_F \left[-{2 \pi^2\over 3} - 2\ln^2(r) - 8\li2 \left(-r\right)\right]
\end{equation}
for the one-loop matching coefficients. This choice is such that the one-loop out-of-jet integrated jet thrust distribution is equal to the full one-loop soft function for two unmeasured jets (see reference \cite{Ellis:2010rwa} for details).

One can check that Eq.\ (\ref{eq:smallRtot}) is equal to the sum of Eqs.\ (\ref{eq:smallrinren})-(\ref{eq:smallrallout}). 
However, to arrive at the above form, one has to rearrange terms in the sum and, clearly, this requires further explanation. By way of motivation, let us consider the in-out
contributions to the $\Ordals L$ integrated jet thrust distribution\footnote{Recall that we have defined the in-out contributions at $\Ordals L$ to be those contributions where $n_{in} > 0$
soft partons get clustered into a single jet, $n_{out} > 0$ soft partons go out of all jets, and the sum $n_{in} + n_{out}$ is equal to the loop order, $L$.},
\bea
\label{eq:inoutL1}
K_{TC}^{{\rm in-out}\,(L)}(\tauo,\omega,r,\mu)
&=& \left( \frac{\mu^2 e^{\gamma_E}}{4\pi}\right)^{{4 - d\over 2}L}\int_0^{\tau_\omega}\dd \tau_\omega^\prime \int_0^\omega
\dd \lambda \int \dd k_L \dd k_R
\el\hspace{-20ex}
\times (-2\pi i)^{n_{out}}\prod_{j = 1}^{n_{out}} \left(\int\frac{\dd^d k_j}{(2\pi)^d} \delta\left(k_j^2\right) \Theta\!\left(k^+_j\right)\Theta\!\left(k^-_j\right)
\Theta\!\left(k_j^- - r k_j^+\right) \Theta\!\left(k_j^+ - r k_j^-\right)\right)
\el\hspace{-20ex}
\times (-2\pi i)^{n_{in}}\prod_{i = 1}^{n_{in}} \left(\int\frac{\dd^d q_i}{(2\pi)^d} \delta\left(q_i^2\right) \Theta\!\left(q^+_i\right)\Theta\!\left(q^-_i\right)
\Theta\!\left(r q_i^- - q_i^+\right)\right)\D\left(k_L\right) \delta\left(k_R - \sum_{i = 1}^{n_{in}} q_i^+\right)
\el\hspace{-20ex}
\times \,I^{(L)}\left(k_j^+,k_j^-,{\bf k}_T^{(j)},q_i^+,q_i^-,{\bf q}_T^{(i)}\right) \delta\left(\lambda - \sum_{j = 1}^{n_{out}} {k_j^- + k_j^+\over 2}\right) 
\delta\left(\tau_\omega^\prime - {k_L + k_R\over Q}\right)\,.
\eea

In order to say something about $K_{TC}^{{\rm in-out}\,(L)}(\tauo,\omega,r,\mu)$ in the small $r$ limit, we first need to put Eq.\ (\ref{eq:inoutL1}) into a form where we can make use of a relation deduced
from the analysis performed in Section \ref{sec:allinpredict}. This is readily accomplished by using the $\delta\big(k_j^2\big)$ and $\delta\left(q_i^2\right)$ to perform the integrals over $|{\bf k}_T^{(j)}|^2$ and $|{\bf q}_T^{(i)}|^2$:
\bea
\label{eq:inoutL2}
K_{TC}^{{\rm in-out}\,(L)}(\tauo,\omega,r,\mu)
&=& \left( \frac{\mu^2 e^{\gamma_E}}{4\pi}\right)^{L\ep}\int_0^{\tau_\omega}\dd \tau_\omega^\prime \int_0^\omega
\dd \lambda \int \dd k_L \dd k_R
\el\hspace{-24ex}
\times (-2\pi i)^{n_{out}}\prod_{j = 1}^{n_{out}} \left(\int\frac{\dd k_j^- \dd k_j^+}{4(2\pi)^{4-2\ep}} \left(k_j^- k_j^+\right)^{-\ep} \Theta\!\left(k^+_j\right)\Theta\!\left(k^-_j\right)
\Theta\!\left(k_j^- - r k_j^+\right) \Theta\!\left(k_j^+ - r k_j^-\right)\right)
\el\hspace{-24ex}
\times (-2\pi i)^{n_{in}}\prod_{i = 1}^{n_{in}} \left(\int\frac{\dd q_i^- \dd q_i^+}{4(2\pi)^{4-2\ep}}  \left(q_i^- q_i^+\right)^{-\ep} \Theta\!\left(q^+_i\right)\Theta\!\left(q^-_i\right)
\Theta\!\left(r q_i^- - q_i^+\right)\right)
\el\hspace{-24ex}
\times \int \dd {\bf \Omega}_\ep~ I^{(L)}\left(k_j^+,k_j^-,q_i^+,q_i^-,{\bf \Omega}\right) \D\left(k_L\right) \delta\left(k_R - \sum_{i = 1}^{n_{in}} q_i^+\right)
\el\hspace{-24ex}
\times \delta\left(\lambda - \sum_{j = 1}^{n_{out}} {k_j^- + k_j^+\over 2}\right) \delta\left(\tau_\omega^\prime - {k_L + k_R\over Q}\right)\,,
\eea
where $\dd {\bf \Omega}_\ep$ represents the angular measure for the $L(L-1)/2$ integrations over the angles specifying the relative orientations of the soft parton momenta. Now, we make the change of variables
\be
k_j^- = {\ell_j^-\over r} \qquad {\rm and} \qquad q_i^- = {\ell_i^-\over r}
\ee
in Eq.\ (\ref{eq:inoutL2}). This leads to
\bea
\label{eq:inoutL3}
K_{TC}^{{\rm in-out}\,(L)}(\tauo,\omega,r,\mu)
&=& r^{(-1+\ep)L} \left( \frac{\mu^2 e^{\gamma_E}}{4\pi}\right)^{L\ep}\int_0^{\tau_\omega}\dd \tau_\omega^\prime \int_0^\omega
\dd \lambda \int \dd k_L \dd k_R
\el\hspace{-24ex}
\times (-2\pi i)^{n_{out}}\prod_{j = 1}^{n_{out}} \left(\int\frac{\dd \ell_j^- \dd k_j^+}{4(2\pi)^{4-2\ep}} \left(\ell_j^- k_j^+\right)^{-\ep} \Theta\!\left(k^+_j\right)\Theta\!\left(\ell^-_j\right)
\Theta\!\left(\ell_j^- - r^2 k_j^+\right) \Theta\!\left(k_j^+ - \ell_j^-\right)\right)
\el\hspace{-24ex}
\times (-2\pi i)^{n_{in}}\prod_{i = 1}^{n_{in}} \left(\int\frac{\dd \ell_i^- \dd q_i^+}{4(2\pi)^{4-2\ep}}  \left(\ell_i^- q_i^+\right)^{-\ep} \Theta\!\left(q^+_i\right)\Theta\!\left(\ell^-_i\right)
\Theta\!\left(\ell_i^- - q_i^+\right)\right)
\el\hspace{-24ex}
\times \int \dd {\bf \Omega}_\ep~ I^{(L)}\left(k_j^+,{\ell_j^-\over r},q_i^+,{\ell_i^-\over r},{\bf \Omega}\right) \D\left(k_L\right) \delta\left(k_R - \sum_{i = 1}^{n_{in}} q_i^+\right)
\delta\left(\lambda - \sum_{j = 1}^{n_{out}} {\ell_j^- + k_j^+ r \over 2 r}\right) 
\el\hspace{-24ex}
\times \delta\left(\tau_\omega^\prime - {k_L + k_R\over Q}\right)
\el\hspace{-25ex}
= r^{\ep L} \left( \frac{\mu^2 e^{\gamma_E}}{4\pi}\right)^{L\ep}\int_0^{\tau_\omega Q}\dd x \int_0^{2 r \omega} \dd y
\el\hspace{-24ex}
\times (-2\pi i)^{n_{out}}\prod_{j = 1}^{n_{out}} \left(\int\frac{\dd \ell_j^- \dd k_j^+}{4(2\pi)^{4-2\ep}} \left(\ell_j^- k_j^+\right)^{-\ep} \Theta\!\left(k^+_j\right)\Theta\left(\ell^-_j\right)
\Theta\left(\ell_j^- - r^2 k_j^+\right) \Theta\left(k_j^+ - \ell_j^-\right)\right)
\el\hspace{-24ex}
\times (-2\pi i)^{n_{in}}\prod_{i = 1}^{n_{in}} \left(\int\frac{\dd \ell_i^- \dd q_i^+}{4(2\pi)^{4-2\ep}}  \left(\ell_i^- q_i^+\right)^{-\ep} \Theta\left(q^+_i\right)\Theta\left(\ell^-_i\right)
\Theta\left(\ell_i^- - q_i^+\right)\right)
\el\hspace{-24ex}
\times \int \dd {\bf \Omega}_\ep~ I^{(L)}\left(k_j^+,\ell_j^-,q_i^+,\ell_i^-,{\bf \Omega}\right) \delta\left(x - \sum_{i = 1}^{n_{in}} q_i^+\right)
\delta\left(y - \sum_{j = 1}^{n_{out}} \Big(\ell_j^- + k_j^+ r\Big)\right)\,,
\eea
where we have used the relation
\begin{equation}
I^{(L)}\left(k_j^+,{k_j^-\over r},q_i^+,{q_i^-\over r},{\bf \Omega}\right) = r^L I^{(L)}\left(k_j^+,k_j^-,q_i^+,q_i^-,{\bf \Omega}\right)\,,
\end{equation}
deduced from the analysis of $K_{TC}^{{\rm all-in}\,(L)}(\tauo,\omega,r,\mu)$ in Section \ref{sec:allinpredict}, in deriving the second line of Eq.\ (\ref{eq:inoutL3}). 
Finally, following the analysis performed in Section 5.2 of reference~\cite{Kelley:2011aa},
we can approximate $\ell_j^- - r^2 k_j^+$ and $\ell_j^- + k_j^+ r$ by $\ell_j^-$  in the small $r$ limit
and then immediately write
\be
\label{eq:smallRinoutL}
K_{TC}^{{\rm in-out}\,(L)}(\tauo,\omega,r\rightarrow 0,\mu)
= 2 \,r^{\ep L} \int_0^{\tau_\omega Q}\dd x \int_0^{2 r \omega}
\dd y~ S_{hemi}^{\big(n_{out},\,n_{in};\, L\big)}(x, y, \mu)\,,
\ee
where $S_{hemi}^{\big(n_{out},\,n_{in};\, L\big)}(x, y, \mu)$ denotes the $\Ordals L$ real-real contributions to the hemisphere soft function with $n_{in}$ soft partons in right hemisphere and $n_{out}$ soft partons in the left hemisphere.

The importance of Eq.\ (\ref{eq:smallRinoutL}) is related to the fact that the in-out contributions, treated above in the abstract,
contribute in an essential way to the non-global logarithms in the $\Ordals L$ integrated $\tauo$ distribution. For generic $r$, the ratio
${\tauo Q \over 2 \omega}$ is what naturally appears in the non-global part of the sector decomposition of these contributions (see {\it e.g.} Eq.\ (\ref{eq:secdecompinout})).
However, in the small $r$ limit, we have effectively just shown that it is actually the ratio ${\tauo Q \over 2 r \omega}$ which comes out of the sector decomposition.
This feature of the calculation is what motivated us to write Eq.\ (\ref{eq:smallRtot}) with non-global logarithms of argument ${\tauo Q \over 2 r \omega}$. Actually, writing things in this way allows one to identify $\ln^2(r)$ terms
analogous to those that appear in the one-loop small $r$ result,
\bea
\hspace{-7ex}K_{TC}^{(1)}(\tauo, \omega, r\rightarrow 0, \mu) &&
\el\hspace{-20ex}=
 C_F \left[-8 \ln^2\left(\frac{\mu }{Q \tau _{\omega }}\right)
-8 \ln (r) \ln \left(\frac{\mu }{Q \tau _{\omega }}\right)
+8\ln (r) \ln \left(\frac{\mu }{2 \omega }\right)
-\frac{\pi ^2}{3}
-4 \ln^2(r)
\right]
\el\hspace{-20ex}
= C_F \left[-8 \ln^2\left(\frac{\mu }{Q \tau _{\omega }}\right)
-8 \ln (r) \ln \left(\frac{\mu }{Q \tau _{\omega }}\right)
+8\ln (r) \ln \left(\frac{\mu }{2 \omega }\right)-\frac{\pi ^2}{3}\right]
-\Gamma_0 \ln^2(r)\,.
\eea
Indeed, from Appendix~\ref{app:refeqs}, we have 
\bea
\label{eq:smallRtot2}
\bar{K}^{(2)}_{TC}(\tau_\omega, \omega, r\rightarrow 0, \mu) &=& C_A C_F \left[-\frac{176}{9} \ln^3\!\left(\frac{\mu }{Q \tau_{\omega}}\right)
+\left(-\frac{88 \ln (r)}{3}+\frac{8 \pi^2}{3}
-\frac{536}{9}\right)\ln^2\!\left(\frac{\mu }{Q \tau_{\omega}}\right)
\right.\el\hspace{-20ex}\left.
+\left(-\frac{44}{3} \ln^2(r) +\frac{8}{3} \pi^2 \ln (r)-\frac{536 \ln (r)}{9}+56 \zeta_3+\frac{44 \pi^2}{9}
-\frac{1616}{27}\right)\ln \left(\frac{\mu }{Q \tau_{\omega}}\right)
\right.\el\hspace{-20ex}\left.
+\left(-\frac{44}{3} \ln^2(r)
-\frac{8}{3} \pi^2 \ln(r)+\frac{536 \ln (r)}{9}-\frac{44 \pi ^2}{9}\right) \ln\left(\frac{\mu }{2 \omega }\right)
+\frac{88}{3} \ln (r) \ln^2\!\left(\frac{\mu }{2 \omega }\right)
\right.\el\hspace{-20ex}\left.
-\frac{8}{3} \pi ^2 \ln^2\!\left(\frac{Q \tau_{\omega }}{2 r \omega }\right)+\left(-16 \zeta_3-\frac{8}{3}+\frac{88 \pi ^2}{9}\right)
\ln \left(\frac{Q \tau_{\omega }}{2 r \omega }\right)
-\frac{682 \zeta_3}{9}+\frac{109 \pi^4}{45}-\frac{1139 \pi ^2}{54}
\right.\el\hspace{-20ex}\left.
-\frac{1636}{81}\right]
+ C_F n_f T_F\left[\frac{64}{9} \ln^3\!\left(\frac{\mu }{Q \tau_{\omega }}\right)+\left(\frac{32 \ln (r)}{3}+\frac{160}{9}\right) \ln^2\!\left(\frac{\mu }{Q \tau_{\omega }}\right)
+\left(\frac{16 \ln^2(r)}{3}
\right.\right.\el\hspace{-20ex}\left.\left.
+\frac{160 \ln (r)}{9}-\frac{16 \pi ^2}{9}+\frac{448}{27}\right) \ln \left(\frac{\mu }{Q \tau_{\omega }}\right)-\frac{32}{3} \ln(r) \ln^2\!\left(\frac{\mu }{2 \omega }\right)
+\left(\frac{16 \ln^2(r)}{3}-\frac{160 \ln (r)}{9}
\right.\right.\el\hspace{-20ex}\left.\left.
+\frac{16 \pi ^2}{9}\right) \ln\left(\frac{\mu }{2 \omega }\right)+\left(\frac{16}{3}-\frac{32 \pi ^2}{9}\right) \ln\left(\frac{Q \tau_{\omega }}{2 r \omega }\right)
+\frac{248 \zeta_3}{9}+\frac{218 \pi ^2}{27}-\frac{928}{81}\right] - \G_1 \ln^2(r)\,;
\nonumber\\ &&
\eea
the logarithms of $r$ which cannot be absorbed into the non-global logarithms in the way suggested by Eq.\ (\ref{eq:smallRinoutL}) are simply expressed in terms
of the $\Ordals 2$ coefficient in the perturbative expansion of the cusp anomalous dimension. Certainly,
this structure would have been difficult, if not impossible, to see if we had insisted that the small $r$ non-global logarithms have argument ${\tauo Q\over 2\omega}$. This supports the conjecture made in reference~\cite{Kelley:2011aa}
that the resummation of the logarithms of $r$ is tied up with the resummation of the logarithms of ${\tauo Q\over 2\omega}$; the form of Eq.\ (\ref{eq:smallRtot2}) suggests that these two classes of potentially large logarithms
not controlled by the factorization theorem for jet thrust are naturally intertwined. However, the presence of the term $- \G_1 \ln^2(r)$ in Eq.\ (\ref{eq:smallRtot2}) 
signals that there is further structure in the integrated distribution which remains to be understood. It would be very interesting to further explore the resummation properties of the $\ln(r)$ terms in future work
since, so far, they have not been discussed in the SCET literature.

\section{Conclusions}
\label{sec:conclusions}
In this work, we completed the study of the soft part of the two-loop integrated jet thrust distribution begun in reference~\cite{Kelley:2011aa}. In fact, we found it useful to recompute all of the contributions to the integrated distribution
derived in previous work to highlight the fact that the thrust cone jet algorithm used to define the jet thrust has an associated soft function with some very appealing theoretical properties. In particular, in the small $r$ limit,
{\it all} potentially large logarithms in the two-loop result are straightforwardly obtainable from either the one-loop result or contributions to known observables defined using the simpler hemisphere jet algorithm. Remarkably, for the
all-in\footnote{Actually, as was shown in Section \ref{sec:allinpredict}, the relation that we derived for the all-in contributions
has nothing to do with the small $r$ limit. In that particular case, the $L$-loop relation we derived is a simple consequence of Lorentz covariance and therefore exact.}, in-out, and all-out contributions we were even able to generalize
the discussion from $\Ordals 2$ to $\Ordals L$ under appropriate assumptions.

At this juncture, one might be tempted to speculate, as was done in reference~\cite{Kelley:2011aa}, that, for sufficiently small $r$, all contributions to the soft part of the integrated $\tauo$ distribution
at all loop orders are straightforwardly obtainable from lower-loop contributions, related to contributions to hemisphere observables in a simple way, or power-suppressed.
However, we find it by no means obvious that such a simple picture should hold. Rather, our success at two loops is most probably related to the fact that, 
at $\Ordals 2$, only two of the three regions of the final state phase space are simultaneously accessible. At three-loop order and beyond, it is possible to simultaneously probe all three regions of the 
final state phase space with soft radiation and there is no natural reason to expect contributions to the $\tauo$ soft function such as the one depicted in Figure 5 to be power-suppressed in the small $r$ limit. This is
unfortunate since it was shown in~\cite{Kelley:2011aa} that, at two-loop order, a small $r$ approximation to the full result based on the integrated hemisphere soft function works surprising well 
for realistic values of the jet cone size.
\begin{figure}[!t]
\begin{center}
  \includegraphics[width=0.7\textwidth]{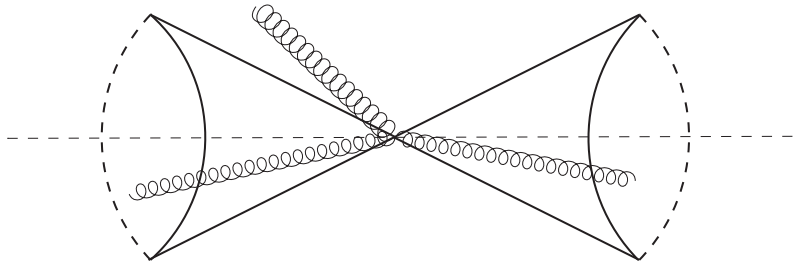}
  \caption{Starting at $\Ordals 3$, there will be contributions to the $\tauo$ soft function which simultaneously probe all three regions of the final state phase space.}
\end{center}
\end{figure}

Another speculation made in reference~\cite{Kelley:2011aa} (see {\it e.g.} Eqs. (5.16) and (6.1) of that work) was that all of the large logarithms of $r$ in the small $r$ limit are intertwined with the non-global logarithms.
In other words, it was shown that the potentially large logarithms in the small $r$ limit not controlled by the factorization theorem for jet thrust have the form
\bea
&& C_A C_F \left(-\frac{8}{3} \pi ^2 \ln^2\left(\frac{Q \tau_{\omega }}{2 r \omega }\right)+\left(-16 \zeta_3-\frac{8}{3}+\frac{88 \pi ^2}{9}\right) \ln \left(\frac{Q \tau_{\omega }}{2 r \omega }\right)\right)
\el
+  C_F n_f T_F \left(\frac{16}{3}-\frac{32 \pi ^2}{9}\right) \ln\left(\frac{Q \tau_{\omega }}{2 r \omega }\right) + f(r)
\eea
and it was suggested that the function $f(r)$ might be identically zero. In Section \ref{sec:lnrstruct} it was shown that, actually, 
\bea
f(r) &=& C_A C_F \left(\frac{4\pi^2}{3} - \frac{268}{9}\right) \ln^2(r) + C_F n_f T_F \frac{80}{9} \ln^2(r)
\\
&=& -\Gamma_1 \ln^2(r)\,,
\eea
where $\Gamma_1$ is the $\Ordals 2$ coefficient in the perturbative expansion of the cusp anomalous dimension.
It is curious to note that this follows the pattern of the one-loop result,
\bea
\hspace{-7ex}K_{TC}^{(1)}(\tauo, \omega, r\rightarrow 0, \mu) &=&
\el\hspace{-20ex}
 C_F \left(-8 \ln^2\left(\frac{\mu }{Q \tau _{\omega }}\right)
-8 \ln (r) \ln \left(\frac{\mu }{Q \tau _{\omega }}\right)
+8\ln (r) \ln \left(\frac{\mu }{2 \omega }\right)-\frac{\pi ^2}{3}\right)
-\Gamma_0 \ln^2(r)\,.
\eea
If this pattern generalizes to higher loop orders, the $\ln(r)$ terms would not necessarily complicate the problem of non-global logarithm resummation.
Rather, the situation could then be comparable to the case of the integrated doubly differential dijet invariant mass distribution defined using a hemisphere jet algorithm (see \cite{Kelley:2011ng}).

One point worth emphasizing concerns the techniques employed in our calculation and their potential application to future problems.
In earlier work, it was pointed out that an analytical understanding of the non-global logarithms which arise in the integrated hemisphere soft function
beyond the two-loop level is an important outstanding problem.
In this paper, we employed dedicated integration routines and an extension of the coproduct calculus for multiple polylogarithms.
We extended these methods to cover not only standard multivariate multiple polylogarithms, but also generalized univariate multiple polylogarithms with
analytical structures beyond what have been discussed in the literature so far.
An important feature of our formalism, discussed in Section \ref{sec:coprod}, is that it allowed us to carry out our calculation in a highly automated way.
We anticipate it will be useful both for the computation of the leading three-loop non-global logarithms and in other contexts. 

Of course, it may well be the case that further technical progress needs to be
made in order to compute the remaining terms in the integrated hemisphere soft function at $\Ordals 3$. At first sight, it may not be obvious that one would learn anything new from such a calculation. However, 
there is at least one reason why such a computation might be useful. Let us begin by discussing the factorization theorem for jet thrust since that was the focus of the present paper. In reference~\cite{Kelley:2011tj},
it was made clear that, in order to predict the potentially large logarithms of ${\mu \over \tauo Q}$ and $\mu \over 2\omega$ in the soft part of the integrated jet thrust distribution at two-loop order,
one must fix not one, but two matching coefficients at one-loop order, $c_{in}^{(1)}$ and $c_{out}^{(1)}$. Even at one-loop,
it is not entirely clear {\it a priori} what choice one should make. In fact, it turns out that one can first choose $c_{out}^{(1)}$ to be exactly the scale-independent part
of the out-of-jet contribution to the soft part of the integrated distribution and then choose $c_{in}^{(1)}$ such that the soft part of the complete one-loop integrated distribution is correctly reproduced.
Unfortunately, it is even less clear what one should do with the matching coefficients at
$\Ordals 2$ because, starting at two-loop order, one also has in-out contributions. Therefore, it would be of interest to obtain at least the scale-dependent part of the three-loop integrated hemisphere soft function
since we expect that similar ambiguities ought to arise in that case as well. Such a calculation would help clarify to what extent resummation beyond the next-to-leading logarithms is possible for exclusive event shape observables.
Quite apart from the specific applications discussed in this paper, it would certainly be of interest to develop more general computational methods for multi-scale problems and we hope to return to this subject in the future as well. 

\acknowledgments
RMS would like to thank M. D. Schwartz and J. R. Walsh for useful discussions.
The authors are also very grateful to S.-O. Moch and M. Neubert for comments on earlier drafts of this work.
The research of AvM is supported in part by the
Research Center {\em Elementary Forces and Mathematical Foundations (EMG)} of the
Johannes Gutenberg University of Mainz and by the
German Research Foundation (DFG).
The research of RMS is supported in part by the DFG under grant NE~398/3-1.
The research of HXZ is supported by the United States Department of Energy, under Contract DE-AC02-76SF00515.
\appendix

\section{Analytic Expressions for $\mbox{\large $\Lambda$}_{C_A}(x,r)$ and $\mbox{\large $\Lambda$}_{n_f}(x,r)$}
\label{app:chi}
For the $C_A C_F$ color structure we have
\bea
\mbox{\large $\Lambda$}_{C_A}(x,r) &=& -\frac{32 G\left(0,-\frac{1}{r+1};x\right) r^2}{3 \left(r-1\right)^2 \left(r+1\right)^2}+\frac{32 r^2}{3 \left(r+1\right) \left(x r+r+x\right)}
-\frac{32 G\left(-\frac{r}{r+1},-\frac{1}{r+1};x\right) r}{3 \left(r+1\right)^2}
\el\hspace{-13.2ex}
+\left(\frac{40}{3 \left(x+1\right)^2}-\frac{40}{3 \left(x+1\right)}\right) G\left(0;r\right)^2-\frac{440}{3} G\left(-1;x\right) G\left(0;r\right)^2+\left(-\frac{32 \left(r-1\right) \left(r+1\right)}{3 r \left(x+1\right)}
\right.\el\hspace{-13.2ex}\left.
-\frac{32 \left(r+1\right)}{3 r \left(\left(r+1\right) x+1\right)}
+\frac{32 r \left(r+1\right)}{3 \left(r+\left(r+1\right) x\right)}\right) G\left(-1;r\right)
+\frac{704}{3} G\left(-1;r\right) G\left(-1;x\right) G\left(0;r\right)
\el\hspace{-13.2ex}
+\left(\frac{32 \left(r+1\right)}{3 r \left(\left(r+1\right) x+1\right)}
+\frac{16 \left(3 r^2-r-2\right)}{3 r \left(x+1\right)}-\frac{16 \left(3 r^4+9 r^3+3 r^2+r\right)}{3 \left(r+\left(r+1\right) x\right) \left(r+1\right)^2}\right) G\left(0;r\right)
\el\hspace{-13.2ex}
+\left(-\frac{16 \left(r-1\right) \left(r+1\right)}{3 r \left(x+1\right)}-\frac{16 \left(r+1\right)}{3 r \left(\left(r+1\right) x+1\right)}
+\frac{16 r \left(r+1\right)}{3 \left(r+\left(r+1\right) x\right)}-\frac{16 \left(r^2+1\right)}{3 \left(r-1\right) \left(r+1\right)}\right)
\el\hspace{-13.2ex}
 \times G\left(0;x\right)
 +\frac{8}{3} \pi ^2 G\left(-1;x\right) G\left(0;x\right)
+40 G\left(-1;x\right) G\left(0;r\right)^2 G\left(0;x\right)
-\frac{176}{9} \pi ^2 G\left(0;x\right)
\el\hspace{-13.2ex}
+\left(\frac{64 r^2}{3 \left(r-1\right)^2 \left(r+1\right)^2}+\frac{32}{3 \left(x+1\right)}-\frac{32}{3 \left(x+1\right)^2}\right) G\left(0;r\right) G\left(0;x\right)
-\frac{352}{3} G\left(-1;r\right) G\left(0;r\right)
\el\hspace{-13.2ex}
\times G\left(0;x\right)-32 G\left(-1;r\right) G\left(-1;x\right) G\left(0;r\right) G\left(0;x\right)
+\frac{352}{3} G\left(0;r\right) G\left(0;x\right) G\left(1;r\right)
\el\hspace{-13.2ex}
+\left(-\frac{16 \left(r-1\right) \left(r+1\right)}{3 r \left(x+1\right)}+\frac{16 r \left(r+1\right)}{3 \left(r+\left(r+1\right) x\right)}-\frac{16 \left(r+1\right)}{3 r \left(\left(r+1\right) x+1\right)}\right) G\left(1;r\right)
 + \left(\frac{32}{3 \left(x+1\right)^2}
 \right.\el\hspace{-13.2ex}\left.
-\frac{32}{3 \left(x+1\right)}\right) G\left(-1;r\right) G\left(-\frac{1}{r+1};x\right)
+\left(\frac{16 r \left(3 r+1\right)}{3 \left(r-1\right) \left(r+1\right)^2}
-\frac{16 \left(3 r^2+r\right)}{3 \left(r+1\right)^2 \left(x r+r+x\right)}\right)
\el\hspace{-13.2ex}
\times  G\left(-\frac{1}{r+1};x\right)
+\left(\frac{32}{3 \left(x+1\right)}
-\frac{32}{3 \left(x+1\right)^2}\right) G\left(0;r\right)
G\left(-\frac{1}{r+1};x\right)
 +\left(\frac{16 \left(r-1\right)}{3 \left(r+1\right)}
 \right.\el\hspace{-13.2ex}\left.
 +\frac{16}{3 \left(x+1\right)}\right) G\left(-\frac{r}{r+1};x\right)
+\left(\frac{16}{3 \left(x+1\right)^2}-\frac{16}{3 \left(x+1\right)}\right) G\left(1;r\right) G\left(-\frac{1}{r+1};x\right)
\el\hspace{-13.2ex}
+\left(\frac{32}{3 \left(x+1\right)}-\frac{32}{3 \left(x+1\right)^2}\right) 
G\left(-1;r\right) G\left(-\frac{r}{r+1};x\right)
+\left(\frac{64 r}{3 \left(r+1\right)^2}-\frac{16}{x+1}
+\frac{16}{\left(x+1\right)^2}\right) 
\el\hspace{-13.2ex}
\times G\left(0;r\right) G\left(-\frac{r}{r+1};x\right)
+32 G\left(-1;x\right) G\left(0;x\right) G\left(-1,0;r\right)
-40 G\left(0;r\right)^2 G\left(-1,0;x\right)
\el\hspace{-13.2ex}
+\left(-\frac{16 r}{3 \left(r+1\right)^2}-\frac{16 \left(r^4+3 r^3\right)}{3 \left(r+1\right)^2 \left(r+\left(r+1\right) x\right)}
+\frac{16 \left(r^2-r-1\right)}{3 \left(x+1\right) r}
+\frac{16 \left(r+1\right)}{3 \left(\left(r+1\right) x+1\right) r}\right)
\el\hspace{-13.2ex}
\times  G\left(-\frac{r^2}{r+1};x\right)
+\left(\frac{16}{3 \left(x+1\right)}-\frac{16}{3 \left(x+1\right)^2}\right) G\left(1;r\right) G\left(-\frac{r}{r+1};x\right)
+\left(\frac{16}{3 \left(x+1\right)}
\right.\el\hspace{-13.2ex}\left.
-\frac{16}{3 \left(x+1\right)^2}\right) G\left(0;x\right) G\left(-\frac{r}{r+1};x\right)
-\frac{8}{3} \pi ^2 G\left(-1,0;x\right)
+\frac{352}{3} G\left(0;r\right) G\left(-1,0;x\right)
\el\hspace{-13.2ex}
+\frac{704}{3} G\left(0;x\right)G\left(-1,0;r\right)
-\frac{704}{3} G\left(-1;x\right) G\left(-1,0;r\right)
+32 G\left(-1;r\right) G\left(0;r\right) G\left(-1,0;x\right)
\el\hspace{-13.2ex}
-32 G\left(-1,0;r\right) G\left(-1,0;x\right)
+\left(\frac{64}{3 \left(x+1\right)}-\frac{64}{3 \left(x+1\right)^2}\right) G\left(0,-1;r\right)
-\frac{352}{3} G\left(-1;r\right)
\el\hspace{-13.2ex}
\times G\left(-1,-\frac{1}{r+1};x\right)
+\frac{352}{3} G\left(0;r\right) G\left(-1,-\frac{1}{r+1};x\right)
-\frac{176}{3} G\left(1;r\right) G\left(-1,-\frac{1}{r+1};x\right)
\el\hspace{-13.2ex}
+\frac{352}{3} G\left(-1;r\right) G\left(-1,-\frac{r}{r+1};x\right)
-176 G\left(0;r\right) G\left(-1,-\frac{r}{r+1};x\right)
-16 G\left(-1;r\right) G\left(0;x\right)
\el\hspace{-13.2ex}
\times  G\left(-1,-\frac{r}{r+1};x\right)
+48 G\left(0;r\right) G\left(0;x\right) G\left(-1,-\frac{r}{r+1};x\right)
+\frac{176}{3} G\left(1;r\right)
\el\hspace{-13.2ex}
 \times G\left(-1,-\frac{r}{r+1};x\right)
 +\left(\frac{32}{3 \left(x+1\right)}-\frac{32}{3 \left(x+1\right)^2}\right) G\left(0,1;r\right)+\frac{352}{3} G\left(-1;x\right) G\left(0,1;r\right)
\el\hspace{-13.2ex}
+\frac{352}{3} G\left(-1;r\right) G\left(0,-\frac{1}{r+1};x\right)-\frac{352}{3} G\left(0;r\right) G\left(0,-\frac{1}{r+1};x\right)
-\frac{704}{3} G\left(0;x\right) G\left(0,1;r\right)
\el\hspace{-13.2ex}
+\frac{176}{3} G\left(0;x\right) G\left(0,-\frac{1}{r+1};x\right)+\frac{176}{3} G\left(1;r\right) G\left(0,-\frac{1}{r+1};x\right)
+\left(\frac{32}{3 \left(x+1\right)^2} 
\right.\el\hspace{-13.2ex}\left.
-\frac{32}{3 \left(x+1\right)}\right) G\left(0,-\frac{r}{r+1};x\right)-\frac{352}{3} G\left(-1;r\right) G\left(0,-\frac{r}{r+1};x\right)
+\frac{176}{3} G\left(0;r\right) 
\el\hspace{-13.2ex}
\times G\left(0,-\frac{r}{r+1};x\right)-\frac{176}{3} G\left(0;x\right) G\left(0,-\frac{r}{r+1};x\right)+\left(\frac{32 r^2}{3 \left(r-1\right)^2 \left(r+1\right)^2}
+\frac{16}{3 \left(x+1\right)}
\right.\el\hspace{-13.2ex}\left.
-\frac{16}{3 \left(x+1\right)^2}\right) G\left(0,-\frac{r^2}{r+1};x\right)
+\left(\frac{16}{3 \left(x+1\right)^2}-\frac{16}{3 \left(x+1\right)}\right)  G\left(-\frac{1}{r+1},0;x\right)
\el\hspace{-13.2ex}
-\frac{176}{3} G\left(1;r\right) G\left(0,-\frac{r}{r+1};x\right)
+16 G\left(-1;r\right) G\left(-1,0,-\frac{r}{r+1};x\right)
+\left(\frac{32 r}{3 \left(r+1\right)^2}
\right.\el\hspace{-13.2ex}\left.
-\frac{16}{3 \left(x+1\right)}
+\frac{16}{3 \left(x+1\right)^2}\right) G\left(-\frac{r}{r+1},-\frac{r^2}{r+1};x\right)
+\left(\frac{16}{3 \left(x+1\right)}-\frac{16}{3 \left(x+1\right)^2}\right) 
\el\hspace{-13.2ex}
\times G\left(-\frac{1}{r+1},-\frac{r^2}{r+1};x\right)-48 G\left(0;r\right) G\left(-1,0,-\frac{r}{r+1};x\right)
-\frac{176}{3} G\left(-1,0,-\frac{r}{r+1};x\right)
\el\hspace{-13.2ex}
+\frac{176}{3} G\left(-1,0,-\frac{r^2}{r+1};x\right)-\frac{176}{3} G\left(-1,-\frac{1}{r+1},0;x\right)
+\frac{176}{3} G\left(-1,-\frac{1}{r+1},-\frac{r^2}{r+1};x\right)
\el\hspace{-13.2ex}
+16 G\left(-1;r\right) G\left(-1,-\frac{r}{r+1},0;x\right)
-48 G\left(0;r\right) G\left(-1,-\frac{r}{r+1},0;x\right)
\el\hspace{-13.2ex}
+\frac{176}{3} G\left(-1,-\frac{r}{r+1},0;x\right)
-64 G\left(0;x\right) G\left(0,-1,0;r\right)
-\frac{176}{3} G\left(-1,-\frac{r}{r+1},-\frac{r^2}{r+1};x\right)
\el\hspace{-13.2ex}
-32 G\left(0;r\right) G\left(0,-1,0;x\right)+16 G\left(-1;r\right) G\left(0,-1,-\frac{1}{r+1};x\right)
+32 G\left(0;r\right) G\left(0,0,-r;x\right)
\el\hspace{-13.2ex}
-32 G\left(0;r\right) G\left(0,-1,-\frac{1}{r+1};x\right)
-16 G\left(-1;r\right) G\left(0,0,-\frac{1}{r+1};x\right)
+32 G\left(0;r\right)
\el\hspace{-13.2ex}
\times G\left(0,0,-\frac{1}{r+1};x\right)-16 G\left(0;x\right) G\left(0,0,-\frac{1}{r+1};x\right)-176 G\left(0,0,-\frac{1}{r+1};x\right)
\el\hspace{-13.2ex}
+16 G\left(-1;r\right) G\left(0,0,-\frac{r}{r+1};x\right)-48 G\left(0;r\right) G\left(0,0,-\frac{r}{r+1};x\right)
+16 G\left(0;x\right) 
\el\hspace{-13.2ex}
\times G\left(0,0,-\frac{r}{r+1};x\right)+176 G\left(0,0,-\frac{r}{r+1};x\right)
-64 G\left(0;x\right) G\left(0,1,0;r\right)
\el\hspace{-13.2ex}
-\frac{176}{3} G\left(0,-\frac{1}{r+1},-\frac{r^2}{r+1};x\right)
+\frac{176}{3} G\left(0,-\frac{r}{r+1},-\frac{1}{r+1};x\right)
\el\hspace{-13.2ex}
+16 G\left(0,-1,0,-\frac{1}{r+1};x\right)
-16 G\left(0,-1,0,-\frac{r^2}{r+1};x\right)
+16 G\left(0,-1,-\frac{1}{r+1},0;x\right)
\el\hspace{-13.2ex}
-16 G\left(0,-1,-\frac{1}{r+1},-\frac{r^2}{r+1};x\right)
-16 G\left(0,-1,-\frac{r}{r+1},0;x\right)
\el\hspace{-13.2ex}
-16 G\left(0,-1,-\frac{r}{r+1},-\frac{1}{r+1};x\right)
+16 G\left(0,-1,-\frac{r}{r+1},-\frac{r}{r+1};x\right)
\el\hspace{-13.2ex}
+16 G\left(0,-1,-\frac{r}{r+1},-\frac{r^2}{r+1};x\right)
+32 G\left(0,0,0,-\frac{1}{r+1};x\right)
-32 G\left(0,0,0,-\frac{r}{r+1};x\right)
\el\hspace{-13.2ex}
-16 G\left(0,0,-r,-\frac{r}{r+1};x\right)
+16 G\left(0,0,-r,-\frac{r^2}{r+1};x\right)
+16 G\left(0,0,-\frac{1}{r+1},-\frac{r^2}{r+1};x\right)
\el\hspace{-13.2ex}
+16 G\left(0,0,-\frac{r}{r+1},-\frac{1}{r+1};x\right)
-16 G\left(0,0,-\frac{r}{r+1},-\frac{r}{r+1};x\right)
\el\hspace{-13.2ex}
-16 G\left(0,0,-\frac{r}{r+1},-\frac{r^2}{r+1};x\right)
+32 G\left(0;x\right) \zeta_3
\eea
and for the $C_F n_f T_F$ color structure we have
\bea
\mbox{\large $\Lambda$}_{n_f}(x,r) &=& \frac{64 G\left(0,-\frac{1}{r+1};x\right) r^2}{3 \left(r-1\right)^2 \left(r+1\right)^2}-\frac{64 r^2}{3 \left(r+1\right) \left(x r+r+x\right)}
+\frac{64 G\left(-\frac{r}{r+1},-\frac{1}{r+1};x\right) r}{3 \left(r+1\right)^2}
\el\hspace{-12.8ex}
+\left(\frac{80}{3 \left(x+1\right)}-\frac{80}{3 \left(x+1\right)^2}\right) G\left(0;r\right)^2+\frac{160}{3} G\left(-1;x\right) G\left(0;r\right)^2+\left(\frac{64 \left(r-1\right) \left(r+1\right)}{3 r \left(x+1\right)}
\right.\el\hspace{-12.8ex}\left.
+\frac{64 \left(r+1\right)}{3 r \left(\left(r+1\right) x+1\right)}-\frac{64 r \left(r+1\right)}{3 \left(r+\left(r+1\right) x\right)}\right) G\left(-1;r\right)+\left(-\frac{64 \left(r+1\right)}{3 r \left(\left(r+1\right) x+1\right)}
\right.\el\hspace{-12.8ex}\left.
-\frac{32 \left(3 r^2-r-2\right)}{3 r \left(x+1\right)}+\frac{32 \left(3 r^4+9 r^3+3 r^2+r\right)}{3 \left(r+\left(r+1\right) x\right) \left(r+1\right)^2}\right) G\left(0;r\right)
-\frac{256}{3} G\left(-1;r\right) G\left(-1;x\right) G\left(0;r\right)
\el\hspace{-12.8ex}
+\left(\frac{32 \left(r-1\right) \left(r+1\right)}{3 r \left(x+1\right)}+\frac{32 \left(r+1\right)}{3 r \left(\left(r+1\right) x+1\right)}-\frac{32 r \left(r+1\right)}{3 \left(r+\left(r+1\right) x\right)}
+\frac{32 \left(r^2+1\right)}{3 \left(r-1\right) \left(r+1\right)}\right) G\left(0;x\right)
\el\hspace{-12.8ex}
+\left(-\frac{128 r^2}{3 \left(r-1\right)^2 \left(r+1\right)^2}-\frac{64}{3 \left(x+1\right)}+\frac{64}{3 \left(x+1\right)^2}\right) G\left(0;r\right) G\left(0;x\right)+\frac{64}{9} \pi ^2 G\left(0;x\right)
\el\hspace{-12.8ex}
+\frac{128}{3} G\left(-1;r\right) G\left(0;r\right) G\left(0;x\right)+\left(\frac{32 \left(r-1\right) \left(r+1\right)}{3 r \left(x+1\right)}+\frac{32 \left(r+1\right)}{3 r \left(\left(r+1\right) x+1\right)}
\right.\el\hspace{-12.8ex}\left.
-\frac{32 r \left(r+1\right)}{3 \left(r+\left(r+1\right) x\right)}\right) G\left(1;r\right)-\frac{128}{3} G\left(0;r\right) G\left(0;x\right) G\left(1;r\right)+\left(\frac{32 \left(3 r^2+r\right)}{3 \left(r+1\right)^2 \left(x r+r+x\right)}
\right.\el\hspace{-12.8ex}\left.
-\frac{32 r \left(3 r+1\right)}{3 \left(r-1\right) \left(r+1\right)^2}\right) G\left(-\frac{1}{r+1};x\right)+\left(\frac{64}{3 \left(x+1\right)}-\frac{64}{3 \left(x+1\right)^2}\right) G\left(-1;r\right) G\left(-\frac{1}{r+1};x\right)
\el\hspace{-12.8ex}
+\left(\frac{64}{3 \left(x+1\right)^2}-\frac{64}{3 \left(x+1\right)}\right) G\left(0;r\right) G\left(-\frac{1}{r+1};x\right)+\left(\frac{32}{3 \left(x+1\right)}
-\frac{32}{3 \left(x+1\right)^2}\right) G\left(1;r\right)
\el\hspace{-12.8ex}
\times  G\left(-\frac{1}{r+1};x\right)+\left(-\frac{32 \left(r-1\right)}{3 \left(r+1\right)}-\frac{32}{3 \left(x+1\right)}\right) G\left(-\frac{r}{r+1};x\right)
+\left(\frac{64}{3 \left(x+1\right)^2}
\right.\el\hspace{-12.8ex}\left.
-\frac{64}{3 \left(x+1\right)}\right)  G\left(-1;r\right) G\left(-\frac{r}{r+1};x\right)+\left(-\frac{128 r}{3 \left(r+1\right)^2}+\frac{32}{x+1}
-\frac{32}{\left(x+1\right)^2}\right) G\left(0;r\right)
\el\hspace{-12.8ex}
\times G\left(-\frac{r}{r+1};x\right)+\left(\frac{32}{3 \left(x+1\right)^2}-\frac{32}{3 \left(x+1\right)}\right) G\left(0;x\right) G\left(-\frac{r}{r+1};x\right)
+\left(\frac{32}{3 \left(x+1\right)^2}
\right.\el\hspace{-12.8ex}\left.
-\frac{32}{3 \left(x+1\right)}\right) G\left(1;r\right) G\left(-\frac{r}{r+1};x\right)
+\left(\frac{32 r}{3 \left(r+1\right)^2}
+\frac{32 \left(r^4+3 r^3\right)}{3 \left(r+1\right)^2 \left(r+\left(r+1\right) x\right)}
\right.\el\hspace{-12.8ex}\left.
-\frac{32 \left(r^2-r-1\right)}{3 \left(x+1\right) r}
-\frac{32 \left(r+1\right)}{3 \left(\left(r+1\right) x+1\right) r}\right) G\left(-\frac{r^2}{r+1};x\right)
+\frac{256}{3} G\left(-1;x\right) G\left(-1,0;r\right)
\el\hspace{-12.8ex}
-\frac{256}{3} G\left(0;x\right) G\left(-1,0;r\right)-\frac{128}{3} G\left(0;r\right) G\left(-1,0;x\right)
+\frac{128}{3} G\left(-1;r\right) G\left(-1,-\frac{1}{r+1};x\right)
\el\hspace{-12.8ex}
-\frac{128}{3} G\left(0;r\right) G\left(-1,-\frac{1}{r+1};x\right)
+\frac{64}{3} G\left(1;r\right) G\left(-1,-\frac{1}{r+1};x\right)-\frac{128}{3} G\left(-1;r\right) 
\el\hspace{-12.8ex}
\times G\left(-1,-\frac{r}{r+1};x\right)
+64 G\left(0;r\right) G\left(-1,-\frac{r}{r+1};x\right)-\frac{64}{3} G\left(1;r\right) G\left(-1,-\frac{r}{r+1};x\right)
\el\hspace{-12.8ex}
+\left(\frac{128}{3 \left(x+1\right)^2}-\frac{128}{3 \left(x+1\right)}\right) G\left(0,-1;r\right)+\left(\frac{64}{3 \left(x+1\right)^2}-\frac{64}{3 \left(x+1\right)}\right) G\left(0,1;r\right)
\el\hspace{-12.8ex}
-\frac{128}{3} G\left(-1;x\right) G\left(0,1;r\right)
+\frac{256}{3} G\left(0;x\right) G\left(0,1;r\right)-\frac{128}{3} G\left(-1;r\right) G\left(0,-\frac{1}{r+1};x\right)
\el\hspace{-12.8ex}
+\frac{128}{3} G\left(0;r\right) G\left(0,-\frac{1}{r+1};x\right)-\frac{64}{3} G\left(0;x\right) G\left(0,-\frac{1}{r+1};x\right)
-\frac{64}{3} G\left(1;r\right) 
\el\hspace{-12.8ex}
\times G\left(0,-\frac{1}{r+1};x\right)+\left(\frac{64}{3 \left(x+1\right)} -\frac{64}{3 \left(x+1\right)^2}\right) G\left(0,-\frac{r}{r+1};x\right)
+\frac{128}{3} G\left(-1;r\right)
\el\hspace{-12.8ex}
\times G\left(0,-\frac{r}{r+1};x\right)-\frac{64}{3} G\left(0;r\right) G\left(0,-\frac{r}{r+1};x\right)
+\frac{64}{3} G\left(0;x\right) G\left(0,-\frac{r}{r+1};x\right)
\el\hspace{-12.8ex}
+\frac{64}{3} G\left(1;r\right) G\left(0,-\frac{r}{r+1};x\right)
+\left(-\frac{64 r^2}{3 \left(r-1\right)^2 \left(r+1\right)^2}-\frac{32}{3 \left(x+1\right)}
+\frac{32}{3 \left(x+1\right)^2}\right) 
\el\hspace{-12.8ex}
\times G\left(0,-\frac{r^2}{r+1};x\right)
+\left(\frac{32}{3 \left(x+1\right)}
-\frac{32}{3 \left(x+1\right)^2}\right) G\left(-\frac{1}{r+1},0;x\right)
+\left(\frac{32}{3 \left(x+1\right)^2}
\right.\el\hspace{-12.8ex}\left.
-\frac{32}{3 \left(x+1\right)}\right) G\left(-\frac{1}{r+1},-\frac{r^2}{r+1};x\right)
+\left(-\frac{64 r}{3 \left(r+1\right)^2}+\frac{32}{3 \left(x+1\right)}-\frac{32}{3 \left(x+1\right)^2}\right)
\el\hspace{-12.8ex}
\times G\left(-\frac{r}{r+1},-\frac{r^2}{r+1};x\right)
+\frac{64}{3} G\left(-1,0,-\frac{r}{r+1};x\right)-\frac{64}{3} G\left(-1,0,-\frac{r^2}{r+1};x\right)
\el\hspace{-12.8ex}
+\frac{64}{3} G\left(-1,-\frac{1}{r+1},0;x\right)
-\frac{64}{3} G\left(-1,-\frac{1}{r+1},-\frac{r^2}{r+1};x\right)-\frac{64}{3} G\left(-1,-\frac{r}{r+1},0;x\right)
\el\hspace{-12.8ex}
+64 G\left(0,0,-\frac{1}{r+1};x\right)
+\frac{64}{3} G\left(-1,-\frac{r}{r+1},-\frac{r^2}{r+1};x\right)+\frac{64}{3} G\left(0,-\frac{1}{r+1},-\frac{r^2}{r+1};x\right)
\el\hspace{-12.8ex}
-64 G\left(0,0,-\frac{r}{r+1};x\right)-\frac{64}{3} G\left(0,-\frac{r}{r+1},-\frac{1}{r+1};x\right)\,.
\eea
These functions are straightforwardly related to the functions $\mbox{\Large $\chi$}_{C_A}(x,r)$ and $\mbox{\Large $\chi$}_{n_f}(x,r)$ defined in reference~\cite{Kelley:2011aa}. 
The difference between the $\mbox{\large $\Lambda$}$ and $\mbox{\Large $\chi$}$ functions is that $\mbox{\large $\Lambda$}_{C_A}(x,r)$ and $\mbox{\large $\Lambda$}_{n_f}(x,r)$ are free of terms which depend on $r$ but not $x$.

\section{Individual Contributions to the Exact Result}
\label{app:indcontrib}

In this appendix, we give the individual contributions to the finite part of the
integrated $\tauo$ distributions in terms of $G$ functions.
We choose to split logarithms of $\mu$, $Q$, $\tauo$ or $\omega$ according to the
presentation of~\cite{Kelley:2011aa} and use the symbol $\bar{K}_{TC}$ to denote the
finite part of $K_{TC}$.

From Eq.\ (\ref{eq:intconerenin1}) we obtain by expanding in $\ep$
\bea
\label{eq:intconerenin2}
\bar{K}^{{\rm in\,Ren}}_{TC}(\tauo,\omega,r,\mu)
 &=& C_A C_F \left[\frac{176}{9} \ln^3\left(\frac{\mu }{\tauo Q }\right)+\frac{88}{3} G(0;r) \ln^2\left(\frac{\mu }{\tauo Q }\right)+\left(\frac{44}{3} G(0;r)^2\right.\right.
\el\hspace{-20ex}
\left.\left.-\frac{22 \pi^2}{9}\right) \ln \left(\frac{\mu }{\tauo Q }\right)+\frac{22}{9}G(0;r)^3-\frac{11}{9} \pi^2 G(0;r)-\frac{44 \zeta_3}{9}\right]
\el\hspace{-20ex}
+C_F n_f T_F \left[-\frac{64}{9} \ln^3\left(\frac{\mu }{\tauo Q }\right)-\frac{32}{3} G(0;r) \ln^2\left(\frac{\mu }{\tauo Q }\right)
+\left(\frac{8 \pi^2}{9}-\frac{16}{3} G(0;r)^2\right) \ln \left(\frac{\mu }{\tauo Q }\right)
\right.
\el\hspace{-20ex}
\left.-\frac{8}{9} G(0;r)^3+\frac{4}{9} \pi^2 G(0;r)+\frac{16 \zeta_3}{9}\right]
\,.
\eea
From Eq.\ (\ref{eq:intconeallinals2}) we obtain by expanding in $\ep$
\bea
\label{eq:intconeallinals2fin}
\bar{K}^{{\rm all-in}\,(2)}_{TC}(\tauo,\omega,r,\mu)
 &=& C_A C_F \left[-\frac{64}{3} \ln^4\left(\frac{\mu }{Q \tau_{\omega}}\right) + \left(-\frac{128}{3} G(0;r)-\frac{352}{9}\right) \ln^3\left(\frac{\mu }{Q \tau_{\omega}}\right)
   \right.\el\hspace{-22ex}\left.
   +\left(-32 G(0;r)^2-\frac{176}{3} G(0;r)+\frac{16 \pi^2}{3}
   -\frac{536}{9}\right) \ln^2\left(\frac{\mu }{Q \tau_{\omega}}\right)+\left(-\frac{32}{3} G(0;r)^3 
   \right.\right.\el\hspace{-22ex}\left.\left.
   -\frac{88}{3} G(0;r)^2 +\frac{16}{3} \pi^2 G(0;r)-\frac{536}{9} G(0;r)+\frac{232
   \zeta_3}{3}-\frac{22 \pi^2}{9}-\frac{1544}{27}\right) \ln \left(\frac{\mu }{Q \tau_{\omega}}\right)
   \right.\el\hspace{-22ex}\left.
   +\frac{116}{3} \zeta_3 G(0;r)-\frac{4}{3} G(0;r)^4-\frac{44}{9}
   G(0;r)^3+\frac{4}{3} \pi^2 G(0;r)^2-\frac{134}{9} G(0;r)^2-\frac{11}{9} \pi^2 G(0;r)
   \right.\el\hspace{-22ex}\left.
   -\frac{772}{27} G(0;r)-\frac{242 \zeta_3}{9}+\frac{137 \pi^4}{180}-\frac{67 \pi^2}{54}-\frac{2392}{81}\right]
   \el\hspace{-22ex}
   +C_F n_f T_F \left[\frac{128}{9} \ln^3\left(\frac{\mu }{Q \tau_{\omega}}\right) + \left(\frac{64}{3} G(0;r)+\frac{160}{9}\right) \ln^2\left(\frac{\mu }{Q \tau _{\omega
   }}\right)+\left(\frac{32}{3} G(0;r)^2
   \right.\right.\el\hspace{-22ex}\left.\left.
   +\frac{160}{9} G(0;r)+\frac{8 \pi^2}{9}+\frac{304}{27}\right) \ln \left(\frac{\mu }{Q \tau_{\omega}}\right)+\frac{16}{9} G(0;r)^3+\frac{40}{9}
   G(0;r)^2+\frac{4}{9} \pi^2 G(0;r)
   \right.\el\hspace{-22ex}\left.
   +\frac{152}{27} G(0;r)+\frac{88 \zeta_3}{9}+\frac{10 \pi^2}{27}+\frac{476}{81}\right]
\,.
\eea
From Eq.\ (\ref{eq:intconervin1}) we obtain by expanding in $\ep$
\bea
\label{eq:intconervin2}
\bar{K}^{{\rm in\,R-V}}_{TC}(\tauo,\omega,r,\mu)
 &=& C_A C_F \left[\frac{64}{3}
   \ln^4\left(\frac{\mu }{Q \tau_{\omega}}\right) + \frac{128}{3} G(0;r) \ln^3\left(\frac{\mu }{Q \tau_{\omega}}\right)+\Big(32 G(0;r)^2
   \right.\el\hspace{-20.5ex}\left.
   -8 \pi^2\Big) \ln^2\left(\frac{\mu }{Q \tau_{\omega}}\right)+\left(\frac{32}{3}
   G(0;r)^3-8 \pi^2 G(0;r)-\frac{64 \zeta_3}{3}\right) \ln \left(\frac{\mu }{Q \tau_{\omega}}\right)-\frac{32}{3} \zeta_3 G(0;r)
   \right.\el\hspace{-20.5ex}\left.
   +\frac{4}{3} G(0;r)^4-2 \pi^2 G(0;r)^2+\frac{\pi^4}{60}\right]
\,.
\eea
From Eq.\ (\ref{eq:intconerenout1}) we obtain by expanding in $\ep$
\bea
\label{eq:intconerenout2}
\bar{K}^{{\rm out\,Ren}}_{TC}(\tauo,\omega,r,\mu)
 &=& C_A C_F \left[-\frac{88}{3} G(0;r) \ln ^2\left(\frac{\mu }{2 \omega }\right)+\left(\frac{44}{3} G(0;r)^2-\frac{176}{3} G(0,-1;r)
 \right.\right.\el\hspace{-20.7ex}\left.\left.
 +\frac{44 \pi ^2}{9}\right) \ln \left(\frac{\mu }{2 \omega
   }\right)-\frac{22}{9} G(0;r)^3+\frac{88}{3} G(0,-1;r) G(0;r)+\frac{11}{9} \pi ^2 G(0;r)
 -\frac{88}{3} G(-1;r) 
   \right.\el\hspace{-20.7ex}\left.
   \times G(0,-1;r)+\frac{88}{3} G(-1,0,-1;r)-\frac{88}{3} G(0,0,-1;r)+\frac{88 \zeta
   _3}{3}\right]
  \el\hspace{-20.5ex}
   +C_F n_f T_F \left[\frac{32}{3} G(0;r) \ln ^2\left(\frac{\mu }{2 \omega }\right)+\left(-\frac{16}{3} G(0;r)^2+\frac{64}{3} G(0,-1;r)-\frac{16 \pi ^2}{9}\right) \ln
   \left(\frac{\mu }{2 \omega }\right)
   \right.\el\hspace{-20.7ex}\left.
   +\frac{8}{9} G(0;r)^3-\frac{32}{3} G(0,-1;r) G(0;r)-\frac{4}{9} \pi ^2 G(0;r)+\frac{32}{3} G(-1;r) G(0,-1;r)
   \right.\el\hspace{-20.7ex}\left.
   -\frac{32}{3} G(-1,0,-1;r)+\frac{32}{3} G(0,0,-1;r)-\frac{32 \zeta _3}{3}\right]
\,,
\eea
where we employ the package {\tt HypExp\,\,2}~\cite{Huber:2005yg,Huber:2007dx} for expansions
of the ${}_2F_1$ functions.
From Eq.\ (\ref{eq:intconervout1}) we obtain by expanding in $\ep$
\bea
\label{eq:intconervout2}
\bar{K}^{{\rm out\,R-V}}_{TC}(\tauo,\omega,r,\mu)
 &=& C_A C_F \left[-\frac{128}{3} G(0;r) \ln ^3\left(\frac{\mu }{2 \omega }\right)+\bigg(32 G(0;r)^2-128 G(0,-1;r)
 \right.\el\hspace{-21.5ex}\left.
 +\frac{32 \pi ^2}{3}\right) \ln
   ^2\left(\frac{\mu }{2 \omega }\bigg) + \bigg(-\frac{32}{3} G(0;r)^3+128 G(0,-1;r) G(0;r)+8 \pi ^2 G(0;r)
   \right.\el\hspace{-21.5ex}\left.
   -256 G(-1;r) G(0,-1;r)+256 G(-1,0,-1;r)+256 G(0,-1,-1;r)-128
   G(0,0,-1;r)
   \right.\el\hspace{-21.5ex}\left.
   +128 \zeta _3\bigg)\ln \left(\frac{\mu }{2 \omega }\right) +\frac{32}{3} \zeta _3 G(0;r)+\frac{4}{3} G(0;r)^4-32 G(0,-1;r) G(0;r)^2
   +64 G(-1;r) 
   \right.\el\hspace{-21.5ex}\left.
   \times  G(0,-1;r)G(0;r)-64 G(-1,0,-1;r) G(0;r)+64 G(0,0,-1;r) G(0;r)
   \right.\el\hspace{-21.5ex}\left.
   -32 G(0,-1;r)^2-\frac{128}{3}
   G(-1;r)^2 G(0,-1;r)+ 8 \pi ^2 G(0,-1;r)-2 \pi ^2 G(0;r)^2
   \right.\el\hspace{-21.5ex}\left.
   +\frac{256}{3} G(-1;r) G(-1,0,-1;r)-\frac{256}{3} G(-1,-1,0,-1;r)+64 G(0,-1,0,-1;r)
   \right.\el\hspace{-21.5ex}\left.
   -64 G(0,0,0,-1;r)+\frac{2 \pi ^4}{15}\right]
\,.
\eea
From Eq.\ \eqref{eq:secdecompinin} we obtain with the methods of Section~\ref{sec:coprod}
\bea
\bar{K}_{TC}^{{\rm opp\,in-in}}(\tauo,\omega,r,\mu) &=& 
C_A C_F \left[(-64 G(0,-1;r)-64 G(0,1;r)) \ln
   ^2\left(\frac{\mu }{Q \tau _{\omega }}\right)
\right.\el\hspace{-24.7ex}\left. 
   +\left(\frac{32 G(0;r) r^2}{3 \left(r^2-1\right)^2}+\frac{16 r^2}{3-3 r^2}+\frac{176}{3} G(-1;r) G(0;r) +\frac{176}{3} G(0;r) G(1;r)-\frac{352}{3} G(0,-1;r)
\right.\right.\el\hspace{-24.7ex}\left.\left.   
  -96 G(0;r) G(0,-1;r)-96 G(0;r)
   G(0,1;r)-\frac{352}{3} G(0,1;r)+64 G(0,-1,-1;r)
\right.\right.\el\hspace{-24.7ex}\left.\left.      
   +64 G(0,-1,1;r)+128 G(0,0,-1;r)+128 G(0,0,1;r)+64 G(0,1,-1;r)
+64 G(0,1,1;r)\!\vphantom{\frac{32 G(0;r) r^2}{3 \left(r^2-1\right)^2}}\right)
\right.\el\hspace{-24.7ex}\left.
    \times \ln \left(\frac{\mu }{Q \tau _{\omega }}\right)-\frac{44}{3} G(0;r) G(-1;r)^2-\frac{16}{3} G(0,-1;r) G(-1;r)^2
+\frac{44}{3} G(0;r)^2 G(-1;r)
\right.\el\hspace{-24.7ex}\left.
-\frac{268}{9} G(0;r) G(-1;r)+\frac{88}{3} G(0;r) G(1;r) G(-1;r)
+16 G(0;r) G(0,-1;r) G(-1;r)
\right.\el\hspace{-24.7ex}\left.
+\frac{88}{3} G(0,-1;r)
   G(-1;r)+\frac{352}{3} G(0,1;r) G(-1;r)
   +\frac{32}{3} G(-1,0,-1;r) G(-1;r)
\right.\el\hspace{-24.7ex}\left.  
-96 G(0,0,1;r) G(-1;r)-\frac{4}{3} G(-1;r)+\frac{44}{3} G(0;r) G(1;r)^2
   -8 G(0,-1;r)^2+40 G(0,1;r)^2
\right.\el\hspace{-24.7ex}\left.     
+8 G(0;r)-\frac{88\pi ^2}{9} G(-1;r)-\frac{88\pi ^2}{9} G(1;r)
   +\frac{44}{3} G(0;r)^2 G(1;r)-\frac{268}{9} G(0;r) G(1;r)
 \right.\el\hspace{-24.7ex}\left.  
-\frac{4}{3} G(1;r)-24 G(0;r)^2 G(0,-1;r)
-\frac{176}{3} G(0;r) G(0,-1;r)-\frac{352}{3} G(1;r) G(0,-1;r)
 \right.\el\hspace{-24.7ex}\left.   
   +\frac{536}{9} G(0,-1;r)-24 G(0;r)^2 G(0,1;r)
   +\frac{1}{(r+1)^2}\bigg(\frac{2}{3} G(0;r)^2-\frac{4}{3} G(-1;r) G(0;r)
 \right.\el\hspace{-24.7ex}\left.
  +\frac{4}{3} G(1;r) G(0;r)-\frac{4}{9} G(0;r)-\frac{8}{3}
   G(0,1;r)-\frac{4 \pi ^2}{9}\bigg)
   +\frac{1}{1-r}\bigg(-\frac{2}{3} G(0;r)^2
   +\frac{20}{3} G(-1;r)
 \right.\el\hspace{-24.7ex}\left.
  \times G(0;r)-\frac{4}{3} G(1;r) G(0;r)-\frac{68}{9} G(0;r)+4 G(-1;r)
   +4 G(1;r)-\frac{32}{3} G(0,-1;r)
 \right.\el\hspace{-24.7ex}\left.
  -\frac{8}{3} G(0,1;r)+\frac{4
   \pi ^2}{9}-\frac{4}{9}\bigg)-\frac{176}{3} G(0;r) G(0,1;r)
   -\frac{352}{3} G(1;r) G(0,1;r)+64 G(0,-1;r) 
   \right.\el\hspace{-24.7ex}\left.
\times G(0,1;r)+\frac{536}{9} G(0,1;r)+\frac{1}{r+1}\left(-\frac{2}{3} G(0;r)^2
   +\frac{4}{3} G(-1;r)
   G(0;r)-\frac{4}{3} G(1;r) G(0;r)
   \right.\right.\el\hspace{-24.7ex}\left.\left.
   -\frac{20}{9} G(0;r)-\frac{4}{3} G(-1;r)-\frac{4}{3} G(1;r)+\frac{8}{3} G(0,1;r)
   +\frac{4 \pi ^2}{9}-\frac{4}{9}\right)
   +\frac{1}{(1-r)^2}\left(\frac{2}{3} G(0;r)^2
   \right.\right.\el\hspace{-24.7ex}\left.\left.
  -\frac{20}{3} G(-1;r)
   G(0;r)+\frac{4}{3} G(1;r) G(0;r)+\frac{20}{9} G(0;r)
  +\frac{32}{3} G(0,-1;r)+\frac{8}{3} G(0,1;r)-\frac{4 \pi ^2}{9}\right)
   \right.\el\hspace{-24.7ex}\left.
   -16 G(0;r) G(0,1,1;r)+12 \pi ^2 G(0,1;r)
   -16 G(0;r) G(-1,0,-1;r)-\frac{88}{3} G(-1,0,-1;r)
    \right.\el\hspace{-24.7ex}\left.
  -176 G(-1,0,1;r)+\frac{352}{3} G(0,0,-1;r)
   +96 G(0;r) G(0,0,-1;r)+96 G(0;r) G(0,0,1;r)
   \right.\el\hspace{-24.7ex}\left.
   +\frac{352}{3} G(0,0,1;r)+12 \pi ^2 G(0,-1;r)
   +\frac{704}{3} G(0,1,1;r)-\frac{32}{3} G(-1,-1,0,-1;r)
  \right.\el\hspace{-24.7ex}\left.
   +96 G(-1,0,0,1;r)-32
   G(0,-1,-1,1;r)
   +48 G(0,-1,0,-1;r)+96 G(0,-1,0,1;r)
   \right.\el\hspace{-24.7ex}\left.
   -32 G(0,-1,1,-1;r)-32 G(0,-1,1,1;r)
   -192 G(0,0,0,-1;r)-192 G(0,0,0,1;r)
   \right.\el\hspace{-24.7ex}\left.
   -96 G(0,0,1,1;r)-32 G(0,1,-1,-1;r)
   -32 G(0,1,-1,1;r)-32 G(0,1,1,-1;r)
   \right.\el\hspace{-24.7ex}\left.
   -32 G(0,1,1,1;r)+\frac{8}{9}\right]
   \el\hspace{-24.7ex}
   +C_F n_f T_F\left[\left(-\frac{64 G(0;r)
   r^2}{3 \left(r^2-1\right)^2}+\frac{32 r^2}{3 \left(r^2-1\right)}-\frac{64}{3} G(-1;r) G(0;r)-\frac{64}{3} G(0;r) G(1;r)
   \right.\right.\el\hspace{-24.7ex}\left.\left.
   +\frac{128}{3} G(0,-1;r)+\frac{128}{3} G(0,1;r)\right) \ln \left(\frac{\mu }{Q \tau _{\omega
   }}\right) + \frac{16}{3} G(0;r) G(-1;r)^2-\frac{16}{3} G(0;r)^2 G(-1;r)
   \right.\el\hspace{-24.7ex}\left.
   -\frac{32}{3} G(0;r) G(1;r) G(-1;r)-\frac{32}{3} G(0,-1;r) G(-1;r)-\frac{128}{3} G(0,1;r) G(-1;r)+\frac{8}{3}
   G(-1;r)
   \right.\el\hspace{-24.7ex}\left.
   -\frac{16}{3} G(0;r) G(1;r)^2-16 G(0;r)-\frac{16}{3} G(0;r)^2 G(1;r)+\frac{128}{9} G(0;r) G(1;r)+\frac{8}{3} G(1;r)
   \right.\el\hspace{-24.7ex}\left.
   + \frac{32\pi ^2}{9} G(-1;r)+\frac{32\pi ^2}{9} G(1;r)+\frac{64}{3} G(0;r)
   G(0,-1;r)+\frac{128}{3} G(1;r) G(0,-1;r)
   \right.\el\hspace{-24.7ex}\left.
   -\frac{256}{9} G(0,-1;r)+\frac{1}{r+1}\left(\frac{4}{3} G(0;r)^2-\frac{8}{3} G(-1;r) G(0;r)+\frac{8}{3} G(1;r) G(0;r)+\frac{52}{9} G(0;r)
   \right.\right.\el\hspace{-24.7ex}\left.\left.
   +\frac{8}{3} G(-1;r)+\frac{8}{3}
   G(1;r)-\frac{16}{3} G(0,1;r)-\frac{8 \pi ^2}{9}-\frac{4}{9}\right)+\frac{1}{(1-r)^2}\left(-\frac{4}{3} G(0;r)^2
   +\frac{40}{3} G(-1;r)
\right.\right.\el\hspace{-24.7ex}\left.\left.
   \times G(0;r)-\frac{8}{3} G(1;r) G(0;r)-\frac{52}{9} G(0;r)-\frac{64}{3} G(0,-1;r)-\frac{16}{3}
   G(0,1;r)
  +\frac{8 \pi ^2}{9}\right)
   \right.\el\hspace{-24.7ex}\left.
   +\frac{128}{9} G(0;r) G(-1;r)+\frac{64}{3} G(0;r) G(0,1;r)+\frac{128}{3} G(1;r) G(0,1;r)-\frac{256}{9} G(0,1;r)
   \right.\el\hspace{-24.7ex}\left.
   +\frac{1}{(r+1)^2}\bigg(-\frac{4}{3} G(0;r)^2+\frac{8}{3} G(-1;r) G(0;r)-\frac{8}{3} G(1;r) G(0;r)-\frac{4}{9}
   G(0;r)
  +\frac{16}{3} G(0,1;r)
   \right.\el\hspace{-24.7ex}\left.
 +\frac{8 \pi ^2}{9}\bigg)
   +\frac{1}{1-r}\left(\frac{4}{3} G(0;r)^2-\frac{40}{3} G(-1;r) G(0;r)
  +\frac{8}{3} G(1;r) G(0;r)+\frac{148}{9} G(0;r)
   \right.\right.\el\hspace{-24.7ex}\left.\left.
   -8 G(-1;r) -8 G(1;r)+\frac{64}{3}
   G(0,-1;r)+\frac{16}{3} G(0,1;r)-\frac{8 \pi ^2}{9}-\frac{4}{9}\right)+\frac{32}{3} G(-1,0,-1;r)
   \right.\el\hspace{-24.7ex}\left.
   +64 G(-1,0,1;r)-\frac{128}{3} G(0,0,-1;r)-\frac{128}{3} G(0,0,1;r)-\frac{256}{3} G(0,1,1;r)+\frac{8}{9}\right]
\,.
\label{eq:resultinin}
\eea
From Eq.\ \eqref{eq:secdecompinout} we obtain with the methods of Section~\ref{sec:coprod}
\bea
\bar{K}_{TC}^{{\rm in-out}\,(2)}(\tauo,\omega,r,\mu) &=&
C_A C_F \left[\left(-32 G(0,-1;r)-32 G(0,1;r)-\frac{8 \pi ^2}{3}\right) \ln^2 \left(\frac{Q \tau _{\omega }}{2 \omega }\right)
\right.\el\hspace{-25ex}\left.
+\left(-\frac{32}{3} G(-1;r)^3+32 G(0;r) G(-1;r)^2-64 G(0,-1;r) G(-1;r)-32 G(0;r) G(1;r)^2
\right.\right.\el\hspace{-25ex}\left.\left.
-64 G(0;r) G(-1,-1;r)+32 G(0;r) G(0,-1;r)+64 G(1;r) G(0,1;r)+64 G(0;r) G(1,1;r)
\right.\right.\el\hspace{-25ex}\left.\left.
+64 G(-1,-1,-1;r)+64 G(-1,0,-1;r)+64
   G(0,-1,-1;r)-96 G(0,0,-1;r)
\right.\right.\el\hspace{-25ex}\left.\left.
   -32 G(0,0,1;r)-64 G(0,1,1;r)-64 G(1,0,1;r)-16 \zeta _3\vphantom{-\frac{32}{3} G(-1;r)^3}\right) \ln \left(\frac{Q \tau _{\omega }}{2 \omega }\right)
  +\left(\vphantom{\frac{16 \pi ^2}{3}}64 G(0,-1;r)
  \right.\right.\el\hspace{-25ex}\left.\left.
   +64 G(0,1;r)+\frac{16 \pi ^2}{3}\right) \ln^2
   \left(\frac{\mu }{Q \tau _{\omega }}\right)+\left(\vphantom{\frac{16 \pi ^2}{3}}64 G(0,-1;r)+64 G(0,1;r)
  +\frac{16 \pi ^2}{3}\right) \ln^2 \left(\frac{\mu }{2 \omega }\right) 
  \right.\el\hspace{-25ex}\left.
  +\left(-\frac{64 r^2 G(0;r)}{3
   \left(r^2-1\right)^2}+\frac{64}{3} G(-1;r)^3-32 G(0;r) G(-1;r)^2
   -\frac{352}{3} G(0;r) G(-1;r)
   \right.\right.\el\hspace{-25ex}\left.\left.
   +128 G(0,-1;r) G(-1;r)-\frac{352}{3} G(0;r) G(1;r)+\frac{704}{3} G(0,1;r)
   +\frac{704}{3} G(0,-1;r)
  \right.\right.\el\hspace{-25ex}\left.\left.
   +64 G(0;r) G(0,1;r)-128
   G(1;r) G(0,1;r)+64 G(0;r) G(-1,-1;r)
   -128 G(-1,-1,-1;r)
   \right.\right.\el\hspace{-25ex}\left.\left.
   -128 G(-1,0,-1;r)-128 G(0,-1,-1;r)+64 G(0,0,-1;r)
  -64 G(0,0,1;r)
   \right.\right.\el\hspace{-25ex}\left.\left.
   +128 G(0,1,1;r)+128 G(1,0,1;r)+\frac{16 \left(r^2+1\right)}{3 \left(r^2-1\right)}+\frac{176
   \pi ^2}{9}\right) \ln \left(\frac{\mu }{2 \omega }\right)
   -16 \zeta _3 G(0;r)
   \right.\el\hspace{-25ex}\left.
   +\frac{88}{3} G(0;r) G(-1;r)^2+\frac{32}{3} G(0,-1;r) G(-1;r)^2-\frac{176}{3} G(0;r)^2 G(-1;r)
  +\frac{536}{9} G(0;r) 
    \right.\el\hspace{-25ex}\left.
   \times G(-1;r)+\frac{176}{3} G(0;r) G(1;r) G(-1;r)-\frac{176}{3} G(0,-1;r) G(-1;r)
  -\frac{352}{3} G(0,1;r) 
    \right.\el\hspace{-25ex}\left.
   \times G(-1;r)-\frac{64}{3} G(-1,0,-1;r) G(-1;r)+128 G(0,0,1;r) G(-1;r)
   +\frac{88}{3} G(0;r)
    \right.\el\hspace{-25ex}\left.
   \times  G(1;r)^2+8 G(0,-1;r)^2-32 G(0,1;r)^2+\frac{88}{9} \pi ^2 G(0;r)-\frac{8}{3} G(0;r)
   -\frac{176}{3} G(0;r)^2
    \right.\el\hspace{-25ex}\left.
  \times G(1;r)
  +\frac{536}{9} G(0;r) G(1;r)-\frac{8}{3} G(1;r)+\frac{1}{r+1}\left(\frac{8}{3} G(0;r)^2
   -\frac{8}{3} G(-1;r) G(0;r)
  \right.\right.\el\hspace{-25ex}\left.\left.
   -\frac{8}{3} G(1;r) G(0;r)-\frac{248}{9} G(0;r)+\frac{8}{3} G(-1;r)+\frac{8}{3}G(1;r)+\frac{104}{9}\right)
  +\frac{1}{1-r}\left(\frac{8}{3} G(0;r)^2
  \right.\right.\el\hspace{-25ex}\left.\left.
   -\frac{8}{3} G(-1;r) G(0;r)-\frac{8}{3} G(1;r) G(0;r)+\frac{40}{9} G(0;r)+\frac{8}{3} G(-1;r)
   +\frac{8}{3} G(1;r)+\frac{8}{9}\right)
   \right.\el\hspace{-25ex}\left.
   +\frac{1}{(1-r)^2}\left(-\frac{8}{3}
   G(0;r)^2+\frac{8}{3} G(-1;r) G(0;r)+\frac{8}{3} G(1;r) G(0;r)
   +\frac{8}{9} G(0;r)\right)
   \right.\el\hspace{-25ex}\left.
   +\frac{1}{(r+1)^2}\left(-\frac{8}{3} G(0;r)^2+\frac{8}{3} G(-1;r) G(0;r)+\frac{8}{3} G(1;r) G(0;r)
   +\frac{200}{9} G(0;r)\right)
   \right.\el\hspace{-25ex}\left.
   +24 G(0;r)^2 G(0,-1;r)+\frac{352}{3} G(0;r) G(0,-1;r)+\frac{8}{3} \pi ^2 G(0,-1;r)
   +32 G(0;r)^2 G(0,1;r)
   \right.\el\hspace{-25ex}\left.
   +\frac{352}{3} G(0;r) G(0,1;r)-64 G(0,-1;r) G(0,1;r)-\frac{8}{3} \pi ^2
   G(0,1;r)
   -\frac{1072}{9} G(0,1;r)
  \right.\el\hspace{-25ex}\left.
   +\frac{176}{3} G(-1,0,-1;r)+\frac{352}{3} G(-1,0,1;r)-144 G(0;r) G(0,0,-1;r)
   -\frac{352}{3} G(0,0,-1;r)
   \right.\el\hspace{-25ex}\left.
   -160 G(0;r) G(0,0,1;r)-\frac{352}{3} G(0,0,1;r)-\frac{352}{3}
   G(0,1,1;r)
   +\frac{64}{3} G(-1,-1,0,-1;r)
  \right.\el\hspace{-25ex}\left.
   -128 G(-1,0,0,1;r)-80 G(0,-1,0,-1;r)-96 G(0,-1,0,1;r)
   +336 G(0,0,0,-1;r)
   \right.\el\hspace{-25ex}\left.
   +336 G(0,0,0,1;r)+96 G(0,0,1,1;r)+64 G(0,1,1,1;r)-{8\over 3} G(-1;r)
   -\frac{1072}{9} G(0,-1;r)
   \right.\el\hspace{-25ex}\left.
   +\frac{88 \zeta _3}{3}+\frac{34 \pi ^4}{45}-\frac{268 \pi ^2}{27}-\frac{32}{9}
   +\mbox{\large $\Lambda$}_{C_A}\left(\frac{Q \tau _{\omega }}{2 \omega },r\vphantom{\frac{Q \tau _{\omega }}{2 \omega }}\right)
   \right]
   \el\hspace{-25ex}
   +C_F n_f T_F\left[ 
   \left(\frac{128 r^2 G(0;r)}{3 \left(r^2-1\right)^2}+\frac{128}{3} G(-1;r) G(0;r)+\frac{128}{3} G(0;r) G(1;r)
    -\frac{256}{3} G(0,-1;r) 
   \right.\right.\el\hspace{-25ex}\left.\left.
   -\frac{256}{3} G(0,1;r)-\frac{32 \left(r^2+1\right)}{3 \left(r^2-1\right)}-\frac{64
   \pi ^2}{9}\right) \ln \left(\frac{\mu }{2 \omega }\right) - \frac{32}{3} G(0;r) G(-1;r)^2
   +\frac{64}{3} G(0;r)^2 
  \right.\el\hspace{-25ex}\left.
   \times G(-1;r)
   -\frac{256}{9} G(0;r) G(-1;r)
   -\frac{64}{3} G(0;r) G(1;r) G(-1;r)
   +\frac{64}{3} G(0,-1;r) G(-1;r)
   \right.\el\hspace{-25ex}\left.
   +\frac{128}{3} G(0,1;r) G(-1;r)+\frac{16}{3}
   G(-1;r)-\frac{32}{3} G(0;r) G(1;r)^2
   -\frac{32}{9} \pi ^2 G(0;r)+\frac{16}{3} G(0;r)
  \right.\el\hspace{-25ex}\left.
   +\frac{64}{3} G(0;r)^2 G(1;r)-\frac{256}{9} G(0;r) G(1;r)+\frac{16}{3} G(1;r)
   +\frac{1}{(r+1)^2}\left(\frac{16}{3} G(0;r)^2
   \right.\right.\el\hspace{-25ex}\left.\left.
    -\frac{16}{3} G(-1;r) G(0;r)
    -\frac{16}{3} G(1;r) G(0;r)-\frac{376}{9} G(0;r)\right)
    +\frac{1}{(1-r)^2}\left(\frac{16}{3} G(0;r)^2
   \right.\right.\el\hspace{-25ex}\left.\left.
  -\frac{16}{3} G(-1;r) G(0;r)-\frac{16}{3} G(1;r) G(0;r)+\frac{8}{9} G(0;r)\right)
   +\frac{1}{1-r}\left(-\frac{16}{3}
   G(0;r)^2
   \right.\right.\el\hspace{-25ex}\left.\left.
   +\frac{16}{3} G(-1;r) G(0;r)+\frac{16}{3} G(1;r) G(0;r)-\frac{104}{9} G(0;r)
   -\frac{16}{3} G(-1;r)-\frac{16}{3} G(1;r)+\frac{8}{9}\right)
   \right.\el\hspace{-25ex}\left.
   +\frac{1}{r+1}\left(-\frac{16}{3} G(0;r)^2+\frac{16}{3} G(-1;r) G(0;r)+\frac{16}{3}
   G(1;r) G(0;r)+\frac{472}{9} G(0;r)
   \right.\right.\el\hspace{-25ex}\left.\left.
  -\frac{16}{3} G(-1;r)
   -\frac{16}{3} G(1;r)-\frac{184}{9}\right)-\frac{128}{3} G(0;r) G(0,-1;r)
   +\frac{512}{9} G(0,-1;r)
   \right.\el\hspace{-25ex}\left.
  -\frac{128}{3} G(0;r) G(0,1;r)
   +\frac{512}{9} G(0,1;r)-\frac{64}{3}
   G(-1,0,-1;r)-\frac{128}{3} G(-1,0,1;r)
   \right.\el\hspace{-25ex}\left.
  +\frac{128}{3} G(0,0,-1;r)
   +\frac{128}{3} G(0,0,1;r)+\frac{128}{3} G(0,1,1;r)-\frac{32 \zeta _3}{3}+\frac{128 \pi ^2}{27}+\frac{40}{9}
   \right.\el\hspace{-25ex}\left.
   +\mbox{\large $\Lambda$}_{n_f}\left(\frac{Q \tau _{\omega }}{2 \omega },r\vphantom{\frac{Q \tau _{\omega }}{2 \omega }}\right)
\right]
\,,
\label{eq:resultinout}
\eea
where $\mbox{\large $\Lambda$}_{C_A}\left(x,r\right)$ and $\mbox{\large $\Lambda$}_{n_f}\left(x,r\right)$ are defined in Appendix \ref{app:chi}.
Finally, from Eq.\ \eqref{eq:secdecompoutout} we obtain with the methods of Section~\ref{sec:coprod}
\bea
\bar{K}_{TC}^{{\rm all-out}\,(2)}(\tauo,\omega,r,\mu) &=& 
C_A C_F \left[\frac{128}{3} G(0;r)\ln^3 \left(\frac{\mu }{2\omega }\right)+\left(-32 G(0;r)^2+\frac{176}{3} G(0;r)
\right.\right.\el\hspace{-25.7ex}\left.\left.
+64 G(0,-1;r)-64 G(0,1;r)-16 \pi ^2\vphantom{-{32\over3} G(0;r)^2}\right) \ln^2 \left(\frac{\mu }{2 \omega }\right) +
   \left(\frac{32}{3} G(0;r)^3-\frac{88}{3} G(0;r)^2
   \right.\right.\el\hspace{-25.7ex}\left.\left.
   +\frac{176}{3} G(-1;r) G(0;r)+\frac{176}{3} G(1;r) G(0;r)-32 G(0,-1;r) G(0;r)+32 G(0,1;r) G(0;r)
   \right.\right.\el\hspace{-25.7ex}\left.\left.
   +\frac{32 r^2 G(0;r)}{3 \left(r^2-1\right)^2}-\frac{16}{3} \pi ^2
   G(0;r)+\frac{536}{9} G(0;r)+\frac{8 r^2+8}{3-3 r^2}-\frac{352}{3} G(0,1;r)+64 G(0,-1,-1;r)
   \right.\right.\el\hspace{-25.7ex}\left.\left.
   -64 G(0,-1,1;r)-64 G(0,0,-1;r)-64 G(0,0,1;r)-64 G(0,1,-1;r)+64 G(0,1,1;r)
   \right.\right.\el\hspace{-25.7ex}\left.\left.
   -128 \zeta _3-\frac{176 \pi^2}{9}\right)\ln \left(\frac{\mu }{2 \omega }\right)
  -\frac{4}{3} G(0;r)^4+\frac{44}{9} G(0;r)^3-44 G(-1;r) G(0;r)^2
   -44 G(1;r)
  \right.\el\hspace{-25.7ex}\left.
 \times G(0;r)^2+16 G(0,-1;r) G(0;r)^2
 +8 G(0,1;r) G(0;r)^2+\frac{4}{3} \pi ^2 G(0;r)^2
 -\frac{134}{9} G(0;r)^2
   \right.\el\hspace{-25.7ex}\left.
 +\frac{44}{3} G(-1;r)^2 G(0;r)+\frac{44}{3} G(1;r)^2 G(0;r)+\frac{268}{9} G(-1;r) G(0;r)
 +\frac{88}{3} G(-1;r) G(1;r) 
 \right.\el\hspace{-25.7ex}\left.
 \times G(0;r)+\frac{268}{9} G(1;r) G(0;r)+\frac{352}{3} G(0,-1;r) G(0;r)
 +\frac{352}{3} G(0,1;r) G(0;r)
    \right.\el\hspace{-25.7ex}\left.
   -16 G(0,0,-1;r) G(0;r)-16 G(0,0,1;r) G(0;r)-\frac{116}{3} \zeta _3 G(0;r)
   -64 G(0,1,1;r) G(0;r)
  \right.\el\hspace{-25.7ex}\left.
   +\frac{11}{9} \pi ^2 G(0;r)+\frac{916}{27} G(0;r)-32 G(0,-1;r)^2+32 G(0,1;r)^2
   -\frac{88}{9} \pi ^2 G(-1;r)
   -\frac{4}{3} G(-1;r)
   \right.\el\hspace{-25.7ex}\left.
   -\frac{88}{9} \pi ^2
   G(1;r)+\frac{4}{3} G(1;r)+\frac{16}{3} G(-1;r)^2 G(0,-1;r)
   -\frac{4}{3} \pi ^2 G(0,-1;r)
   \right.\el\hspace{-25.7ex}\left.
   +\frac{1}{(1-r)^2}\left(-2 G(0;r)^2+\frac{4}{3} G(-1;r) G(0;r)+\frac{4}{3} G(1;r) G(0;r)
   -\frac{8}{9} G(0;r)-\frac{8}{3} G(0,1;r)
   \right.\right.\el\hspace{-25.7ex}\left.\left.
   -\frac{4 \pi^2}{9}\right)+\frac{1}{(r+1)^2}\left(-2 G(0;r)^2+\frac{4}{3} G(-1;r) G(0;r)
   +\frac{4}{3} G(1;r) G(0;r)+\frac{16}{9} G(0;r)
   \right.\right.\el\hspace{-25.7ex}\left.\left.
   -\frac{8}{3} G(0,1;r)-\frac{4 \pi ^2}{9}\right)
   -\frac{176}{3} G(-1;r) G(0,1;r)
   -\frac{176}{3} G(1;r)
   G(0,1;r)+32 G(0,-1;r)
    \right.\el\hspace{-25.7ex}\left.
   \times G(0,1;r) +12 \pi ^2 G(0,1;r)-\frac{536}{9} G(0,1;r)
   +\frac{1}{r+1}\left(2 G(0;r)^2-\frac{4}{3} G(-1;r) G(0;r)
   \right.\right.\el\hspace{-25.7ex}\left.\left.
   -\frac{4}{3} G(1;r) G(0;r)-\frac{40}{9} G(0;r)+\frac{4}{3} G(-1;r)
   -\frac{4}{3} G(1;r)+\frac{8}{3}
   G(0,1;r)+\frac{4 \pi ^2}{9}+\frac{16}{9}\right)
   \right.\el\hspace{-25.7ex}\left.
  +\frac{1}{1-r}\left(2 G(0;r)^2-\frac{4}{3} G(-1;r) G(0;r)
 -\frac{4}{3} G(1;r) G(0;r)-\frac{16}{9} G(0;r)+\frac{4}{3} G(-1;r)   
 \right.\right.\el\hspace{-25.7ex}\left.\left.
   -\frac{4}{3} G(1;r)+\frac{8}{3} G(0,1;r)+\frac{4 \pi
   ^2}{9}+\frac{16}{9}\right)
   -\frac{32}{3} G(-1;r) G(-1,0,-1;r)
  -176 G(0,0,-1;r)
 \right.\el\hspace{-25.7ex}\left.
   -64 G(-1;r) G(0,0,1;r)
   -\frac{352}{3} G(0,0,1;r)+\frac{352}{3} G(0,1,1;r)+\frac{32}{3} G(-1,-1,0,-1;r)
   \right.\el\hspace{-25.7ex}\left.
  +64 G(-1,0,0,1;r)
   -32 G(0,-1,-1,1;r)+32 G(0,-1,0,-1;r)+32 G(0,-1,0,1;r)
   \right.\el\hspace{-25.7ex}\left.
  -32 G(0,-1,1,-1;r) +32 G(0,-1,1,1;r)+16 G(0,0,0,-1;r)-32 G(0,1,-1,-1;r)
   \right.\el\hspace{-25.7ex}\left.
   +32 G(0,1,-1,1;r)+32 G(0,1,1,-1;r)-32 G(0,1,1,1;r)
   -\frac{308 \zeta _3}{3}+\frac{34 \pi ^4}{45}-\frac{268 \pi ^2}{27}
   \right.\el\hspace{-25.7ex}\left.
  -\frac{28}{9}\right]
   \el\hspace{-25.7ex}
   +C_F n_f T_F\left[-\frac{64}{3} G(0;r) \ln^2 \left(\frac{\mu }{2 \omega }\right) + \left(\frac{32}{3} G(0;r)^2-\frac{64 r^2 G(0;r)}{3 \left(1-r^2\right)^2}-\frac{64}{3} G(-1;r) G(0;r)
   \right.\right.\el\hspace{-25.7ex}\left.\left.
   -\frac{160}{9} G(0;r)-\frac{64}{3} G(0;r) G(1;r)+\frac{128}{3} G(0,1;r)-\frac{16 \left(r^2+1\right)}{3
   \left(1-r^2\right)}+\frac{64 \pi ^2}{9}\right) \ln \left(\frac{\mu }{2 \omega }\right)
   \right.\el\hspace{-25.7ex}\left.
   -\frac{16}{9} G(0;r)^3+16 G(-1;r) G(0;r)^2+16 G(1;r)
   G(0;r)^2+\frac{40}{9} G(0;r)^2-\frac{16}{3} G(-1;r)^2 G(0;r)
   \right.\el\hspace{-25.7ex}\left.
   -\frac{16}{3} G(1;r)^2 G(0;r)-\frac{80}{9} G(-1;r) G(0;r)-\frac{32}{3} G(-1;r) G(1;r) G(0;r)-\frac{80}{9} G(1;r) G(0;r)
   \right.\el\hspace{-25.7ex}\left.
   -\frac{128}{3} G(0,-1;r)
   G(0;r)-\frac{128}{3} G(0,1;r) G(0;r)-\frac{440}{27} G(0;r)+\frac{8}{3} G(-1;r)-\frac{8}{3} G(1;r)
   \right.\el\hspace{-25.7ex}\left.
   + \frac{32\pi ^2}{9} G(-1;r)-\frac{4\pi ^2}{9} G(0;r)+\frac{32\pi ^2}{9} G(1;r)+\frac{1}{1-r}\left(-4 G(0;r)^2+\frac{8}{3}G(-1;r) G(0;r)
   \right.\right.\el\hspace{-25.7ex}\left.\left.
   +\frac{8}{3} G(1;r) G(0;r)+\frac{44}{9} G(0;r)-\frac{8}{3} G(-1;r)+\frac{8}{3} G(1;r)-\frac{16}{3} G(0,1;r)-\frac{8 \pi ^2}{9}-\frac{44}{9}\right)
   \right.\el\hspace{-25.7ex}\left.
   +\frac{1}{r+1}\left(-4 G(0;r)^2+\frac{8}{3} G(-1;r) G(0;r)+\frac{8}{3} G(1;r) G(0;r)+\frac{92}{9} G(0;r)-\frac{8}{3} G(-1;r)
    \right.\right.\el\hspace{-25.7ex}\left.\left.
   +\frac{8}{3} G(1;r)-\frac{16}{3} G(0,1;r)-\frac{8 \pi ^2}{9}-\frac{44}{9}\right)+\frac{64}{3} G(-1;r) G(0,1;r)+\frac{64}{3} G(1;r)
   G(0,1;r)
    \right.\el\hspace{-25.7ex}\left.
   +\frac{160}{9} G(0,1;r)+\frac{1}{(r+1)^2}\left(4 G(0;r)^2-\frac{8}{3} G(-1;r) G(0;r)-\frac{8}{3} G(1;r) G(0;r)-\frac{44}{9} G(0;r)
   \right.\right.\el\hspace{-25.7ex}\left.\left.
   +\frac{16}{3} G(0,1;r)+\frac{8 \pi ^2}{9}\right) +\frac{1}{(1-r)^2}\left(4 G(0;r)^2-\frac{8}{3} G(-1;r)
   G(0;r)-\frac{8}{3} G(1;r) G(0;r)
    \right.\right.\el\hspace{-25.7ex}\left.\left.
   +\frac{4}{9} G(0;r)+\frac{16}{3} G(0,1;r)+\frac{8 \pi ^2}{9}\right)+64 G(0,0,-1;r)+\frac{128}{3} G(0,0,1;r)-\frac{128}{3} G(0,1,1;r)
   \right.\el\hspace{-25.7ex}\left.
   +\frac{112 \zeta _3}{3}+\frac{80 \pi
   ^2}{27}+\frac{68}{9}\right]\,.
\label{eq:resultoutout}
\eea

\section{Known One- and Two-Loop Results}
\label{app:refeqs}

The hard function, $H(Q^2, \mu)$, is related to the finite part of the time-like quark-antiquark form factor normalized to one at tree level. To two-loop accuracy, $H(Q^2, \mu)$ is given by
\bea
\label{eq:hardfunc}
H(Q^2, \mu) &=& 1 + C_F \left[-2 \ln ^2\left(\frac{\mu^2 }{Q^2}\right)-6 \ln \left(\frac{\mu^2 }{Q^2}\right)+\frac{7 \pi ^2}{3}-16\right]\left({\als\over 4\pi}\right) 
\el
+ \Bigg\{ C_F^2\left[2 \ln ^4\left(\frac{\mu^2 }{Q^2}\right)+12 \ln^3\left(\frac{\mu^2 }{Q^2}\right)+\left(50-\frac{14 \pi ^2}{3}\right) \ln ^2\left(\frac{\mu^2 }{Q^2}\right)
\right.\el\left.
+\left(93 - 10 \pi ^2 - 48 \zeta_3 \right) \ln \left(\frac{\mu^2 }{Q^2}\right)-60 \zeta _ 3+\frac{67 \pi ^4}{30}-\frac{83 \pi^2}{3}+\frac{511}{4}\right]
\el
+C_A C_F \left[-\frac{22}{9} \ln ^3\left(\frac{\mu^2 }{Q^2}\right)+\left(\frac{2\pi ^2}{3}  -\frac{233}{9}\right) \ln ^2\left(\frac{\mu^2 }{Q^2}\right)
\right.\el\left.
+\left(\frac{44\pi ^2}{9}  - \frac{2545}{27} + 52 \zeta_3 \right) \ln \left(\frac{\mu^2 }{Q^2}\right)+\frac{626 \zeta_3}{9}-\frac{8 \pi ^4}{45}+\frac{1061 \pi^2}{54}-\frac{51157}{324}\right]
\el
+C_F n_f T_F \left[\frac{8}{9} \ln ^3\left(\frac{\mu^2 }{Q^2}\right)+\frac{76}{9} \ln ^2\left(\frac{\mu^2 }{Q^2}\right)+\left(-\frac{16\pi ^2}{9}  + \frac{836}{27} \right) \ln \left(\frac{\mu^2 }{Q^2}\right)
\right.\el\left.
+\frac{8 \zeta_3}{9}-\frac{182 \pi ^2}{27}+\frac{4085}{81}\right]\Bigg\}\left({\als\over 4\pi}\right)^2 + {\Ordals 3}\,.
\eea
Although this result was derived long ago~\cite{Gonsalves:1983nq,Kramer:1986sg,Matsuura:1987wt}, reference~\cite{Becher:2008cf}
presents the one- and two-loop results in a way which makes the validation of Eq.\ (\ref{eq:hardfunc}) particularly straightforward.

To two-loop accuracy, the finite part of the fully-inclusive integrated jet function is given by
\bea
\label{eq:intjetfunc}
{\mathbf j}(q^2, \mu) &=& 1 + C_F \Big[2 \ln ^2\left(\frac{\mu^2}{q^2}\right)+3 \ln \left(\frac{\mu^2}{q^2}\right)-\pi ^2+7\Big] \left({\als\over 4\pi}\right)
\el
+ \Bigg\{ C_F^2 \left[2 \ln^4\left(\frac{\mu ^2}{q^2}\right)+6 \ln ^3\left(\frac{\mu ^2}{q^2}\right)+\left(\frac{37}{2}-\frac{10 \pi ^2}{3}\right) \ln ^2\left(\frac{\mu^2}{q^2}\right)
\right.\el\left.
+\left(8 \zeta _ 3-7 \pi^2+\frac{45}{2}\right) \ln \left(\frac{\mu ^2}{q^2}\right)-18 \zeta _ 3+\frac{14 \pi ^4}{15}-\frac{67 \pi^2}{6}+\frac{205}{8}\right]
\el
+C_A C_F \left[\frac{22}{9} \ln ^3\left(\frac{\mu ^2}{q^2}\right)+\left(\frac{367}{18}-\frac{2 \pi ^2}{3}\right) \ln^2\left(\frac{\mu ^2}{q^2}\right)
\right.\el\left.
+\left(-40 \zeta _ 3-\frac{22 \pi^2}{9}+\frac{3155}{54}\right) \ln \left(\frac{\mu ^2}{q^2}\right)-\frac{206 \zeta _ 3}{9}-\frac{17 \pi ^4}{180}-\frac{208 \pi^2}{27}+\frac{53129}{648}\right]
\el
+C_F n_f T_F \left[-\frac{8}{9} \ln^3\left(\frac{\mu ^2}{q^2}\right)-\frac{58}{9} \ln ^2\left(\frac{\mu^2}{q^2}\right)+\left(\frac{8 \pi ^2}{9}-\frac{494}{27}\right) \ln\left(\frac{\mu ^2}{q^2}\right)
\right.\el\left.
+\frac{16 \zeta _ 3}{9}+\frac{68 \pi^2}{27}-\frac{4057}{162}\right]\Bigg\}\left({\als\over 4\pi}\right)^2 + {\Ordals 3}\,.
\eea
This result can be derived by starting with the expression for the bare two-loop integrated jet function, $j(q^2, \mu)$, given in reference~\cite{Becher:2006qw}.
As mentioned in the introduction, the thrust cone jet function essentially reduces to the fully-inclusive jet function in the limit of small $\tauo$ and $\omega/Q$ where the event shape distribution factorizes. 
It is worth emphasizing that this is why it is the fully-inclusive jet function that appears in factorization formula (\ref{eq:factauo1}).

Finally, to two-loop accuracy, the cusp anomalous dimension is~\cite{Kodaira:1981nh}
\be
\G = \G_0 \left({\als\over 4\pi}\right) + \G_1 \left({\als\over 4\pi}\right)^2 + {\Ordals 3}\,,
\ee
where
\bea
\G_0 &=& 4 C_F\,,\\
\G_1 &=& \left(-\frac{4 \pi^2}{3}+\frac{268}{9}\right) C_A C_F
-\frac{80}{9} C_F n_f T_F\,.
\eea 
\bibliographystyle{JHEP3}
\bibliography{tauomega}
\end{document}